\documentclass[aps,twocolumn,showpacs,preprintnumbers,prl,amsmath,amssymb,amsfonts,superscriptaddress,floatfix]{revtex4-1}

\usepackage[latin1]{inputenc}
\usepackage{graphics}
\usepackage{color}
\usepackage{amssymb}
\usepackage{amsmath}
\usepackage{overpic}
\usepackage{epstopdf}
\usepackage{bm}
\usepackage{graphicx}
\usepackage{subfigure}
\usepackage{hyperref}
\usepackage{url}

\begin{document}

\title{Functionality-directed Screening of Pb-free Hybrid Organic-inorganic Perovskites with Desired Intrinsic Photovoltaic Functionalities}

\author{Dongwen Yang}
\altaffiliation{These authors contributed equally.}
\affiliation{College of Materials Science and Engineering and Key Laboratory of Automobile Materials of MOE, Jilin University, Changchun 130012, China}
\author{Jian Lv}
\altaffiliation{These authors contributed equally.}
\affiliation{College of Materials Science and Engineering and Key Laboratory of Automobile Materials of MOE, Jilin University, Changchun 130012, China}
\author{Xingang Zhao}
\author{Qiaoling Xu}
\author{Yuhao Fu}
\affiliation{College of Materials Science and Engineering and Key Laboratory of Automobile Materials of MOE, Jilin University, Changchun 130012, China}
\author{Yiqiang Zhan}
\affiliation{State Key Laboratory of ASIC and System, Department of Microelectronics, SIST, Fudan University, Shanghai 200433, China}
\author{Alex Zunger}
\email{Corresponding author: alex.zunger@gmail.com}
\affiliation{University of Colorado and Renewable and Sustainable Energy Institute, Boulder, Colorado 80309, USA}
\author{Lijun Zhang}
\email{Corresponding author: lijun\_zhang@jlu.edu.cn}
\affiliation{College of Materials Science and Engineering and Key Laboratory of Automobile Materials of MOE, Jilin University, Changchun 130012, China}

\date{\today}

\begin{abstract}
The material class of hybrid organic-inorganic perovskites has risen rapidly 
from a virtually unknown material in photovoltaic applications a short 7 years ago into a $\sim$ 20\% 
efficient thin-film solar cell material. 
As promising as this class of materials is, however, 
there are limitations associated with its poor long-term stability, 
non-optimal band gap, presence of environmentally-toxic Pb element, etc. 
We herein apply a functionality-directed theoretical materials selection approach 
as a filter for initial screening of the compounds that satisfy 
the desired intrinsic photovoltaic functionalities and 
might overcome the above limitations. 
First-principles calculations are employed to systemically study thermodynamic stability and 
photovoltaic-related properties of hundred of candidate hybrid perovskites. 
We have identified in this materials selection process fourteen Ge and Sn-based materials 
with potential superior bulk-material-intrinsic photovoltaic performance. 
A distinct class of compounds containing NH$_3$COH$^+$ with the organic molecule 
derived states intriguingly emerging at band-edges is found. 
Comparison of various candidate materials offers insights on 
how composition variation and microscopic structural changes 
affect key photovoltaic relevant properties in this family of materials.
\end{abstract}

\maketitle

\noindent{\textbf{1. Introduction}}
\vspace{0.5cm}

The material class of hybrid organic-inorganic halide perovskites AM$^{IV}$X$^{VII}_3$, 
where A is a small organic molecule, M$^{IV}$ is a low-valent (+2) group-IVA metalloid 
and X$^{VII}$ is a halogen, 
has recently become a rising star in the field of solar-energy conversion materials. 
With intense research efforts in the past few years,\cite{kojima2009organometal, lee2012efficient, 
heo2013efficient, liu2013efficient, burschka2013sequential, jeon2014methoxy, malinkiewicz2014perovskite, zhou2014interface, jeon2014solvent, jeon2015compositional, yang2015high} 
the power conversion efficiency of thin film photovoltaic (PV) solar cells 
based on this class of materials has rapidly progressed from the initial value of 3.8\% in 2009,\cite{kojima2009organometal} 
to a stellar high value of 22.1\% only 7 years later.\cite{picture2015} 
Although most of them are limited to very small cell area ($\sim$ 0.1-0.2 cm$^2$),\cite{sum2014advancements}
such efficiencies have become competitive with the record efficiency of the conventional thin film solar cells 
based on crystalline Si,\cite{scandale2007high, zeng2006efficiency} CdTe,\cite{wu2004high} Cu(In,Ga)Se$_2$,\cite{chirilua2011highly} etc. 
that have been studied for several decades. 
Quite recently, large area ($>$ 1 cm$^2$) solar cells have been fabricated showing conversion efficiency $>$ 15\%.
\cite{chen2015efficient} 
Furthermore, hybrid perovskites can be synthesized by inexpensive room temperature solution processing,
\cite{malinkiewicz2014perovskite}, \cite{jeon2014solvent}, \cite{liu2014perovskite, chen2013planar, docampo2013efficient}
making them ideal for low-cost commercial device applications.
\vspace{0.3cm}

\textbf{Limitations of the currently discovered compounds in solar cell applications.} 
Though exhibiting high power conversion efficiency, 
the AM$^{IV}$X$^{VII}_3$ family of solar materials is far from optimal, 
and there exist a number of scientific challenges and practical factors 
impeding their large-scale device application.\cite{green2014emergence, berry2015hybrid}
 These include (i) the devices made by the currently used materials show poor long-term stability under high temperatures and outdoor illumination, 
as well as in the presence of moisture or oxygen,\cite{niu2015review} 
which may be attributed to the intrinsic thermodynamic instability of the materials discovered thus far;\cite{zhang2015intrinsic, conings2015intrinsic} 
(ii) the use of Pb-containing materials is an environmental concern, 
therefore there is a strong desire to replace toxic Pb by a benign element 
without affecting drastically the conversion efficiency; 
(iii) some key properties affecting PV performance need to be further optimized, 
such as reducing the optical band gap of [CH$_3$NH$_3$]PbI$_3$ (1.5 eV)\cite{kim2012lead, stoumpos2013semiconducting}
 and [CH$_3$NH$_3$]PbBr$_3$ (2.35 eV)\cite{kitazawa2002optical} 
to the optimal value of 1.34 eV from the Shockley-Queisser limit,\cite{shockley1961detailed} 
further enhancing light capture efficiency by making the organic molecule contribute in some way to absorption in the solar range, 
turning material parameters to eliminate anomalous hysteresis in the current-voltage curves,\cite{snaith2014anomalous} etc.
\vspace{0.3cm}

\textbf{Can the current limitations be overcome by identifying other compounds from the same general group?}
 Identification of the alternative AM$^{IV}$X$^{VII}_3$ materials 
offers a possible solution to overcome above limitations. 
Given that this group of compounds is much broader (see Figure 1a) than 
the most common selection of A = CH$_3$NH$_3^{+}$, CH(NH$_2$)$_2^{+}$, M$^{IV}$ = Pb and X$^{VII}$ = I, Br,\cite{jeon2015compositional, yang2015high}, \cite{sum2014advancements}, \cite{green2014emergence}, 
\cite{niu2015review}, \cite{lee2014high, eperon2014formamidinium, koh2013formamidinium, loi2013hybrid, yin2015halide}
one also wonders if other members might have advantages 
and solve the problems encountered by the currently used ones. 
A few alternative materials have been experimentally synthesized and implemented into solar cells, 
but showing so far limited conversion efficiency.\cite{hao2014lead, safdari2015structure, noel2014lead} 
In this aspect, materials screening via computational simulations is 
of valuable help to avoid the expensive trial-and-error process of experimental laboratory exploration. 
To perform the materials screening non-phenomenologically, 
one needs to establish current understanding of what are the critical, 
materials-specific `design principles' (DPs) that render such materials superior in solar-energy conversion. 
To proceed beyond this one needs to formulate the `design metrics' (DMs), 
the embodiment of corresponding DPs, which consist of a series of computable or measurable quantities. 
In the area of chalcogenide PV materials, 
successful screening was recently accomplished by applying key properties oriented DMs, 
computed via first principles electronic structure theory.\cite{yu2012identification, yu2013inverse} 
In principle, the DMs are not limited to intrinsic materials properties closely relevant to PV performance such as light absorption, defect tolerance, 
carrier mobility, recombination lifetime, charge extraction, etc., 
but light capture engineering via larger films thickness or better design of optics 
(making Si a successful PV platform). However, 
the application of such methodologies to the hybrid organic-inorganic perovskite family 
has been historically overlooked, and is currently at its early stages.\cite{filip2014steric, filip2015computational,
castelli2014bandgap, zheng2015material}
\vspace{0.3cm}

Here we apply a functionality-directed material discovery approach 
to screen via systemic first-principles calculations potential 
high-performance PV hybrid AM$^{IV}$X$^{VII}_3$ perovskites. 
Such a non-statistical, science-based inverse-design approach has been recently 
applied successfully by some of us in theoretical identification 
and laboratory realization of previously unknown 18-electron Half-Heusler functional compounds.\cite{yan2015design, gautier2015prediction} 
We design the principles/filters that enable selection of the compounds with target photovoltaic performance. 
The data set of compounds considered for materials screening 
according to these criteria includes both existing compounds 
(many not considered yet for photovoltaic applications) as well as new compounds. 
Particularly, we first (i) formulate the relevant DPs and DMs 
that would guide us effectively to the compounds with desired optimal PV functionalities, 
(ii) search on this basis the promising solar hybrid perovskites 
satisfying the key compound-intrinsic DMs including thermodynamic stability, 
as well as the critical properties controlling PV performance such as band gaps, 
effective masses, dopability, excitons bonding, etc. 
Figure 1a shows the candidate materials space for screening, 
whereas Figure 1b provides a summary and conclusion of the screening process 
where red squares indicate the winning materials that survive the filter of the respective DM. 
In total, the materials space consists of about 100 compounds, 
of which as far as we know from literature just 22 ones were previously reported as being made 
and 15 have been reported as being considered for PV applications. 
(There is certainly the possibility that other materials have been experimented with but not reported). 
The goal of our materials screening is dual: 
to identify which materials own potentially superior photovoltaic performance, 
and at the same time to establish the physical-understanding-based principles that 
can help experimentalists comprehend their both successes and failures. 
The materials screening process we described led to identification of fourteen optimal AM$^{IV}$X$^{VII}_3$ 
perovskites, 
as highlighted in the last row of Figure 1b. 
By comparing various candidate materials, 
we offer new insights on how composition variation and 
microscopic structural changes affect key PV relevant properties. 
Additionally, we find an exotic class of materials with the organic molecule 
derived state emerging in proximity to band-edges; 
they are distinct from all the known materials where the molecular state lies far away from band-edges.
\vspace{0.3cm}

\begin{figure}[t]
\includegraphics[width=3.0in]{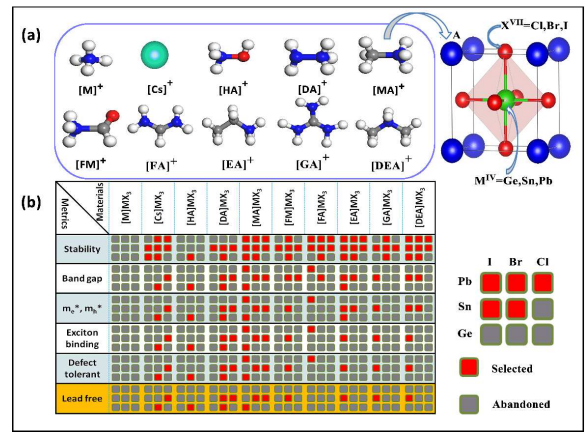}
\centering
\caption{
 (a) Constitution space of candidate AM$^{IV}$X$^{VII}_3$ perovskites 
for the materials screening, where ten cations (see the main text) 
were chosen for the A site, three group-IVA metalloids (Ge/Sn/Pb) for the M$^{IV}$ site, 
and three halogen (Cl/Br/I) anions for the X$^{VII}$ site. 
(b) Step-by-step screening process with the more and more DMs applied (different rows). 
Each column corresponds to one class of the 9 AM$^{IV}$X$^{VII}_3$ compounds with fixed A, 
whose arrangement coordinates are shown in the right panel. 
The red squares mean the materials passing the screening (Selected) 
and the gray ones mean the materials not passing (Abandoned). 
The last row of finally winning solar materials is highlighted in orange background.}
\end{figure}

We are aware of the fact that even though only 22 of the 100 compounds studied here 
have been previously reported in the published literature, 
it is likely that synthetic materials chemists have screened additional compounds 
that remain unpublished for a variety of possible reasons, 
including failures to produce good devices at current stage. 
We wish to emphasize that the purpose of our theoretical materials screening 
based on functionality-directed design principles is not just to inform experimentalists 
on which chemical formulas of such materials they should attempt synthetically next. 
Ever since the attractive PV performance of Pb-based hybrid perovskites was first noted, 
this field has progressed mostly by a trial-and-error Edisonian approach of combinatorially 
testing a large number of possible related compounds derived by substitution from a parent compound. 
Relative successes or failures were sometimes briefly discussed ex post facto. 
Our aim is actually also to develop understanding-based 
insights on the physical factors (i.e., the design metrics listed in Sec. 2) 
that control both success and failure of compound members in the hybrid perovskite family, 
whether they were already tried or new ones. 
We wish that such understanding-based tools can be used as effective guidelines 
on how to extend the results presented here to search of other PV material systems.
\vspace{0.5cm}

\noindent{\textbf{2. Design principles required for an optimum AM$^{IV}$X$^{VII}_3$ PV material 
and their embodiment as quantitative design metrics.}}
\vspace{0.5cm}

The DMs can be divided into (a) the metrics that characterize the AM$^{IV}$X$^{VII}_3$ material itself 
(`compound-intrinsic DMs'), as well as (b) the `device-level DMs' specific 
to the AM$^{IV}$X$^{VII}_3$ material configured within a solar cell device structure 
that includes components other than this material [such as the hole-blocking layer (HBL) 
and the electron-blocking layer (EBL), 
various interfaces and possible metal contacts, etc.]. 
An instance of the category (b) is the experimentally observed high efficiency 
of charge collection at the [CH$_3$NH$_3$]PbBr$_3$ absorber/HBL as well as the [CH$_3$NH$_3$]PbBr$_3$/EBL interfaces, 
creating a natural p-i-n device where a large fraction of 
the band gap energy E$_g$ is converted to the open-circuit voltage V$_{oc}$.\cite{edri2013high}
As important as this observation is, 
it however does not reveal why this special absorber material and 
no other has this superior charge separation in conjunction with the HBL and EBL components. 
We believe that it makes sense to use the compound-intrinsic DMs 
as the first filter for pre-selecting promising candidates. 
Afterward, the more practical device-level DMs should be used as subsequent filter, 
applied, however, 
only to a far smaller group of materials than those screened by the first filter.
\vspace{0.3cm}

We next discuss the basic understanding of the pertinent physical mechanisms 
controlling PV performance (DPs) that characterize the AM$^{IV}$X$^{VII}_3$ material itself, 
and the corresponding computable quantities (DMs) 
that will be used for screening procedures. 
These compound-intrinsic DMs are divided into those that 
correspond to the structurally ideal AM$^{IV}$X$^{VII}_3$ compounds 
[items (1)-(5) below], and those that relate to defects and other inhomogeneities 
embedded in the compounds [items (6)-(7)].
\vspace{0.3cm}

\textbf{(1) Thermodynamic immunity towards the main decomposition channel into competing phases.} 
The need for a stable structure is one of the first, most obvious conditions required. 
In the context of the AM$^{IV}$X$^{VII}_3$ perovskite family, 
there are at least two DPs pertaining to stability issues. 
First, thermodynamic stability requirement needs the eligible PV materials to be stable 
against all the possible decomposition pathways at solar cell working conditions. 
The most important decomposition channel of AM$^{IV}$X$^{VII}_3$ is AM$^{IV}$X$^{VII}_3$ 
$\to$  AX$^{VII}$ + M$^{IV}$X$^{VII}_2$. The more stable material, 
the larger positive magnitude of the decomposition enthalpy 
(energy difference between the decomposed products and AM$^{IV}$X$^{VII}_3$) is. 
We thus will use the computed decomposition enthalpy (Figure 2) 
as the DM here to screen stable compounds. 
Note that although the decomposition enthalpy is not a direct indication of the materials stability 
at high temperature, 
under illumination or moisture, etc. 
required for solar materials, it in principle relates to such stabilities. 
Additionally, apart from the thermodynamic stability, 
whether the perovskite materials are formed can be judged by the criterion of crystallographic stability. 
It can be expressed by two dimensionless parameters: 
the Goldschmidt tolerance factor t (t=(r$_A$+r$_X$)/$\sqrt2$(r$_M$+r$_X$)) 
and the octahedral factor $\mu$ ($\mu$=r$_M$/r$_X$), 
where r$_A$, r$_M$, r$_X$ are ionic radii of corresponding ions. 
Depending on temperature the AM$^{IV}$X$^{VII}_3$ family 
may be stabilized in a simple cubic perovskite or its distorted pseudocubic structure,\cite{baikie2013synthesis} 
and the t and $\mu$ describe the stabilities of cubic (or pseudocubic) perovskite framework and octahedral M$^{IV}$X$^{VII}_6$ units, respectively. 
To address this crystallographic stability, 
we will analyze all the candidate materials with respect to t and $\mu$ (Figure 3) 
and delineate whether and how feasibly they can be stabilized in the perovskite structure.
\vspace{0.3cm}

\textbf{(2) Direct band gap matching solar spectrum and strong absorption near threshold due to the p-p effect.} 
Given a stable material, the solar cell application requires a high solar-energy 
absorption efficiency, i.e., an optical band gap (absorption threshold) 
with the suitable magnitude of 1.0-1.7 eV and strong take-off 
(slope of absorption versus energy) near threshold. 
So far all the reported AM$^{IV}$X$^{VII}_3$ materials show strong optical transition 
at the optical band gaps.\cite{yin2015halide} 
The physical mechanism responsible for this condition is attributed to the existence 
of a valence band p (VB-p) to a conduction band p (CB-p) channel of interband absorption.\cite{yin2014unique} 
For the high-Z element M$^{IV}$, the M$^{IV}$-s state moves down in energy (forming lone-pair states), 
separating from M$^{IV}$-p state (Mass-Darwin relativistic effect).\cite{Dirac1992}
This makes the +2 low oxidation state of M$^{IV}$ favorable in the AM$^{IV}$X$^{VII}_3$ family. 
As the result, the conduction bands have M$^{IV}$-p character not s character, 
the valence bands are mainly from the anti-bonding hybridization 
between halogen X$^{VII}$-p and M$^{IV}$-s states (see Figure 5). 
Such a VB-p to CB-p band transition channel has the higher intensity 
(due to the larger degeneracy of p states than s states) than 
the VB-p to CB-s transitions characterizing most of conventional solar materials 
such as GaAs or CuInSe$_2$. 
This type of design principle has also been identified in the context of discovery 
of pnictide-chalcogenide absorbers of the Cu-Sb-S family,\cite{yu2013inverse}
where the high-Z element Sb causes its outer valence 5s orbital to become localized and deeper in energy, 
thus leaving the p channel as the main one available for bonding. 
This p-p effect ensures the AM$^{IV}$X$^{VII}_3$ family 
strong absorption intensity at the threshold, which greatly favors solar-energy capture efficiency. 
We will analyze the direct/indirect nature and magnitude of band gaps (Figure 4), 
examine the appearance of the p-p effect (Figure 5) 
that can induce strong band-gap absorptions, and use them as the DMs to select the compounds.
\vspace{0.3cm}

\textbf{(3) Simultaneously low electron and hole effective masses at band edges.}
 In solar cell devices, photo-generated carriers should be mobile 
so that they can be efficiently collected at electrodes. 
One of main factors determining carrier mobility is the carrier effective mass. 
Most of semiconducting materials have low effective mass of electrons (m$_{e}^{*}$), 
but relatively high effective mass of holes (m$_{h}^{*}$). For known AM$^{IV}$X$^{VII}_3$ materials, 
not only m$_{e}^{*}$ , but m$_{h}^{*}$ show quite low values.\cite{giorgi2013small} 
This contributes to ambipolar conductivity and is one of the most important advantages 
of the AM$^{IV}$X$^{VII}_3$ materials showing high PV performance. 
The basic principle operating here is the effect of anti-bonding states at VB edges 
formed by hybridization between halogen X$^{VII}$-p and M$^{IV}$ lone-pair s states (Figure 5). 
Such hybridization renders the VB edges broaden and more dispersive, 
leading to the small m$_{h}^{*}$. 
This is in contrast to the oxides having a rather narrow VB due to localized O-p states, 
or the transition metal compounds where the narrow VB reflects localized d states, 
both of which show large m$_{h}^{*}$ . 
This type of physics has been elucidated earlier for binary lead chalcogenides PbX$^{VI}$.\cite{wei1997electronic} 
The explicit feature of the X$^{VII}$-p and M$^{IV}$-s hybridization in VB 
for all candidate materials is analyzed in Figure 5. 
The results for the computed DMs - m$_{e}^{*}$  and m$_{h}^{*}$ are shown in Figure 7. 
It should be pointed out that we consider here the effective masses 
as the first filter of the charge-carrier mobility metric. 
One needs to keep in mind, however, 
that a more refined mobility metric would also involve the momentum scattering time, 
which, in the case of polar semiconductors is most likely governed by Fr\"{o}hlich interactions. 
These in turn depend on factors such as the high-frequency value 
of dielectric function, which changes with the materials chemical composition.
\vspace{0.3cm}

\textbf{(4) Sufficiently low binding energy and large localization radius for electron-hole pairs.}
 One of the key factors that determine the solar cell performance 
is whether the photo-generated electron-hole pairs (i.e. excitons) 
are effectively separated for collection before their recombination. 
This is to a large extent determined by the properties of excitons. 
Very high binding energy (E$_B$) and too small Bohr radius ($\alpha_{ex}$) 
of excitons correspond to strong Coulomb interaction between electron and hole, 
and thus are basically detrimental to electron-hole separation process. 
The hybrid AM$^{IV}$X$^{VII}_3$ perovskites exhibit rather low E$_B$ and large $\alpha_{ex}$ 
that are comparable to those of conventional semiconductors. 
For instance, E$_B$  of [CH$_3$NH$_3$]PbI$_3$ was reported to depend on temperature 
and vary in a wide range of 2-50 meV,\cite{miyata2015direct, lin2015electro, yamada2015photoelectronic, even2014analysis, menendez2014self}
 and $\alpha_{ex}$ is around 30 $\AA$.\cite{koutselas1996electronic, hirasawa1994exciton} 
Such low E$_B$ values, which are comparable to the room temperature thermal energy of 25 meV, 
are beneficial for a large fraction of photo-generated excitons to 
dissociate spontaneously into free carriers.\cite{sum2014advancements} 
This may be one of predominate facts responsible for the exceptionally long diffusion lengths 
of carriers observed in this family of materials.\cite{xing2013long, stranks2013electron}
It has been established that the excitons in this family 
belong to the three-dimensional Wannier type.\cite{tanaka2003comparative} 
We thus compute excitonic properties following the standard Wannier-Mott hydrogenic model, 
and use the resulted E$_B$  (Figure 8) and $\alpha_{ex}$ (Figure S1, Supporting Information) 
as the DM for materials screening.
\vspace{0.3cm}

\textbf{(5) Defect tolerant structures where bond breaking does not produce deep levels due to anti-bonding character.}
 To guarantee high carrier concentration and efficient carrier transport, 
the qualified solar materials need to be absent of deep defect-derived states 
that are usually trapping centers for carriers' non-radiative recombination. 
One design principle to avoid such deep defect levels 
is exploiting the anti-bonding characters of CB and VB edges. 
In most of semiconducting materials, band gap is formed at the separation of bonding 
and anti-bonding states and thus the CB usually has anti-bonding character. 
In case that the VB also derives from anti-bonding states, 
bond breaking associated with formation of both n-type and p-type defects 
will produce but shallow rather than deep levels in mid-gap region. 
This defect tolerant behavior originating from the anti-bonding characters of VB and CB 
has been confirmed in ternary chalcopyrites (I-III-VI$_{2}$)\cite{zhang1998defect, lany2005anion}
 as well as in AM$^{IV}$X$^{VII}_3$.\cite{yin2014unique, brandt2015identifying}
 Taking [CH$_3$NH$_3$]PbI$_3$ as the example, 
extensive defect calculations indicate most of intrinsic point defects form shallow levels.\cite{yin2015halide, du2014efficient, yin2014unusual} 
To probe this defect tolerant behavior for materials to be screened, 
we provide a thorough analysis of bonding and anti-bonding features (especially for VB) in Figure 5.
\vspace{0.3cm}

In addition to isolated point defects noted above, 
the defect tolerant behavior also occurs in the materials that 
can sustain unusually stable defect pairs. 
In the context of chalcopyrites (\textit{e.g.} CuInSe$_2$), 
it was previously discovered that defect tolerance 
is manifested by the existence of highly stable defect-pairs [2V$_{Cu}^{-}$ +In$_{Cu}^{2+}$] 
that give rise to a series of ordered vacancy structures (\textit{e.g.} CuIn$_5$Se$_8$, 
CuIn$_3$Se$_5$, Cu$_2$In$_4$Se$_7$, etc.).\cite{zhang1998defect} 
The analogous highly stable defect pairs [V$_{A}^{-}$ +M$_{A}^{+}$] in AM$^{IV}$X$^{VII}_3$
are likely, which might exist in high concentrations without producing deep carrier trapping centers, 
giving rise to series of novel compounds 
(\textit{e.g.} AM$^{IV}_2$X$^{VII}_5$, AM$^{IV}_4$X$^{VII}_9$,
 A$_2$M$^{IV}_5$X$^{VII}_{12}$, etc.).
\vspace{0.3cm}

\textbf{(6) Intrinsic defects can lead to both p-type and n-type doping by control of chemical potentials of reactants during growth.}
 Once we established materials stability and absence of spontaneously 
formed detrimental carrier-trapping centers, 
we need to explore if the qualified solar materials can be made both p-type and n-type 
so as to afford for instance, a p-i-n structure that is usually required by solar cells. 
According to recent calculations on [CH$_3$NH$_3$]PbI$_3$,\cite{yin2015halide, yin2014unusual} vacancy of heavy metal M$^{IV}$ (V$_{M}$) is predominant intrinsic defect responsible for hole formation and 
interstitial doping of molecular entity A (A$_i$) responsible for electron formation. 
Depending on relative supply of required reactants during growth 
it is possible to pin the Fermi energy near the VBM (where V$_{M}$ renders the sample p-type) 
or near the CBM (where A$_i$ renders the sample n-type), 
thus accomplishing bipolar conductivity. 
\vspace{0.3cm}

\textbf{(7) Microscopic reason for exceptionally long carrier diffusion length, i.e. 
why electron-hole recombination is low.} 
The long carrier diffusion length is another important factor 
making the hybrid AM$^{IV}$X$^{VII}_3$ 
perovskites good solar materials. 
Though the underlying mechanism remains elusive, 
it may relate closely to intrinsic properties such as small carrier effective masses, 
low exciton binding energy, absence of deep carrier trapping center, 
enhanced Born effective charges and lattice polarization, etc.\cite{du2014efficient} 
Other mechanism slowing electron-hole recombination may also exist. 
For chalcopyrites, it was discovered that the interface between (a) parent compound CuInSe$_2$ 
and (b) the Cu-poor ordered vacancy structures places electrons 
on (a) and holes on (b), 
thus separating them into different real-space regions.\cite{persson2003anomalous} 
This effect minimizes the recombination process and enhances carrier transport. 
The analogous interfacial structure existing in AM$^{IV}$X$^{VII}_3$ 
is likely the interface between parent phase and the highly stable defect-pair 
derived ordered structures as mentioned.
\vspace{0.3cm}

The strategy of the present work is to compute the DMs 
we have distilled from our physical understanding of the relevant DPs at work, 
and then use the compound-intrinsic DMs as the first-level filter 
to screen optimal solar AM$^{IV}$X$^{VII}_3$ perovskites. 
The device-level DMs are postponed to future screening. 
Among the above compound-intrinsic DMs we will use only those that pertain to ideal bulk compounds 
[items (1)-(5)] rather than the others that pertain to defected or inhomogeneous compounds.
\vspace{0.5cm}

\noindent{\textbf{3. The association of design metrics with candidate AM$^{IV}$X$^{VII}_3$ materials.}}
\vspace{0.5cm}

The constitution space of candidate compounds considered. 
Figure 1a shows the constitution space of candidate materials for screening. 
It is known that most of halide perovskites have temperature-dependent phase diagrams, 
i.e., while at high temperatures they form cubic perovskites 
(most likely stabilized by phonon entropy), 
at low temperatures distorted-perovskite or low-dimensional non-perovskite phases occur. 
Taking CsPbI$_3$ as example, the high-temperature cubic perovskite phase transforms to a yellow, 
large-gap, non-perovskite $\delta$ phase at temperatures below 315 $^\circ$C; 
fortunately its cubic perovskite phase can be stabilized to room temperature 
by nanostructuring\cite{swarnkar2016quantum} or composition engineering 
(\textit{e.g.}, via partially replacing I with Br, see Beal \textit{et al.}\cite{beal2016cesium}). 
Experimentally all the high-performance solar cells (based on \textit{e.g.}, 
[MA]PbI$_3$, [FA]PbI$_3$, CsPbI$_3$) are made from the cubic 
(or slightly distorted) perovskite phases that are stabilized at room temperature 
against competing non-perovskite phases. 
Considering this fact, screening of optimal AM$^{IV}$X$^{VII}_3$ 
in cubic perovskite structure is certainly valuable, 
even though some compounds form non-perovskite structures at normal synthesis condition. 
Therefore the cubic perovskite phase is uniformly adopted for all the AM$^{IV}$X$^{VII}_3$ 
to be screened even though some materials may exist with non-perovskite structure 
at normal condition. Totally 10 inorganic/organic cations were considered on the A site, 
3 group-IVA metalloids (Ge, Sn and Pb) on the M$^{IV}$ site, 
and 3 halogens (Cl, Br and I) on the X$^{VII}$ site. 
The chosen monovalent A-site cations have generally comparable ionic sizes, 
including Cs$^{+}$, ammonium NH$_{4}^{+}$ (M), hydroxylamine NH$_3$OH$^{+}$ (HA), 
diamine NH$_{2}$NH$_{3}^{+}$ (DA), methylammonium CH$_{3}$NH$_{3}^{+}$ (MA), 
formamid NH$_{3}$COH$^{+}$ (FM), fromamidinium CH(NH$_{2}$)$_{2}^+$ (FA), 
ethylamine CH$_{3}$CH$_{2}$NH$_{3}^+$ (EA), dimethylamine NH$_{2}$(CH$_{3}$)$_{2}^+$ (DEA) 
and guanidine amine C(NH$_{2}$)$_{3}^+$ (GA). 
In addition to these, we have also considered the cases with the pseudo-halogen anions, i.e., 
[BF$_{4}$]$^-$ and [SCN]$^-$, 
occupying the X$^{VII}$ site. In total, 
the materials space consists of about 100 compounds. 
To our best knowledge, among these candidate materials 
only 22 were previously made (including CsPbCl$_{3}$,\cite{moller1957phase} 
CsPbBr$_{3}$,\cite{stoumpos2013crystal} CsSnCl$_{3}$,\cite{kuok1992raman} CsGeBr$_{3}$,\cite{schwarz1996effect} 
CsGeCl$_{3}$,\cite{schwarz1996effect} [MA]GeCl$_{3}$,\cite{baikie2013synthesis} [MA]SnCl$_{3}$,\cite{baikie2013synthesis} 
CsPbI$_{3}$,\cite{eperon2015inorganic} CsSnBr$_{3}$,\cite{sabba2015impact} 
CsSnI$_{3}$,\cite{kumar2014lead} CsGeI$_{3}$,\cite{krishnamoorthy2015lead, stoumpos2015hybrid} 
[MA]GeI$_{3}$,\cite{krishnamoorthy2015lead, stoumpos2015hybrid} 
[MA]SnBr$_{3}$,\cite{hao2014lead} [MA]SnI$_{3}$,\cite{hao2014lead, noel2014lead} 
[MA]PbBr$_{3}$,\cite{kojima2009organometal} [MA]PbI$_{3}$,\cite{kojima2009organometal} 
[FA]GeI$_{3}$,\cite{krishnamoorthy2015lead, stoumpos2015hybrid} 
[FA]SnI$_{3}$,\cite{koh2015formamidinium} [FA]PbBr$_{3}$,\cite{eperon2014formamidinium}
 [FA]PbI$_{3}$,\cite{yang2015high, eperon2014formamidinium}
 [EA]PbI$_{3}$,\cite{safdari2015structure} [GA]GeI$_{3}$ \cite{stoumpos2015hybrid}), 
and the last 15 were considered for PV applications.
\vspace{0.3cm}

The explicit screening process in terms of the above compound-intrinsic DMs 
is summarized in Figure 1b, 
which clearly shows how the inferior materials were knocked out by considering more and more DMs 
step by step, and which AM$^{IV}$X$^{VII}_3$ 
live through multiple screenings 
as the winning materials. 
The explicit calculated data of corresponding DMs are listed in Table S1-S6 of Supporting Information.
\vspace{0.3cm}

\textbf{The DM (1): good thermodynamic stability and crystallographic stability.}
 Figure 2 shows the DM reflecting thermodynamic stability, 
the decomposition enthalpy $\Delta$H of AM$^{IV}$X$^{VII}_3$ perovskites [except for AGeCl$_{3}$,
 see Experimental Section (ii)]. 
The order of small molecules in the x-axis is sorted in terms of their steric size (i.e. r$_A$). 
Generally one observes a rather large span of $\Delta$H ($\sim$ 1.0 eV). 
This indicates the thermodynamic stability of this class of compounds depends 
strongly on the (A, M$^{IV}$, X$^{VII}$) combination.
 While for Pb (Figure 2a) and Sn (Figure 2b) based compounds the change of $\Delta$H from chlorides, 
bromides to iodides is small, Ge based bromides exhibit the much stronger stability 
than that of iodides (Figure 2c).
 Walking through different cations at the A site, 
we find four cations, i.e. Cs$^{+}$, MA$^{+}$, EA$^{+}$ and DEA$^{+}$, show evidently strong stability. 
The perovskites containing the former two cations have been extensively investigated, 
but the studies of the latter two are rarely reported.\cite{stoumpos2013semiconducting, safdari2015structure,
baikie2013synthesis}
The common feature of the three organic molecules is the appearance of CH$_3$-terminal. 
For the prototype material [MA]PbI$_3$, our calculated $\Delta$H is -0.02 eV/f.u., 
indicating a marginal instability against decomposition. 
The result is qualitatively in accord with the calculation of Zhang \textit{et al.} 
where the even lower $\Delta$H of -0.12 eV/f.u. was reported.\cite{zhang2015intrinsic}
\vspace{0.3cm}

The quantitive discrepancy may originate from the difference 
in specific structures of decomposed products, van der Waals functionals, 
orientation of molecules, etc. adopted in the two calculations. 
Both of the results show consistency with the observed intrinsic thermal instability in experiment, 
i.e., significant decomposition occurring in [MA]PbI$_3$ at $\sim$ 80 $^\circ$C.\cite{conings2015intrinsic} 
For all the AM$^{IV}$X$^{VII}_3$ materials, 
the magnitudes of $\Delta$H (below 0.5 eV/f.u.) 
are in general much smaller than those of oxide perovskites 
(up to $\sim$ 1.5 eV/f.u.).\cite{selbach2008thermodynamic, xu2005perovskite}
By considering the boundary thermodynamic stability feature of this family, 
we used a not strict criterion, i.e., $\Delta$H $>$ -0.10 eV/f.u., 
for the materials screening. 
This filter makes us discard all the compounds containing the smallest molecule NH$_{4}^+$ 
and most of ones with HA$^+$. 
50 compounds with satisfied thermodynamic stability pass the screening to the next step 
(as in the second row of Figure 1b).
\vspace{0.3cm}

A few notes on the meaning of stability/instability with respect to the disproportionation 
need to be made: (i) the compounds that are predicted to be slightly unstable 
with respect to the disproportionation into AX$^{VII}$ + M$^{IV}$X$^{VII}_2$, 
\textit{e.g.}, the simplest compound of CsPbI$_{3}$, 
might still be synthesized from different reactants other than the above binaries. 
However, once such a ternary AM$^{IV}$X$^{VII}_3$ compound is prepared, 
it will be prone to -according to our first-principle calculations -
decompose into the binary components noted. 
(ii) the instability with respect to the AX$^{VII}$ + M$^{IV}$X$^{VII}_2$ 
disproportionation is not the only instability channel possible. 
Other channels will have to be considered for specific cases. 
For example, the compounds containing multivalent elements such as Sn 
(that can exist in either valence Sn$^{4+}$ or Sn$^{2+}$) might be prone for valence instability
transformation of the ASnX$^{VII}_3$ with bivalent Sn to the structures 
affording tetravalent Sn, i.e., destabilization due to oxidation of metallic Sn 
from Sn$^{2+}$ to Sn$^{4+}$.\cite{scragg2011chemical, biswas2010electronic}
(iii) all the calculations here are performed using the standard cubic perovskite structure 
and we thus consider the disproportionation stability of this phase. 
Note however that other disproportionation channels involving non-cubic phases 
can be possible in specific cases such as Ge-based iodide perovskites 
that have been experimentally synthesized recently,\cite{krishnamoorthy2015lead, stoumpos2015hybrid} exhibiting, 
however, substantial distortion from the cubic phase, 
or the [FA]PbI$_{3}$, where the black cubic perovskite phase is prone to transform 
to a yellow phase of different crystallography.\cite{binek2015stabilization} 
\vspace{0.3cm}

\begin{figure}[t]
\includegraphics[width=3.0in]{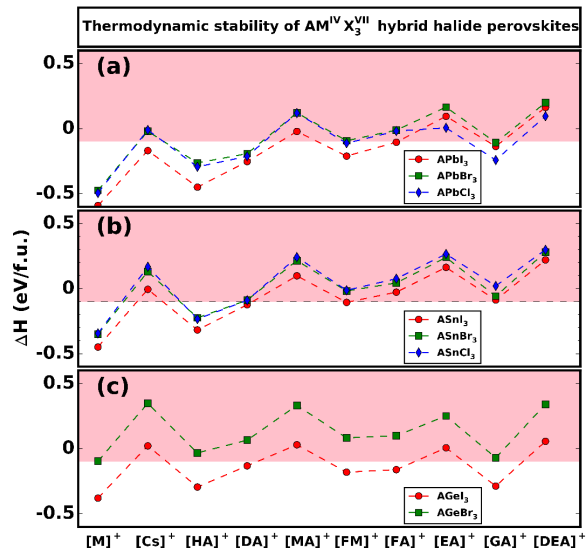}
\centering
\caption{Calculated decomposition enthalpies $\Delta$H of (a) Pb, 
(b) Sn and (c) Ge based AM$^{IV}$X$^{VII}_3$ perovskites with respect to 
decomposed products of AX$^{VII}$ +M$^{IV}$X$^{VII}_2$. Positive $\Delta$H values mean 
no decomposition occurring. 
The compounds located in shaded area (with $\Delta$H $>$ -0.1 eV/f.u.) pass the materials screening.}
\end{figure}

We then map all the candidate materials onto two-dimensional plot 
with the tolerance factor t and the octahedral factor $\mu$ as variables 
(Figure 3) for analyzing crystallographic stability. 
For inorganic AM$^{IV}$X$^{VII}_3$ halides, 
a previous statistic analysis indicated the formability of perovskites 
requires 0.81 $<$ t $<$ 1.11 and 0.44 $<$ $\mu$ $<$ 0.90.\cite{li2008formability} 
We find that almost all the materials fall in the stable range required by t , 
except for [EA,GA,DEA]GeCl$_3$ that are close to the upper boundary of 1.11. 
The three compounds have actually been abandoned because of their low $\Delta$H. 
The inset of Figure 3 shows the $\Delta$H of the AM$^{IV}$X$^{VII}_3$ perovskites as the function of t. 
Generally the results can be divided into two groups: the Pb/Sn based materials with t $<$ 0.97 
(shaded in pink) and the Ge based ones with t $>$ 0.97 (shaded in skyblue). 
For the former group, we observe a general trend of the larger t, 
the stronger thermodynamic stability (i.e., the higher $\Delta$H). 
This resembles the behavior observed in oxide perovskites,\cite{yokokawa1989thermodynamic} 
and suggests that the larger t below 1.0 
is more favorable for stabilizing AM$^{IV}$X$^{VII}_3$ 
in the perovskite structure. At the rather smaller t ($<$ 0.89), 
for instance for [CH$_3$NH$_3$]PbI$_3$, 
the distorted pseudocubic perovskites with tilted M$^{IV}$X$^{VII}_{6}$ octahedrons 
(called orthorhombic $\gamma$ or tetragonal $\beta$ phases) 
become energetically more stable than the standard cubic perovskite (called $\alpha$ phase) 
at room or low temperatures.\cite{baikie2013synthesis} 
If the energy difference between the $\gamma$/$\beta$ phases and
 the $\alpha$ phase is substantial, 
this extra energy gain from structural distortion may stabilize the material 
despite the low $\Delta$H of the $\alpha$ phase. 
This explains many known room-temperature phases of AM$^{IV}$X$^{VII}_3$ 
existing as the less symmetric structures.\cite{stoumpos2013semiconducting, safdari2015structure,
 baikie2013synthesis} 
Turning to the $\mu$ value, 
while Pb-based and Sn-based compounds are safely located within the stable range, 
the too small $\mu$ of Ge-based compounds implies that 
they might not tend to be stabilized in the perovskite form. 
This seems, however, not consistent with the recent experiments,\cite{krishnamoorthy2015lead, stoumpos2015hybrid} 
where Ge-based iodide perovskites have been successfully synthesized, 
though accompanying with substantial distortion from the cubic $\alpha$ phase. 
We thus did not rule out as yet the Ge-based compounds at this step.
\vspace{0.3cm}

\begin{figure}[t]
\includegraphics[width=3.0in]{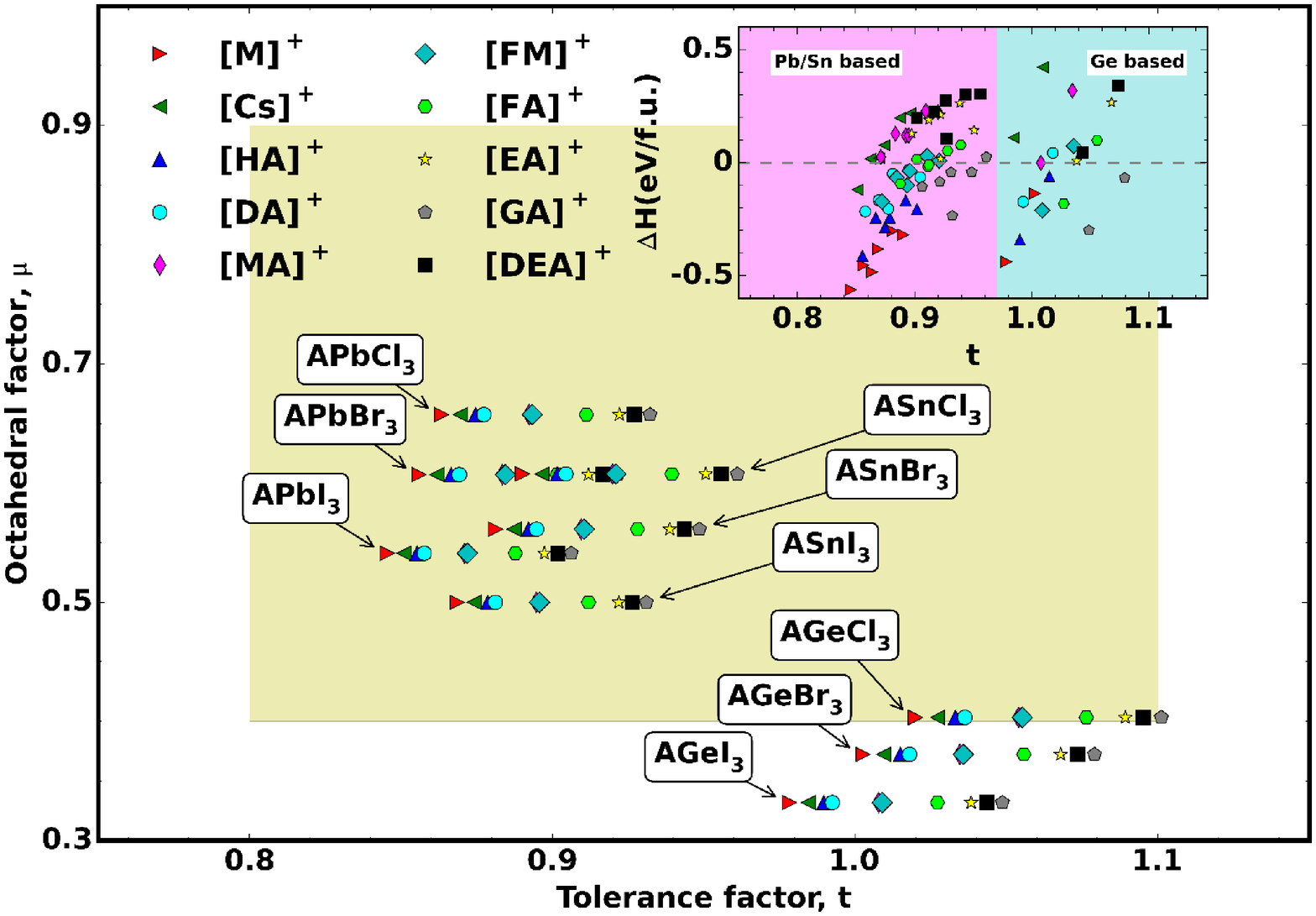}
\centering
\caption{
Mapping of various AM$^{IV}$X$^{VII}_3$ perovskites with respect to the octahedral factor ($\mu$) 
and the tolerance factor (t). 
Shaded area (with 0.44 $<$ $\mu$ $<$ 0.90 and 0.81 $<$ t $<$ 1.11) 
represents the crystallographic stability requirement for the formability 
of halide perovskites.\cite{biswas2010electronic} 
The insert shows the decomposition enthalpies $\Delta$H of the AM$^{IV}$X$^{VII}_3$ perovskites 
as a function of t. 
The regions of Pb/Sn and Ge based compounds are shaded by pink and skyblue, respectively.}
\end{figure}

\textbf{The DM (2): suitable direct band gap and strong absorption near threshold.}
 The problem of obtaining the correct band gap in such complex 
organic-inorganic hybrid structures is a difficult one, 
and we approach this specific point at this time by distilling information 
from a few compounds for which the band gap is known, 
and determining the adjustable parameters (i.e., the percent age of exact Fock exchange component) 
for the employed hybrid functional calculations including the spin-orbit coupling effect. 
We then use the same parameter values for similar compounds to arrive at the band gap values 
for the new compounds (see Experimental Section). 
We found most of the AM$^{IV}$X$^{VII}_3$ perovskites considered 
show direct band gap (E$_{g}^{d}$) at the R point of the cubic Brillouin zone. 
Only sixteen materials show indirect gaps formed by the VBM at the R point 
and the CBM at the point slightly deviating from R (see Figure S2, Supporting Information). 
This indirect-gap feature has been proposed to facilitate enhanced carrier radiative lifetime.\cite{motta2015revealing, zheng2015rashba} 
Such indirect gaps are merely several or several tens meV lower than the E$_{g}^{d}$, 
thus having negligible effect on optical absorption. 
We thus regard all of them as the direct-gap materials in the screening process.
\vspace{0.3cm}

\begin{figure}[t]
\includegraphics[width=3.0in]{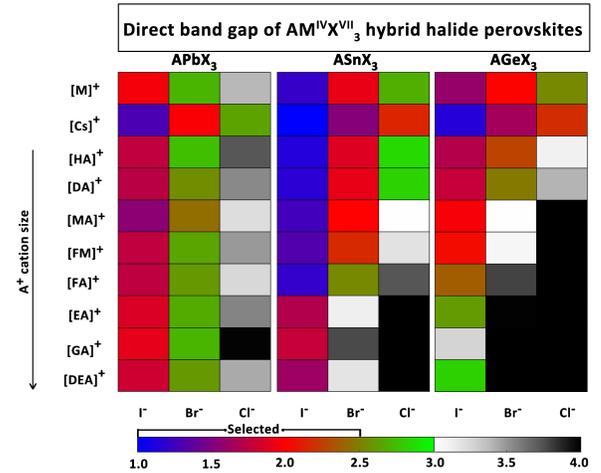}
\centering
\caption{
Calculated direct band gaps E$_{g}^{d}$ of the AM$^{IV}$X$^{VII}_3$ perovskites. 
The A$^+$ cations are sorted by the increasing steric sizes. 
The gap values of 1.0-3.0 eV (promising for solar materials) 
are depicted by the RGB colors, 
and the oversized ones above 3.0 eV are shown in the white-black scale. 
The criterion of E$_{g}^{d}$  $<$ 2.5 eV is used for the materials screening.}
\end{figure}

Figure 4 shows calculated E$_{g}^{d}$ of the AM$^{IV}$X$^{VII}_3$ perovskites 
and Figure 5 depicts how the band gap is formed by various bonding/anti-bonding states 
that are characterized with the analysis of crystal orbital overlap population (COOP).\cite{hoffmann1987chemistry} 
There exists a wide distribution of gap values ranging from about 1.0 to 6.0 eV. 
This wide variation of E$_{g}^{d}$ is attributed to 
(i) the energy difference between the X$^{VII}$-p orbital 
and the M$^{IV}$-p/s orbitals forming band-edges changes with different (M$^{IV}$, X$^{VII}$) combination, 
(ii) the absolute positions of the CBM and the VBM, 
which are both anti-bonding states, are tuned by varied M$^{IV}$-X$^{VII}$ bonding strengths 
(see Figure 5a). With increasing electronegativity of X$^{VII}$, 
the band gaps show significant enlargement, 
which is consistent with 
experimental observations.\cite{kim2012lead, stoumpos2013semiconducting, kitazawa2002optical},\cite{edri2013high},
\cite{noh2013chemical, kulkarni2014band, suarez2014recombination} 
This makes the majority of Pb-based, part of Sn/Ge-based bromides and chlorides not favorable 
as solar materials owing to their too large E$_{g}^{d}$. 
The iodides family thus contains the most numerous optimal solar materials. 
Walking through different organic molecules, 
we found Pb-based compounds show weak dependence of E$_{g}^{d}$ on A$^+$ cations, 
whereas Sn/Ge-based ones have dramatically increased E$_{g}^{d}$ 
with increasing steric sizes of A$^+$. 
Taking the bromides as the instance, 
the E$_{g}^{d}$ of Pb-based compounds varies within a relatively narrow energy range 
of 1.98-2.74 eV, but the E$_{g}^{d}$ of Sn and Ge-based ones 
show the much wider tunability within the range of 1.53-3.77 eV and 1.64-4.15 eV, respectively. 
Since A$^+$ cations have no direct contribution to band-edge states, 
their influence on E$_{g}^{d}$ is exerted though geometric modification 
of the M$^{IV}$-X$^{VII}$ network composed of corner-sharing M$^{IV}$X$^{VII}_{6}$ octahedra.
\vspace{0.3cm}

To unravel how the A$^+$ cations affect the band gaps of AM$^{IV}$X$^{VII}_3$ perovskites, 
we perform further analysis on the obtained structure-property (i.e., E$_{g}^{d}$) data. 
We find a geometric quantity of the M$^{IV}$X$^{VII}_{6}$ octahedron - 
the averaged bond length of six M$^{IV}$-X$^{VII}$ bonds (d$_{ave}$) - 
is a good descriptor for establishing anticipated structure-property relationship. 
Figure 6 shows the dependence of E$_{g}^{d}$ on d$_{ave}$ for all the AM$^{IV}$X$^{VII}_3$ materials. 
One clearly observes that for each (M$^{IV}$, X$^{VII}$) 
combination the magnitude of E$_{g}^{d}$ is linearly proportional to the d$_{ave}$, 
which actually increases with the steric size of A$^+$. 
The trend becomes more evident from Pb, Sn to Ge-based materials. 
The physical mechanism responsible for this trend is revealed by 
projecting band-edge states onto the bonding/anti-bonding orbitals in Figure 5. 
With the increasing sizes of the A$^+$ cations of for instance Ge based compounds (Figure 5c), 
the resulted lattice expansion makes the bond lengths of M$^{IV}$-X$^{VII}$ bonds increase, 
thus reducing the bond strengths. 
This pulls down the energy levels of both of two anti-bonding states, $\sigma$$^{*}$(M$^{IV}$-s/X$^{VII}$-p) (forming VB) and $\sigma$$^{*}$(M$^{IV}$-p/X$^{VII}$-p) (forming CB). 
Owing to the more delocalized nature of M$^{IV}$-s state than that of M$^{IV}$-p state, 
the interaction between the M$^{IV}$-s  and X$^{VII}$-p  states are stronger, 
evidenced by the larger energy splitting between $\sigma$(M$^{IV}$-s/X$^{VII}$-p) 
and $\sigma$$^{*}$(M$^{IV}$-s/X$^{VII}$-p) (about 7 eV) than that between $\sigma$(M$^{IV}$-p/X$^{VII}$-p) 
 and $\sigma$$^{*}$(M$^{IV}$-p/X$^{VII}$-p) (about 5 eV). As the result the $\sigma$$^{*}$(M$^{IV}$-s/X$^{VII}$-p) 
 state is affected more pronouncedly, 
which exhibits the larger downshift. 
This enlarges E$_{g}^{d}$  and gives rise to its nearly linear increase with d$_{ave}$. 
For the Pb-based compounds (Figure 5b), 
since the M$^{IV}$X$^{VII}_{6}$ octahedron is relatively compact with the stronger stiffness 
owing to the larger size of Pb$^{2}$$^{+}$, d$_{ave}$ is less affected by the changes 
of A$^+$ cations. 
Thus E$_{g}^{d}$  shows an inconspicuous increase. 
\vspace{0.3cm}

From Figure 5 (and Figure S3 in Supporting Information) 
we can see that all the AM$^{IV}$X$^{VII}_3$ perovskites 
have the VB from the $\sigma$$^{*}$(M$^{IV}$-s/X$^{VII}$-p)  
states and the CB from the $\sigma$$^{*}$(M$^{IV}$-p/X$^{VII}$-p)  states. 
There always exist the VB-p to CB-p channels of band-gap transition 
as mentioned in favor of strong optical transition intensity (see also Figure 9). 
The magnitude of E$_{g}^{d}$  can thus act as the valid DM to 
select the optimal solar materials with high light absorption efficiency. 
Considering the visible light region of $\sim$ 1.5-3.0 eV, 
the criterion of E$_{g}^{d}$  lower than 2.5 eV are used for the materials screening. 
Totally 41 compounds pass the screening (Table S3, Supporting Information). 
This leads us to further select 22 compounds from the 50 winning compounds 
surviving from the DM (1) (as in the third row of Figure 1b). 
Among them six are Pb-based compounds, five are Ge-based and the remaining ones are Sn-based.
\vspace{0.3cm}

\begin{figure}[t]
\includegraphics[width=3.0in]{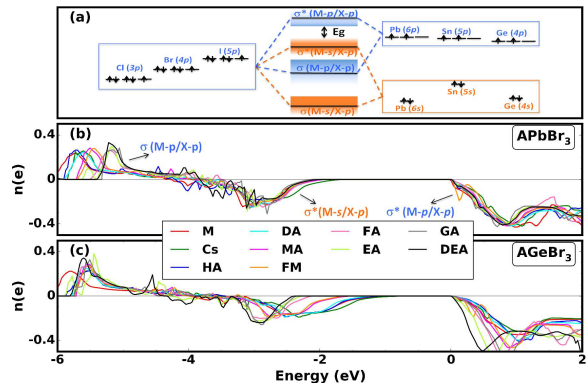}
\centering
\caption{ 
(a) Schematic diagram of constituted atomic orbitals and 
their formed bonding ($\sigma$)/antibonding ($\sigma$$^{*}$) bands in the AM$^{IV}$X$^{VII}_3$ perovskites. 
(b, c) The crystal orbital overlap populations (COOP) analysis 
for APbBr$_3$ and AGeBr$_3$, respectively. 
For comparison, the CBM of each material is set to energy zero.}
\end{figure}

\textbf{The DM (3): low electron and hole effective masses favoring ambipolar conductivity.}
 As demonstrated in Figure 5, 
the VB of the AM$^{IV}$X$^{VII}_3$ perovskites is composed of the anti-bonding states 
hybridized by the M$^{IV}$-s  and X$^{VII}$-p  orbitals ($\sigma$$^{*}$(M$^{IV}$-s/X$^{VII}$-p)). 
This hybridization causes a rather dispersive VB, 
and thus low hole effective mass (m$_{h}^{*}$). 
As in most of semiconducting materials, the CB dominated by cationic M$^{IV}$-p states 
usually gives a small electron effective mass (m$_{e}^{*}$). 
The feature of simultaneously low m$_{h}^{*}$  and m$_{e}^{*}$  
for this family of materials has been indicated in Figure 7. 
Most of compounds except for [FM]M$^{IV}$Cl$_{3}$ (M$^{IV}$=Ge,Sn,Pb) (to be discussed below) 
show comparably small m$_{h}^{*}$  and m$_{e}^{*}$  at the order of several tenths of m$_{0}$ . 
This is beneficial for the ambipolar conductivity favored by PV applications. 
Generally, the values of m$_{e}^{*}$  are larger than those of m$_{h}^{*}$ , 
which is distinct from the usual semiconductors with the lighter m$_{e}^{*}$  
but the heavier m$_{h}^{*}$ . Both m$_{h}^{*}$  and m$_{e}^{*}$  show increasing trends from iodides, bromides to chlorides. 
For Sn and Ge based compounds, we see a clear trend of m$_{h}^{*}$  
increasing with the sizes of A$^{+}$ cations. 
These follow the same trend of E$_{g}^{d}$  (Figure 6), 
which can be understood in terms of the nature of bonding/anti-bonding states 
forming band-edges as above. 
With a screening criterion of m$_{h}^{*}$ $<$ 0.5m$_{0}$  and m$_{e}^{*}$ $<$ 0.5m$_{0}$  
applied, 40 compounds pass the screening (see Table S4, Supporting Information). 
Among the 22 winners passing the DM (1, 2), 
[MA]PbBr$_3$, [HA]SnBr$_3$ and [DA]SnBr$_3$ 
are ruled out and 19 compounds are left (as in the fourth row of Figure 1b).
\vspace{0.3cm}

\begin{figure}[t]
\includegraphics[width=3.0in]{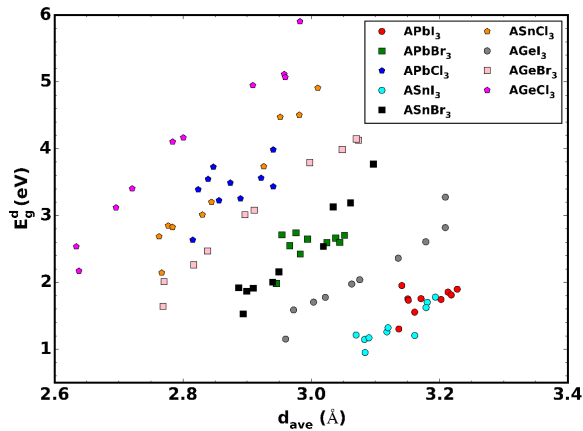}
\centering
\caption{
Dependence of direct band gaps E$_{g}^{d}$  on the averaged M$^{IV}$-X$^{VII}$ bond length (d$_{ave}$) 
of the AM$^{IV}$X$^{VII}_3$ perovskites.}
\end{figure}

\textbf{The DM (4): low exciton binding energy not impeding photon-induced charge separation.}
 The calculated binding energy (E$_B$) and Bohr radius ($\alpha_{ex}$) 
of Wannier-type excitons for the AM$^{IV}$X$^{VII}_3$ perovskites 
are shown in Figure 8 and Figure S1 in Supporting Infomation, respectively.
\vspace{0.3cm}

For [MA]PbI$_{3}$, using the calculated reduced exciton mass $\mu$$^{*}$ (0.13) 
and high-frequency limit of dielectric constant $\epsilon$$_{\varpropto}$  (5.7), 
we obtain E$_B$  and $\alpha_{ex}$  of 52 meV and 24 $\AA$, respectively. 
With the $\epsilon$$_{\varpropto}$  used, 
in principle the excitonic properties are evaluated for the exciton 
appearing immediately after photon absorption 
(i.e., in absence of ion relaxation). 
The value of E$_B$  reasonably falls in the range of the experimental values\cite{miyata2015direct, tanaka2003comparative}, \cite{d2014excitons, hirasawa1994magnetoabsorption} 
and previous calculations.\cite{menendez2014self} 
One sees E$_B$  shows clear increase from iodides, bromides to chlorides. 
The trend is in accordance with the observation in experiments.\cite{tanaka2003comparative} 
This originates from the heavier m$_{h}^{*}$  and m$_{e}^{*}$  (Figure 7, thus the larger $\mu$$^{*}$ ) 
and the lower $\epsilon$$_{\varpropto}$  (corresponding to the larger E$_{g}^{d}$  in Figure 4) 
for the perovskites with the more electronegative halogens. 
For Sn and Ge based compounds, E$_B$  unambiguously increases with the size of the A$^{+}$ 
cation as the result of the increased m$_{h}^{*}$  and the decreased $\epsilon$$_{\varpropto}$. 
To select the materials for facilitating photon-induced electron-hole pairs separation, 
we use the criterion of E$_B$ $<$ 100 meV, which corresponds to 15 $\AA$. 
This results in the abandonment of most of chlorides and 42 compounds 
pass the screening (see Table S5 and S6, Supporting Information). 
Among the superior compounds surviving the above DMs (1-3), 
[EA]PbI$_3$ is ruled out and 18 ones are kept as the optimal solar materials 
(as in the fifth row of Figure 1b).
\vspace{0.3cm}

\begin{figure}[t]
\includegraphics[width=3.0in]{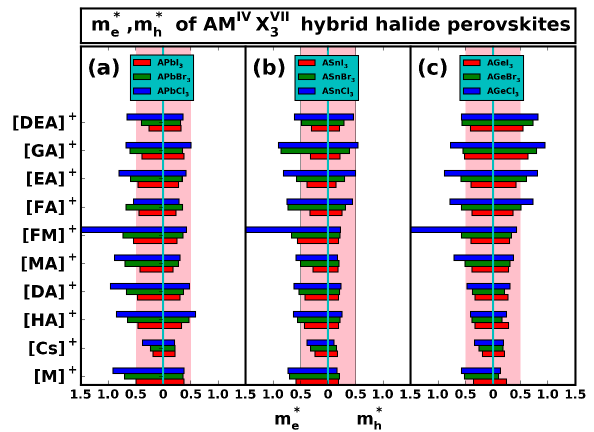}
\centering
\caption{
Calculated electron (m$_{e}^{*}$, left panels) and hole (m$_{h}^{*}$ , right panels)
effective masses of (a) Pb, (b) Sn and (c) Ge based AM$^{IV}$X$^{VII}_3$ perovskites.
Shaded areas indicate the screening criterion applied (m$_{e}^{*}$ $<$ 0.5 m$_{0}$ and m$_{h}^{*}$ $<$ 0.5 m$_{0}$).}
\end{figure}

\textbf{The DM (5): defect tolerant behavior beneficial from the anti-bonding feature of band-edge states.}
 As shown in Figure 5 and Figure S3 in Supporting Information, 
all the studied AM$^{IV}$X$^{VII}_3$ perovskites 
show the VB and CB composed of anti-bonding states, 
i.e. $\sigma$$^{*}$(M$^{IV}$-s/X$^{VII}$-p)  (of VB) and 
 $\sigma$$^{*}$(M$^{IV}$-p/X$^{VII}$-p) (of CB). 
Therefore they all belong to the materials with the defect tolerant feature 
as mentioned and pass the screening in terms of this DM. 
The number of selected superior materials is kept unchanged (as in the sixth row of Figure 1b).
\vspace{0.3cm}

We should mention whereas all the studied AM$^{IV}$X$^{VII}_3$ perovskite compounds 
show the VB and CB to be composed of anti-bonding states, 
so they can all be classified as the defect tolerant materials, 
expecting shallow defect (donor or acceptor) transition levels, 
specific calculations on the defect formation enthalpies for particular compounds 
will be needed in the future. 
This will determine if some charge-carrier trapping defects can form in large quantities 
(i.e., have low formation enthalpies), 
and thus lead to short nonradiative lifetime, which limits photo-induced charge extraction.
\vspace{0.3cm}

In addition to the AM$^{IV}$X$^{VII}_3$ perovskites with halogen anions 
sitting at the X$^{VII}$ site, 
we have also investigated the cases involving pseudo-halogen anions. 
Two types of pseudo-halogen anions, [BF$_4$]$^-$and [SCN]$^-$ are considered. 
Both of them have been experimentally incorporated into [MA]PbI$_3$ to 
partially substitute I$^-$ anions.\cite{jiang2015pseudohalide, nagane2014ch} 
The Pb${2}^{+}$/Sn${2}^{+}$ and Cs$^+$/MA$^+$/FA$^+$ are chosen as 
the M$^{IV}$ and A-site cations, respectively. 
Our calculations (Table S7, Supporting Information) 
indicate that the materials containing [BF$_4$]$^-$ show strong stability with respect to decomposition 
(with $\Delta$H $>$ 0.30 eV/f.u.); 
the stability of [SCN]$^-$ based materials depends on the choice of A-site cations: 
while the ones with the inorganic Cs$^+$ are quite stable (with $\Delta$H $>$ 0.30 eV/f.u.),
 the ones with organic molecular MA$^+$/FA$^+$ exhibit bad stability 
(with $\Delta$H $<$ -0.15 eV/f.u.). The [BF$_4$]$^-$ based materials show extremely large band gaps 
(even above 8 eV), consistent with previous calculation.\cite{hendon2015assessment} 
This may be attributed to the high electronegative nature of F 
that makes the M$^{IV}$-X$^{VII}$ bond predominantly ionic, 
weakens the orbital hybridization between M$^{IV}$ and X$^{VII}$, 
and thus leads to narrow-width bands and enlarged band gaps. 
The band gaps of the [SCN]$^-$ based materials are not as high as those of the [BF$_4$]$^-$ based ones, 
but still much higher than the upper limit of visible spectrum. 
Corresponding to the large band gaps, both [BF$_4$]$^-$ and [SCN]$^-$ 
based materials show heavy m$_{h}^{*}$  and m$_{e}^{*}$  (1.0-8.0m$_{0}$ ).
 Therefore by themselves they are not promising solar materials. 
By considering the robust thermodynamic stability of [BF$_4$]$^-$ and Cs$^{+}$[SCN]$^-$ based materials, 
alloying them with other AM$^{IV}$X$^{VII}_3$ perovskites 
may be a promising strategy to improve the stability of currently used materials.
\vspace{0.3cm}

\begin{figure}[!h]
\includegraphics[width=3.0in]{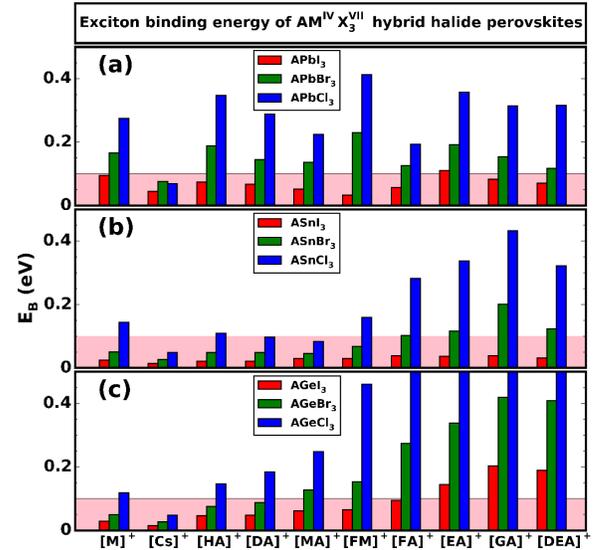}
\centering
\caption{
Calculated exciton binding energies of (a) Pb, (b) Sn and (c) Ge
based AM$^{IV}$X$^{VII}_3$ perovskites with the hydrogen-like Wannier-Mott model.
Shaded areas indicate the criterion applied (E$_B$ $<$ 0.1 eV) for the materials screening.}
\end{figure}

\noindent{\textbf{4. Emerging optimal solar materials with desired target functionalities.}}
\vspace{0.3cm}

The above process leads to 18 superior solar AM$^{IV}$X$^{VII}_3$ perovskites 
that become winners from the materials screening. 
Of them 9 were previously made but not recognized as PV materials, 
and 5 were not made. In particular, 
these materials are chosen in terms of the following criterions: 
(i) relatively high thermodynamic stability: the decomposition enthalpy higher than -0.1 eV/f.u., 
(ii) high light absorption effeciency: the optical band gap sitting below 2.5 eV, 
(iii) advantages for ambipolar carrier conduction: both electron and hole effective masses smaller than 0.5m$_{0}$ , 
(iv) facilitated photon-induced charge dissociation: small exciton bonding energy ($<$ 100 meV), 
(v) defect tolerant feature: appearance of anti-bonding states in both valence and conduction bands.
\vspace{0.3cm}

\begin{figure}[h]
\includegraphics[width=3.0in]{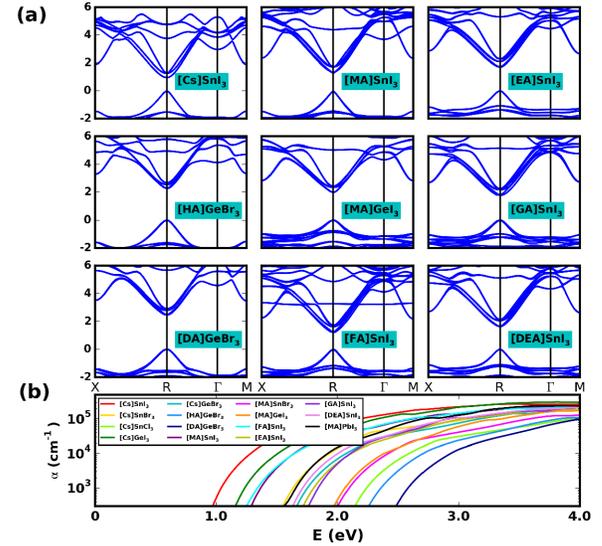}
\centering
\caption{
(a) Calculated band structures of selected finally winning AM$^{IV}$X$^{VII}_3$ perovskites 
passing all the DMs of the materials screening. 
(b) Absorption spectra of 14 Pb-free winning AM$^{IV}$X$^{VII}_3$ perovskites 
compared with that of [MA]PbI$_{3}$ (in black).}
\end{figure}

Four materials are Pb-based, 
among which [MA]PbI$_{3}$ and [FA]PbI$_{3}$ have been experimentally 
identified as high-performance solar materials with the maximum conversion 
efficiencies of 19.3\%\cite{zhou2014interface} 
and 20.1\%\cite{yang2015high}. 
If we further consider the environmental concern of toxic Pb and 
apply the screen criterion of Pb-free, they should be excluded from the optimal materials list. 
Then we obtain 14 optimal materials (as in the last row of Figure 1b), 
which include 5 Ge-based materials (i.e., CsGeI$_{3}$, CsGeBr$_{3}$, [HA]GeBr$_{3}$, [DA]GeBr$_{3}$,
 [MA]GeI$_{3}$) and 9 Sn-based ones (i.e., CsSnI$_{3}$, CsSnBr$_{3}$, CsSnCl$_{3}$, [MA]SnI$_{3}$,
 [MA]SnBr$_{3}$, [FA]SnI$_{3}$, [EA]SnI$_{3}$, 
[GA]SnI$_{3}$ and [DEA]SnI$_{3}$). Among them, CsSnBr$_{3}$, CsSnCl$_{3}$, CsGeBr$_{3}$,
 [MA]SnI$_{3}$, [MA]SnBr$_{3}$, [EA]SnI$_{3}$ and [DEA]SnI$_{3}$ 
own the much higher $\Delta$H (above 0.1 eV/f.u.) than that of [MA]PbI$_{3}$ (-0.02 eV/f.u.),
 indicating their potentially outstanding thermodynamic stability. 
The attempt to synthesize them and use them for PV applications holds promise for 
resolving the issue of poor long-term stability in the solar cell devices based on [MA]PbI$_{3}$.
\cite{niu2015review} 
The band structures and absorption spectra of selected optimal materials are shown in Figure 9.
 As seen all these materials show similar band structures in the band-gap region, \textit{e.g.},
 with the direct gap forming by the dispersive VBM and CBM at the R point (Figure 9a). 
Most of them show comparably strong absorptions at the threshold of between 1.0-2.0 eV (Figure 9b), 
implying potentially high solar-energy capture efficiency. 
Note that although there is instability issue associated with the tendency of Sn$^{2+}$/Ge$^{2+}$
 oxidized to Sn$^{4+}$/Ge$^{4+}$, 
it could be prevented during materials synthesis or device fabrication.
 The solar cells made by [MA]SnI$_{3}$ have been reported with an early-stage efficiency around 6\%.
\cite{hao2014lead, noel2014lead}

\vspace{0.3cm}

\noindent{\textbf{5. Discovery of an exotic class of hybrid AM$^{IV}$X$^{VII}_3$ perovskites with band-edges involving organic molecule derived state. }}
\vspace{0.5cm}

It generally appears that for the known hybrid AM$^{IV}$X$^{VII}_3$ materials 
the role of organic molecular A$^{+}$ cations is stabilizing the perovskite lattice, 
and thus has only indirect effect on the PV relevant properties through 
modifying the M$^{IV}$-X$^{VII}$ framework 
(\textit{e.g.}, the specific orientation of them may cause indirect bandgap and slow carrier recombination
\cite{motta2015revealing}). 
This is because the organic molecule derived electronic states are located far away from band-edges,
\cite{yin2014unusual} 
thus not participating in visible light absorption,
 charge carrier separation and transport, etc.
\vspace{0.3cm}

\begin{figure}[t]
\includegraphics[width=3.0in]{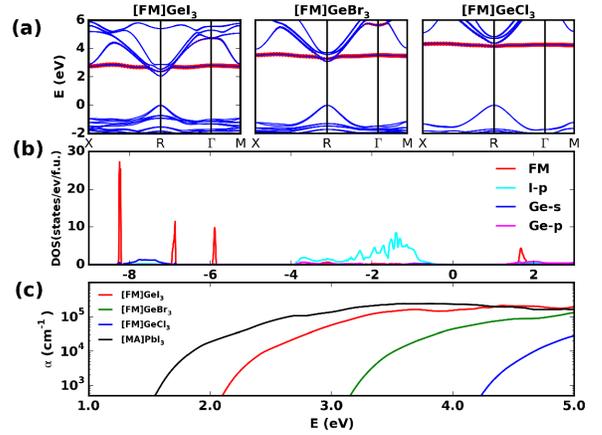}
\centering
\caption{(a) Electronic band structures,
(b) projected density of states (DOS) and (c) absorption spectra of
[FM]Ge(I/Br/Cl)$_{3}$.
In (a) the orbital projections onto the FM molecule are indicated by red circles.
The absorption spectrum of [MA]PbI$_{3}$ is given in (c) for comparison.}
\end{figure}

We find here this picture is not always true. 
Figure 10a shows the band structures of [FM]GeX$^{VII}_{3}$ (X$^{VII}$= I, Br, Cl). 
Clearly there appears a quite flat band (with red circles) close to the CBM 
in [FM]GeI$_{3}$. With the band gap increasing from iodides, bromides to chlorides, 
its energy level moves gradually toward the gap region, 
and becomes the real CBM in [FM]GeCl$_{3}$. 
From the projected density of states of [FM]GeI$_{3}$ (Figure 10b), 
we can see it is derived from the FM molecule (red lines). 
The extremely dispersiveless nature of these localized states is responsible for the abnormally huge m$_{e}^{*}$  in [FM]M$^{IV}$Cl$_{3}$ (M$^{IV}$= Ge, Sn, Pb) (Figure 7). 
Note that the relative position of the FM molecule derived band within the CB 
relates closely to the relativistic effect of spin-orbit coupling (SOC). 
While the SOC does not affect much the VB mainly from the M$^{IV}$-s states, 
it has significant impact on the CB that is dominated by the M$^{IV}$-p states. 
As shown in Figure S4 in Supporting Information, 
with the SOC included the CB formed by the M$^{IV}$-X$^{VII}$ framework shows evident downshift 
and the shift amount decreases from the heavier Pb, Sn, to the lighter Ge based compounds. 
Since the FM-molecule derived states are much less affected by the SOC, 
the energy difference between them and the minimum of the CB formed 
by the M$^{IV}$-X$^{VII}$ framework decreases from Pb, Sn, to Ge based compounds, 
and becomes negative in [FM]GeCl$_{3}$ 
(making the molecular states turn into the actual CBM). 
The appearance of such molecule derived states in proximity to the CB edge 
in [FM]M$^{IV}$X$^{VII}_{3}$ is attributed to the relatively low binding energy 
of the anti-bonding state of C$=$O  bond for the FM molecule. 
These states are strongly localized at the molecule sites. 
Though in [FM]M$^{IV}$X$^{VII}_{3}$ the states have negligible effects 
on the light absorption (Figure 10c) because of the large gap (about 8 eV)
 between their highest occupied and lowest unoccupied molecular orbitals (Figure 10b) 
and the forbidden optical transition between them and the states of M$^{IV}$-X$^{VII}$ framework, 
finding of such molecule derived states around band edges may represent an advance in having organic molecules 
involved in light absorption to increase solar capture efficiency. 
On the other hand, because of the strongly localized characteristic, 
they may behave as carrier trapping centers mimicking deep defect states, 
and thus impede photo-induced carrier separation and transport to electrodes. 
This would be detrimental to the solar-to-electric conversion efficiency. 
Therefore caution needs to be taken in designing the AM$^{IV}$X$^{VII}_3$ perovskites with such states 
in proximity to band edges.
\vspace{0.3cm}


\noindent{\textbf{6. Conclusion and Discussion.}}
\vspace{0.5cm}

In summary, we present via systemic first-principles quantum-mechanical calculations 
a materials-by-design exploration of high-performance hybrid organic-inorganic perovskites 
with desired intrinsic photovoltaic functionalities. 
Based on the knowledge of what makes this family of materials superior in solar cell applications, 
we formulate the scientific, materials-specific design metrics 
that consist of a series of computable quantities (\textit{e.g.}, thermodynamic and crystallographic stability, 
light absorption, carrier effective mass, dopability, exciton binding, etc.), 
and use them to screen the compounds that satisfy explicitly such target functionalities. 
We have identified in this materials screening program 18 winning compounds 
from a materials space composed of $\sim$ 100 candidates, including the most commonly used materials 
of [CH$_3$NH$_3$]PbI$_3$ and [CH(NH$_2$)$_2$]PbI$_3$. 
Of them 14 are Pb-free (including 5 Ge-based materials, i.e., 
CsGeI$_3$, CsGeBr$_3$, [NH$_3$OH]GeBr$_3$, [NH$_2$NH$_3$]GeBr$_3$, 
[CH$_3$NH$_3$]GeI$_3$ and 9 Sn-based materials, i.e., 
CsSnI$_3$, CsSnBr$_3$, CsSnCl$_3$, [CH$_3$NH$_3$]SnI$_3$, [CH$_3$NH$_3$]SnBr$_3$, 
[CH(NH$_2$)$_2$]SnI$_3$, [CH$_3$CH$_2$NH$_3$]SnI$_3$,
[C(NH$_2$)$_3$]SnI$_3$ and [NH$_2$(CH$_3$)$_2$]SnI$_3$), 
and 7 show substantially enhanced thermodynamic stability with respect 
to [CH$_3$NH$_3$]PbI$_3$ 
(i.e., CsSnBr$_{3}$, CsSnCl$_{3}$, CsGeBr$_{3}$, [CH$_{3}$NH$_{3}$]SnI$_{3}$,[CH$_{3}$NH$_{3}$]SnBr$_{3}$,
[CH$_3$CH$_2$NH$_3$]SnI$_3$ and [NH$_2$(CH$_3$)$_2$]SnI$_3$).
 Furthermore, we have discovered a distinct class of hybrid perovskites from the known ones, 
i.e., the compounds containing NH$_3$COH$^{+}$, 
exhibiting the organic molecule derived electronic states in proximity to band-edges. 
Our results also offer useful insights on the structure-property relationship, 
i.e., how microscopic structural changes affect photovoltaic relevant properties in this system. 
The current study provides a roadmap for acceleration of experimental discovery 
of alternative solar-energy hybrid perovskites to further boost the conversion efficiency 
and overcome the currently identified limitations of the existing materials. 
Finally we note that the current work focuses on selection of superior solar hybrid perovskite 
in the single composition. 
In reality the excellent photovoltaic performance can be achieved 
by compositional engineering of multiple perovskite materials, 
such as the conversion efficiency of 18\% reached by mixing  [CH(NH$_2$)$_2$]PbI$_3$ 
with [CH$_3$NH$_3$]PbBr$_3$.\cite{jeon2015compositional} 
In such cases, our results reported here on thethermodynamic stabilities, 
optical band gaps, carrier effective masses, etc. 
of a broad range of hybrid perovskites will essentially provide useful guidance 
for how to choose suitable materials for the compositional engineering.
\vspace{0.3cm}

Note that experimentally thus far [MA]PbI3 has been reported to make better performing solar cells 
than the Sn-and Ge-based alternatives. 
Particularly, even though our calculations predicts Sn-compounds to be more stable than 
the equivalent Pb-compounds as far as the tendency to disproportionation reaction 
AM$^{IV}$X$^{VII}_3$ $\rightarrow$  AX$^{VII}$ +M$^{IV}$X$^{VII}_2$ is concerned (Fig. 2), 
there are credible reports that the Sn-compounds 
often lead to poor-performance solar cells, 
showing intrinsic p-type conductivity, 
decrease in photocurrent density and fill factor.\cite{stoumpos2013semiconducting, hao2014lead, 
noel2014lead}
This suggests that additional criteria beyond intrinsic bulk properties may be at play. 
This is analogous to the replacement in 2CuInS$_{2}$ of 2In by Zn+Sn creating Cu$_{2}$ZnSnS$_{4}$ (CZTS) 
that involves Sn substitution. 
Because of the multivalency nature of Sn (being Sn$^{2+}$ as well as Sn$^{4+}$), 
associated deep defect center detrimental to photovoltaic performance may form.\cite{Biswas2010} 
Among AM$^{IV}$X$^{VII}_3$ perovskites, 
while Pb-compunds contain rather stable Pb$^{2+}$, 
oxidation disproportionation of Sn$^{2+}$ to Sn$^{4+}$ are prone to occur in Sn-compunds. 
The same physics might be at work here and is under investigation.
\vspace{0.3cm}

Whereas indeed some of our selected winning compounds 
may be known to experimentalists and tried for fabricating solar cells, 
we must emphasize that the goal of this work is to supplement the 'try-them-all' 
Edisonian shotgun approach usually used in this field (as well as in combinatorial chemistry) 
by an understanding based approach encompassing list of 'principles/filters' 
needed to make scientific choices, 
and to understand the performance of choices already made. 
Even if our selected compounds did not work experimentally, 
our study is still meaningful to the community - the implication is that more additional factors 
beyond the ones involved here need to be considered to reliably evaluate the photovoltaic performance 
of the AM$^{IV}$X$^{VII}_3$ material system. 
\vspace{0.3cm}

\noindent{\textbf{7. Experimental Section}}
\vspace{0.3cm}

All calculations are performed within the framework of density functional theory (DFT) 
using plane-wave pseudopotential method as implemented in the VASP code.\cite{kresse1996efficient}
The electron-ion interaction is described by means of 
projector-augmented wave (PAW) pseudopotentials.\cite{blochl1994projector} 
Configurations of ns$^2$np$^2$ for Pb/Sn/Ge, ns$^2$np$^5$ for I/Br/Cl, 2s$^2$2p$^2$ for C, 
2s$^2$2p$^3$ for N, 2s$^2$2p$^4$ for O and 1s for H are considered as valence electrons. 
The scalar-relativistic effects (including the mass-velocity and Darwin terms), 
which are especially important to high-Z elements of Pb/Sn, are included. 
The inclusion of semicore d electrons into the PAW pseudopotentials of Pb/Sn/Ge 
is found to have negligible effects on the calculated DMs. 
We use the Perdew-Burke-Ernzerhof (PBE) parameterization of the generalized gradient approximation\cite{perdew1996generalized} 
as the exchange-correlation functional. 
A plane-wave basis set with an energy cutoff of 520 eV and Monkhorst-Pack k-point meshes 
with spacing 2$\pi$$\times$0.05 $\AA$$^{-1}$ or less are used for total energy minimization. 
The convergence of total energy with respect to these parameters is examined 
and found to reach the level of less than 0.001 eV/f.u.. 
To properly take into account the long-range dispersion interactions 
that play important role in the current system involving organic molecules, 
we have carefully assessed performance of various van der Waals (vdWs) 
functionals in describing equilibrium lattice parameter of CH$_3$NH$_3$-containing materials, 
i.e., [CH$_3$NH$_3$]Pb(I/Br/Cl)$_3$ and [CH$_3$NH$_3$](Pb/Sn/Ge)I$_3$ in the cubic $\alpha$ phase, 
as depicted in the upper part of Table I. 
We find that both optB86b\cite{klimevs2009chemical} and D2-vdW\cite{grimme2006semiempirical} functionals 
give satisfied agreements between calculations and experiments 
(with the maximum deviation below 0.8\%). 
The former is chosen for the energetic calculations and structural optimizations. 
The hybrid functional (HSE)\cite{krukau2006influence} approach including the spin-orbit coupling (SOC) effect 
is used for accurately calculating band gaps. 
More specific details on calculating various compound-intrinsic DMs are described as below:
\vspace{0.3cm}

\begin{table*}
\label{Table I}
\caption{Calculated lattice constants and band gaps of [MA]Pb(I/Br/Cl)$_{3}$ and [MA](Pb/Sn/Ge)I$_{3}$ 
at different levels of theory, compared with the available 
experimental values.\cite{poglitsch1987dynamic}, \cite{ryu2014voltage}
For the HSE$^{*}$ + SOC calculations, 55\%, 70\% and 55\% percentage of exact Fock exchange 
are used for Pb, Sn and Ge-based perovskites, respectively 
[see Experimental Section (iii)].}
%
\centering
\begin{tabular}{clcccccc}
\hline\hline
Properties &  Materials & Different levels of theory & \\ 
\hline
                  &               & Exp          & optB86b & D$_2$ & vdw-DF2 & optPBE-vdw & DFT  \\
                  &[MA]PbCl$_{3}$ & 5.67$^{107}$ & 5.69    & 5.67  & 5.83       & 5.76       & 5.81  \\
                  &[MA]PbBr$_{3}$ & 5.90$^{107}$ & 5.95    & 5.91  & 6.13       & 6.03       & 6.07  \\
Lattice Constant  &[MA]PbI$_{3}$  & $6.31^{32}$ & 6.33    & 6.29  & 6.54       & 6.42       & 6.47  \\
a ($\mathrm {\AA}$)&[MA]SnI$_{3}$  & 6.23$^{28}$  & 6.26    & 6.24  & 6.50       & 6.38       & 6.44  \\
                  &[MA]GeI$_{3}$  & --           & 6.10    & 6.08  & 6.34       & 6.22       & 6.33 \\ 
\hline\hline
                  &               & Exp          & HSE$^{*}$+SOC & HSE+SOC & HSE        & PBE+SOC    & PBE \\
                  &[MA]PbCl$_{3}$ & 3.11$^{29}$  & 3.23          & 2.08    & 3.18       & 1.38       & 2.44  \\
                  &[MA]PbBr$_{3}$ & 2.35$^{108}$  & 2.43          & 1.41    & 2.50       & 0.81       & 1.86  \\
Band Gap (eV)     &[MA]PbI$_{3}$  & 1.52$^{32}$  & 1.55          & 0.75    & 1.90       & 0.27       & 1.37  \\
                  &[MA]SnI$_{3}$  & 1.20$^{28}$  & 1.26          & 0.40    & 0.75       & 0.03       & 0.37  \\
                  &[MA]GeI$_{3}$  & 1.90$^{80}$  & 1.98          & 1.31    & 1.41       & 0.84       & 0.96  \\
\hline\hline
%
\end{tabular}
\end{table*}

(i) The \textbf{structure optimization} of all the AM$^{IV}$X$^{VII}_3$ perovskites
is done in the $\alpha$ phase with cubic unit cell.
We have minimized the total energy with respect to the cubic lattice constant
and internal atom positions. The resulted tilting of M$^{IV}$X$^{VII}_{6}$ octahedra
from the broken symmetry caused by non-spherical A molecules is taken into account.
It is generally accepted that the small A molecule freely rotates in the room-temperature $\alpha$ phase
owing to the low rotation barrier.\cite{motta2015revealing, frost2014atomistic, wasylishen1985cation}
We here align the principal axes of small molecules along the $<$111$>$ direction
that is one of energetically favorable directions.\cite{yin2015halide}
The other align direction (\textit{e.g.} $<$100$>$) of small molecules is examined
and found to have negligibly minor effort on the compounds DMs
(see for instance the decomposition enthalpy results in Figure S5, Supporting Information).
\vspace{0.3cm}

(ii) The \textbf{decomposition enthalpy} ($\Delta$H) of an AM$^{IV}$X$^{VII}_3$ perovskite
is defined as the energy gain during the reaction channel of
AM$^{IV}$X$^{VII}_3$ $\rightarrow$  AX$^{VII}$+M$^{IV}$X$^{VII}_2$.
So positive $\Delta$H values indicate no decomposition.
This should be the predominant decomposition channel by considering the experimental synthesis routine
of AM$^{IV}$X$^{VII}_3$ materials usually involving M$^{IV}$X$^{VII}_2$.\cite{niu2015review, stoumpos2013semiconducting, safdari2015structure, baikie2013synthesis}
For AX$^{VII}$,
the cesium chloride structure, which is the $\epsilon$-phase of [CH$_3$NH$_3$]I,\cite{ishida1995exafs, ishida1982pre} is adopted.
For M$^{IV}$X$^{VII}_2$, the actual low-temperature structure from experiments is taken;
this does not include GeCl$_2$ that is experimentally absent,
and thus the $\Delta$H of AGeCl$_3$ is not calculated.
\vspace{0.3cm}

(iii) \textbf{Band gaps:} 
To reliably calculate band gaps of the AM$^{IV}$X$^{VII}_3$ perovskites, 
we examine performance of different levels of theoretical approaches for CH$_3$NH$_3$ containing materials, 
i.e., [CH$_3$NH$_3$]Pb(I/Br/Cl)$_3$ and [CH$_3$NH$_3$](Pb/Sn/Ge)I$_3$, 
as summarized in the lower part of Table I. 
As expected the DFT-PBE approach seriously underestimates the band gaps of all the materials 
due to the self-interaction error. 
The HSE approach (with the standard 25\% non-local Fock exchange) rectifies to some extent this problem 
and gives the larger gaps than experimental values 
for all the Pb-based compounds, but still underestimates those of Sn and Ge-based ones. 
The inclusion of the SOC results in expected significant reduction of gap values for Pb and Sn-based compounds. 
For all the materials, the HSE$+$SOC approach gives the lower gap values than the experimental ones. 
These results are generally consistent with the previous calculations,\cite{yin2015halide} 
and indicate the higher-level GW calculations with the SOC included are necessary 
for accurately calculating band gaps of this system.\cite{umari2014relativistic} 
Unfortunately the SOC-GW method is extremely computationally expensive, 
which is not affordable for the current materials screening involving hundred of materials. 
Taking into account these facts, we use the HSE+SOC method with 
adjustable percentage ($\alpha$) of exact Fock exchange (HSE$^{*}$ $+$ SOC). 
In particular, we adjusted the $\alpha$ values to reproduce the experimental gap values 
of [CH$_3$NH$_3$](Pb/Sn/Ge)I$_3$ and found that $\alpha$ $=$ 55\% 
for [CH$_3$NH$_3$]PbI$_3$, 70\% for [CH$_3$NH$_3$]SnI$_3$ and 55\% for [CH$_3$NH$_3$]GeI$_3$ 
give good agreements between theory and experiment (with the maximum deviation of 5\%). 
We then adopt this set of $\alpha$ values for the calculations of other Pb, Sn and Ge-based materials, respectively. 
This HSE$^{*}$+SOC approach, \textit{e.g.}, using the unified $\alpha$ for the materials 
containing the same M$^{IV}$ cation, 
is validated by the satisfied agreements of gap values between theory and experiment 
for [CH$_3$NH$_3$]PbBr$_3$ and [CH$_3$NH$_3$]PbCl$_3$.
\vspace{0.3cm}

(iv) \textbf{Steric sizes of organic cations:} 
To calculate the tolerance factor t  and the octahedral factor $\mu$ , 
one needs to know the ionic radii of A$^{+}$, M$^{IV2+}$, and X$^{VII-}$ ions.
 While the effective crystal ionic radii of M$^{IV2+}$, and X$^{VII-}$ (r$_M$  and r$_X$ ) 
are readily known,\cite{shannon1976revised} 
evaluation of the sizes of non-spherical organic A$^{+}$ cations (r$_A$ ) is challenging. 
Here we calculate  r$_A$ through the expression of within the idealized solid-sphere model, 
where a is cubic lattice constant from our theoretically optimized structures; 
the ionic radius of inorganic Cs$^{+}$ is used to derive the radii of other organic cations. 
For instance, the ionic radius of MA$^{+}$ (r$_{MA}$ ) is calculated via the relation of
(r$_{MA}$+r$_{X}$) / (r$_{Cs}$+r$_{X}$) = (a$_{[MA]MX_{3}}$) / (a$_{[Cs]MX_{3}}$),
and then the averaged value among all the (M$^{IV}$, X$^{VII}$) combinations is taken (see Figure S6, Supporting Information). 
\vspace{0.3cm}

 (v) The \textbf{effective mass} (m$^{*}$) tensors that relate directly to carrier's electrical conductivity, 
are calculated through the semiclassical Boltzmann transport theory.\cite{madsen2006boltztrap} 
The DFT-PBE derived Kohn-Sham eigenvalues at more dense k-points grid of 2$\pi$ $\times$ 0.016 $\AA^{-1}$ 
are used for these calculations. 
The unified carrier concentration of 1.0$\times$ 10$^{18}$ cm$^{-3}$ and room temperature of 300 K are chosen. 
This takes into account effects of non-parabolicity, 
anisotropy of bands, multiple bands, etc. on carrier transport. 
For all the AM$^{IV}$X$^{VII}_3$ materials in the cubic $\alpha$ phase, 
the anisotropy of m$^{*}$  tensors is small and the averaged m$^{*}$  among the values of 
three principal axes is taken as the DMs.
\vspace{0.3cm}

(vi) The \textbf{exciton bonding energy} 
(E$_B$) and Bohr radius ($\alpha_{ex}$) are calculated in terms of the hydrogen-like Wannier-Mott model 
with the calculated averaged  m$^{*}$ of electron and hole, 
as well as dielectric constant as the input parameters. 
Particularly, the E$_B$  is given by E$_B$ = $\mu^{*}R_{y}$ / $m_{0}\epsilon_{\varpropto}^{2}$ , 
where $\mu^{*}$  is the reduced exciton mass (i.e. 1/$\mu^{*}$ = 1/m$_{e}^{*}$ + 1/m$_{h}^{*}$ ),   
$R_{y}$ is the atomic Rydberg energy, 
and $\epsilon_{r}$  is the relative dielectric constant. 
Here the high-frequency limit of dielectric constant ($\epsilon_{\varpropto}$) 
resulted from electronic polarization, 
which is calculated from the finite-electric field calculations using the Berry-phase technique,\cite{umari2002ab, souza2002first}
is adopted as $\epsilon_{r}$. The resulted E$_B$  describes the excitons generated immediately 
after phonon excitation (without involving lattice relaxation). 
The $\alpha_{ex}$  is given by $\alpha_{ex}$ = $\alpha_{H}$ $\epsilon_{r}$ $m_{0}$/ $\mu^{*}$, 
where $\alpha_{H}$   is the Bohr radius.

\vspace{0.5cm}
\noindent{\textbf{ACKNOWLEDGMENT}}
\vspace{0.5cm}

The authors acknowledge funding support from the Recruitment Program of Global Youth Experts in China, 
National Key Research and Development Program of China (under Grants No. 2016YFB0201204), 
and the Special Fund for Talent Exploitation in Jilin Province of China. 
The work at CU was supported by the Office of Science, 
Basic Energy Science, MSE Division under Grant No. DE-FG02-13ER46959.
 Part of calculations was performed in the high performance computing center of Jilin University 
and on TianHe-1(A) of National Supercomputer Center in Tianjin.

\bibliography{tmp.bib}

\begin{thebibliography}{117}%
\makeatletter
\providecommand \@ifxundefined [1]{%
 \@ifx{#1\undefined}
}%
\providecommand \@ifnum [1]{%
 \ifnum #1\expandafter \@firstoftwo
 \else \expandafter \@secondoftwo
 \fi
}%
\providecommand \@ifx [1]{%
 \ifx #1\expandafter \@firstoftwo
 \else \expandafter \@secondoftwo
 \fi
}%
\providecommand \natexlab [1]{#1}%
\providecommand \enquote  [1]{``#1''}%
\providecommand \bibnamefont  [1]{#1}%
\providecommand \bibfnamefont [1]{#1}%
\providecommand \citenamefont [1]{#1}%
\providecommand \href@noop [0]{\@secondoftwo}%
\providecommand \href [0]{\begingroup \@sanitize@url \@href}%
\providecommand \@href[1]{\@@startlink{#1}\@@href}%
\providecommand \@@href[1]{\endgroup#1\@@endlink}%
\providecommand \@sanitize@url [0]{\catcode `\\12\catcode `\$12\catcode
  `\&12\catcode `\#12\catcode `\^12\catcode `\_12\catcode `\%12\relax}%
\providecommand \@@startlink[1]{}%
\providecommand \@@endlink[0]{}%
\providecommand \url  [0]{\begingroup\@sanitize@url \@url }%
\providecommand \@url [1]{\endgroup\@href {#1}{\urlprefix }}%
\providecommand \urlprefix  [0]{URL }%
\providecommand \Eprint [0]{\href }%
\providecommand \doibase [0]{http://dx.doi.org/}%
\providecommand \selectlanguage [0]{\@gobble}%
\providecommand \bibinfo  [0]{\@secondoftwo}%
\providecommand \bibfield  [0]{\@secondoftwo}%
\providecommand \translation [1]{[#1]}%
\providecommand \BibitemOpen [0]{}%
\providecommand \bibitemStop [0]{}%
\providecommand \bibitemNoStop [0]{.\EOS\space}%
\providecommand \EOS [0]{\spacefactor3000\relax}%
\providecommand \BibitemShut  [1]{\csname bibitem#1\endcsname}%
\let\auto@bib@innerbib\@empty
\bibitem [{\citenamefont {Kojima}\ \emph {et~al.}(2009)\citenamefont {Kojima},
  \citenamefont {Teshima}, \citenamefont {Shirai},\ and\ \citenamefont
  {Miyasaka}}]{kojima2009organometal}%
  \BibitemOpen
  \bibfield  {author} {\bibinfo {author} {\bibfnamefont {A.}~\bibnamefont
  {Kojima}}, \bibinfo {author} {\bibfnamefont {K.}~\bibnamefont {Teshima}},
  \bibinfo {author} {\bibfnamefont {Y.}~\bibnamefont {Shirai}}, \ and\ \bibinfo
  {author} {\bibfnamefont {T.}~\bibnamefont {Miyasaka}},\ }\href@noop {}
  {\bibfield  {journal} {\bibinfo  {journal} {Journal of the American Chemical
  Society}\ }\textbf {\bibinfo {volume} {131}},\ \bibinfo {pages} {6050}
  (\bibinfo {year} {2009})}\BibitemShut {NoStop}%
\bibitem [{\citenamefont {Lee}\ \emph {et~al.}(2012)\citenamefont {Lee},
  \citenamefont {Teuscher}, \citenamefont {Miyasaka}, \citenamefont
  {Murakami},\ and\ \citenamefont {Snaith}}]{lee2012efficient}%
  \BibitemOpen
  \bibfield  {author} {\bibinfo {author} {\bibfnamefont {M.~M.}\ \bibnamefont
  {Lee}}, \bibinfo {author} {\bibfnamefont {J.}~\bibnamefont {Teuscher}},
  \bibinfo {author} {\bibfnamefont {T.}~\bibnamefont {Miyasaka}}, \bibinfo
  {author} {\bibfnamefont {T.~N.}\ \bibnamefont {Murakami}}, \ and\ \bibinfo
  {author} {\bibfnamefont {H.~J.}\ \bibnamefont {Snaith}},\ }\href@noop {}
  {\bibfield  {journal} {\bibinfo  {journal} {Science}\ }\textbf {\bibinfo
  {volume} {338}},\ \bibinfo {pages} {643} (\bibinfo {year}
  {2012})}\BibitemShut {NoStop}%
\bibitem [{\citenamefont {Heo}\ \emph {et~al.}(2013)\citenamefont {Heo},
  \citenamefont {Im}, \citenamefont {Noh}, \citenamefont {Mandal},
  \citenamefont {Lim}, \citenamefont {Chang}, \citenamefont {Lee},
  \citenamefont {Kim}, \citenamefont {Sarkar}, \citenamefont {Nazeeruddin}
  \emph {et~al.}}]{heo2013efficient}%
  \BibitemOpen
  \bibfield  {author} {\bibinfo {author} {\bibfnamefont {J.~H.}\ \bibnamefont
  {Heo}}, \bibinfo {author} {\bibfnamefont {S.~H.}\ \bibnamefont {Im}},
  \bibinfo {author} {\bibfnamefont {J.~H.}\ \bibnamefont {Noh}}, \bibinfo
  {author} {\bibfnamefont {T.~N.}\ \bibnamefont {Mandal}}, \bibinfo {author}
  {\bibfnamefont {C.-S.}\ \bibnamefont {Lim}}, \bibinfo {author} {\bibfnamefont
  {J.~A.}\ \bibnamefont {Chang}}, \bibinfo {author} {\bibfnamefont {Y.~H.}\
  \bibnamefont {Lee}}, \bibinfo {author} {\bibfnamefont {H.-j.}\ \bibnamefont
  {Kim}}, \bibinfo {author} {\bibfnamefont {A.}~\bibnamefont {Sarkar}},
  \bibinfo {author} {\bibfnamefont {M.~K.}\ \bibnamefont {Nazeeruddin}},  \emph
  {et~al.},\ }\href@noop {} {\bibfield  {journal} {\bibinfo  {journal} {Nature
  Photonics}\ }\textbf {\bibinfo {volume} {7}},\ \bibinfo {pages} {486}
  (\bibinfo {year} {2013})}\BibitemShut {NoStop}%
\bibitem [{\citenamefont {Liu}\ \emph {et~al.}(2013)\citenamefont {Liu},
  \citenamefont {Johnston},\ and\ \citenamefont {Snaith}}]{liu2013efficient}%
  \BibitemOpen
  \bibfield  {author} {\bibinfo {author} {\bibfnamefont {M.}~\bibnamefont
  {Liu}}, \bibinfo {author} {\bibfnamefont {M.~B.}\ \bibnamefont {Johnston}}, \
  and\ \bibinfo {author} {\bibfnamefont {H.~J.}\ \bibnamefont {Snaith}},\
  }\href@noop {} {\bibfield  {journal} {\bibinfo  {journal} {Nature}\ }\textbf
  {\bibinfo {volume} {501}},\ \bibinfo {pages} {395} (\bibinfo {year}
  {2013})}\BibitemShut {NoStop}%
\bibitem [{\citenamefont {Burschka}\ \emph {et~al.}(2013)\citenamefont
  {Burschka}, \citenamefont {Pellet}, \citenamefont {Moon}, \citenamefont
  {Humphry-Baker}, \citenamefont {Gao}, \citenamefont {Nazeeruddin},\ and\
  \citenamefont {Gr{\"a}tzel}}]{burschka2013sequential}%
  \BibitemOpen
  \bibfield  {author} {\bibinfo {author} {\bibfnamefont {J.}~\bibnamefont
  {Burschka}}, \bibinfo {author} {\bibfnamefont {N.}~\bibnamefont {Pellet}},
  \bibinfo {author} {\bibfnamefont {S.-J.}\ \bibnamefont {Moon}}, \bibinfo
  {author} {\bibfnamefont {R.}~\bibnamefont {Humphry-Baker}}, \bibinfo {author}
  {\bibfnamefont {P.}~\bibnamefont {Gao}}, \bibinfo {author} {\bibfnamefont
  {M.~K.}\ \bibnamefont {Nazeeruddin}}, \ and\ \bibinfo {author} {\bibfnamefont
  {M.}~\bibnamefont {Gr{\"a}tzel}},\ }\href@noop {} {\bibfield  {journal}
  {\bibinfo  {journal} {Nature}\ }\textbf {\bibinfo {volume} {499}},\ \bibinfo
  {pages} {316} (\bibinfo {year} {2013})}\BibitemShut {NoStop}%
\bibitem [{\citenamefont {Jeon}\ \emph
  {et~al.}(2014{\natexlab{a}})\citenamefont {Jeon}, \citenamefont {Lee},
  \citenamefont {Kim}, \citenamefont {Seo}, \citenamefont {Noh}, \citenamefont
  {Lee},\ and\ \citenamefont {Seok}}]{jeon2014methoxy}%
  \BibitemOpen
  \bibfield  {author} {\bibinfo {author} {\bibfnamefont {N.~J.}\ \bibnamefont
  {Jeon}}, \bibinfo {author} {\bibfnamefont {H.~G.}\ \bibnamefont {Lee}},
  \bibinfo {author} {\bibfnamefont {Y.~C.}\ \bibnamefont {Kim}}, \bibinfo
  {author} {\bibfnamefont {J.}~\bibnamefont {Seo}}, \bibinfo {author}
  {\bibfnamefont {J.~H.}\ \bibnamefont {Noh}}, \bibinfo {author} {\bibfnamefont
  {J.}~\bibnamefont {Lee}}, \ and\ \bibinfo {author} {\bibfnamefont {S.~I.}\
  \bibnamefont {Seok}},\ }\href@noop {} {\bibfield  {journal} {\bibinfo
  {journal} {Journal of the American Chemical Society}\ }\textbf {\bibinfo
  {volume} {136}},\ \bibinfo {pages} {7837} (\bibinfo {year}
  {2014}{\natexlab{a}})}\BibitemShut {NoStop}%
\bibitem [{\citenamefont {Malinkiewicz}\ \emph {et~al.}(2014)\citenamefont
  {Malinkiewicz}, \citenamefont {Yella}, \citenamefont {Lee}, \citenamefont
  {Espallargas}, \citenamefont {Graetzel}, \citenamefont {Nazeeruddin},\ and\
  \citenamefont {Bolink}}]{malinkiewicz2014perovskite}%
  \BibitemOpen
  \bibfield  {author} {\bibinfo {author} {\bibfnamefont {O.}~\bibnamefont
  {Malinkiewicz}}, \bibinfo {author} {\bibfnamefont {A.}~\bibnamefont {Yella}},
  \bibinfo {author} {\bibfnamefont {Y.~H.}\ \bibnamefont {Lee}}, \bibinfo
  {author} {\bibfnamefont {G.~M.}\ \bibnamefont {Espallargas}}, \bibinfo
  {author} {\bibfnamefont {M.}~\bibnamefont {Graetzel}}, \bibinfo {author}
  {\bibfnamefont {M.~K.}\ \bibnamefont {Nazeeruddin}}, \ and\ \bibinfo {author}
  {\bibfnamefont {H.~J.}\ \bibnamefont {Bolink}},\ }\href@noop {} {\bibfield
  {journal} {\bibinfo  {journal} {Nature Photonics}\ }\textbf {\bibinfo
  {volume} {8}},\ \bibinfo {pages} {128} (\bibinfo {year} {2014})}\BibitemShut
  {NoStop}%
\bibitem [{\citenamefont {Zhou}\ \emph {et~al.}(2014)\citenamefont {Zhou},
  \citenamefont {Chen}, \citenamefont {Li}, \citenamefont {Luo}, \citenamefont
  {Song}, \citenamefont {Duan}, \citenamefont {Hong}, \citenamefont {You},
  \citenamefont {Liu},\ and\ \citenamefont {Yang}}]{zhou2014interface}%
  \BibitemOpen
  \bibfield  {author} {\bibinfo {author} {\bibfnamefont {H.}~\bibnamefont
  {Zhou}}, \bibinfo {author} {\bibfnamefont {Q.}~\bibnamefont {Chen}}, \bibinfo
  {author} {\bibfnamefont {G.}~\bibnamefont {Li}}, \bibinfo {author}
  {\bibfnamefont {S.}~\bibnamefont {Luo}}, \bibinfo {author} {\bibfnamefont
  {T.-b.}\ \bibnamefont {Song}}, \bibinfo {author} {\bibfnamefont {H.-S.}\
  \bibnamefont {Duan}}, \bibinfo {author} {\bibfnamefont {Z.}~\bibnamefont
  {Hong}}, \bibinfo {author} {\bibfnamefont {J.}~\bibnamefont {You}}, \bibinfo
  {author} {\bibfnamefont {Y.}~\bibnamefont {Liu}}, \ and\ \bibinfo {author}
  {\bibfnamefont {Y.}~\bibnamefont {Yang}},\ }\href@noop {} {\bibfield
  {journal} {\bibinfo  {journal} {Science}\ }\textbf {\bibinfo {volume}
  {345}},\ \bibinfo {pages} {542} (\bibinfo {year} {2014})}\BibitemShut
  {NoStop}%
\bibitem [{\citenamefont {Jeon}\ \emph
  {et~al.}(2014{\natexlab{b}})\citenamefont {Jeon}, \citenamefont {Noh},
  \citenamefont {Kim}, \citenamefont {Yang}, \citenamefont {Ryu},\ and\
  \citenamefont {Seok}}]{jeon2014solvent}%
  \BibitemOpen
  \bibfield  {author} {\bibinfo {author} {\bibfnamefont {N.~J.}\ \bibnamefont
  {Jeon}}, \bibinfo {author} {\bibfnamefont {J.~H.}\ \bibnamefont {Noh}},
  \bibinfo {author} {\bibfnamefont {Y.~C.}\ \bibnamefont {Kim}}, \bibinfo
  {author} {\bibfnamefont {W.~S.}\ \bibnamefont {Yang}}, \bibinfo {author}
  {\bibfnamefont {S.}~\bibnamefont {Ryu}}, \ and\ \bibinfo {author}
  {\bibfnamefont {S.~I.}\ \bibnamefont {Seok}},\ }\href@noop {} {\bibfield
  {journal} {\bibinfo  {journal} {Nature materials}\ }\textbf {\bibinfo
  {volume} {13}},\ \bibinfo {pages} {897} (\bibinfo {year}
  {2014}{\natexlab{b}})}\BibitemShut {NoStop}%
\bibitem [{\citenamefont {Jeon}\ \emph {et~al.}(2015)\citenamefont {Jeon},
  \citenamefont {Noh}, \citenamefont {Yang}, \citenamefont {Kim}, \citenamefont
  {Ryu}, \citenamefont {Seo},\ and\ \citenamefont
  {Seok}}]{jeon2015compositional}%
  \BibitemOpen
  \bibfield  {author} {\bibinfo {author} {\bibfnamefont {N.~J.}\ \bibnamefont
  {Jeon}}, \bibinfo {author} {\bibfnamefont {J.~H.}\ \bibnamefont {Noh}},
  \bibinfo {author} {\bibfnamefont {W.~S.}\ \bibnamefont {Yang}}, \bibinfo
  {author} {\bibfnamefont {Y.~C.}\ \bibnamefont {Kim}}, \bibinfo {author}
  {\bibfnamefont {S.}~\bibnamefont {Ryu}}, \bibinfo {author} {\bibfnamefont
  {J.}~\bibnamefont {Seo}}, \ and\ \bibinfo {author} {\bibfnamefont {S.~I.}\
  \bibnamefont {Seok}},\ }\href@noop {} {\bibfield  {journal} {\bibinfo
  {journal} {Nature}\ }\textbf {\bibinfo {volume} {517}},\ \bibinfo {pages}
  {476} (\bibinfo {year} {2015})}\BibitemShut {NoStop}%
\bibitem [{\citenamefont {Yang}\ \emph {et~al.}(2015)\citenamefont {Yang},
  \citenamefont {Noh}, \citenamefont {Jeon}, \citenamefont {Kim}, \citenamefont
  {Ryu}, \citenamefont {Seo},\ and\ \citenamefont {Seok}}]{yang2015high}%
  \BibitemOpen
  \bibfield  {author} {\bibinfo {author} {\bibfnamefont {W.~S.}\ \bibnamefont
  {Yang}}, \bibinfo {author} {\bibfnamefont {J.~H.}\ \bibnamefont {Noh}},
  \bibinfo {author} {\bibfnamefont {N.~J.}\ \bibnamefont {Jeon}}, \bibinfo
  {author} {\bibfnamefont {Y.~C.}\ \bibnamefont {Kim}}, \bibinfo {author}
  {\bibfnamefont {S.}~\bibnamefont {Ryu}}, \bibinfo {author} {\bibfnamefont
  {J.}~\bibnamefont {Seo}}, \ and\ \bibinfo {author} {\bibfnamefont {S.~I.}\
  \bibnamefont {Seok}},\ }\href@noop {} {\bibfield  {journal} {\bibinfo
  {journal} {Science}\ }\textbf {\bibinfo {volume} {348}},\ \bibinfo {pages}
  {1234} (\bibinfo {year} {2015})}\BibitemShut {NoStop}%
\bibitem [{pic()}]{picture2015}%
  \BibitemOpen
  \href@noop {} {\ }\bibinfo {note} {\url{http://www.nrel.gov/ncpv/images/
  efficiency_chart.jpg}}\BibitemShut {NoStop}%
\bibitem [{\citenamefont {Sum}\ and\ \citenamefont
  {Mathews}(2014)}]{sum2014advancements}%
  \BibitemOpen
  \bibfield  {author} {\bibinfo {author} {\bibfnamefont {T.~C.}\ \bibnamefont
  {Sum}}\ and\ \bibinfo {author} {\bibfnamefont {N.}~\bibnamefont {Mathews}},\
  }\href@noop {} {\bibfield  {journal} {\bibinfo  {journal} {Energy \&
  Environmental Science}\ }\textbf {\bibinfo {volume} {7}},\ \bibinfo {pages}
  {2518} (\bibinfo {year} {2014})}\BibitemShut {NoStop}%
\bibitem [{\citenamefont {Scandale}\ \emph {et~al.}(2007)\citenamefont
  {Scandale}, \citenamefont {Still}, \citenamefont {Carnera}, \citenamefont
  {Della~Mea}, \citenamefont {De~Salvador}, \citenamefont {Milan},
  \citenamefont {Vomiero}, \citenamefont {Baricordi}, \citenamefont {Dalpiaz},
  \citenamefont {Fiorini} \emph {et~al.}}]{scandale2007high}%
  \BibitemOpen
  \bibfield  {author} {\bibinfo {author} {\bibfnamefont {W.}~\bibnamefont
  {Scandale}}, \bibinfo {author} {\bibfnamefont {D.~A.}\ \bibnamefont {Still}},
  \bibinfo {author} {\bibfnamefont {A.}~\bibnamefont {Carnera}}, \bibinfo
  {author} {\bibfnamefont {G.}~\bibnamefont {Della~Mea}}, \bibinfo {author}
  {\bibfnamefont {D.}~\bibnamefont {De~Salvador}}, \bibinfo {author}
  {\bibfnamefont {R.}~\bibnamefont {Milan}}, \bibinfo {author} {\bibfnamefont
  {A.}~\bibnamefont {Vomiero}}, \bibinfo {author} {\bibfnamefont
  {S.}~\bibnamefont {Baricordi}}, \bibinfo {author} {\bibfnamefont
  {P.}~\bibnamefont {Dalpiaz}}, \bibinfo {author} {\bibfnamefont
  {M.}~\bibnamefont {Fiorini}},  \emph {et~al.},\ }\href@noop {} {\bibfield
  {journal} {\bibinfo  {journal} {Physical review letters}\ }\textbf {\bibinfo
  {volume} {98}},\ \bibinfo {pages} {154801} (\bibinfo {year}
  {2007})}\BibitemShut {NoStop}%
\bibitem [{\citenamefont {Zeng}\ \emph {et~al.}(2006)\citenamefont {Zeng},
  \citenamefont {Yi}, \citenamefont {Hong}, \citenamefont {Liu}, \citenamefont
  {Feng}, \citenamefont {Duan}, \citenamefont {Kimerling},\ and\ \citenamefont
  {Alamariu}}]{zeng2006efficiency}%
  \BibitemOpen
  \bibfield  {author} {\bibinfo {author} {\bibfnamefont {L.}~\bibnamefont
  {Zeng}}, \bibinfo {author} {\bibfnamefont {Y.}~\bibnamefont {Yi}}, \bibinfo
  {author} {\bibfnamefont {C.}~\bibnamefont {Hong}}, \bibinfo {author}
  {\bibfnamefont {J.}~\bibnamefont {Liu}}, \bibinfo {author} {\bibfnamefont
  {N.}~\bibnamefont {Feng}}, \bibinfo {author} {\bibfnamefont {X.}~\bibnamefont
  {Duan}}, \bibinfo {author} {\bibfnamefont {L.}~\bibnamefont {Kimerling}}, \
  and\ \bibinfo {author} {\bibfnamefont {B.}~\bibnamefont {Alamariu}},\
  }\href@noop {} {\bibfield  {journal} {\bibinfo  {journal} {Applied Physics
  Letters}\ }\textbf {\bibinfo {volume} {89}},\ \bibinfo {pages} {111111}
  (\bibinfo {year} {2006})}\BibitemShut {NoStop}%
\bibitem [{\citenamefont {Wu}(2004)}]{wu2004high}%
  \BibitemOpen
  \bibfield  {author} {\bibinfo {author} {\bibfnamefont {X.}~\bibnamefont
  {Wu}},\ }\href@noop {} {\bibfield  {journal} {\bibinfo  {journal} {Solar
  energy}\ }\textbf {\bibinfo {volume} {77}},\ \bibinfo {pages} {803} (\bibinfo
  {year} {2004})}\BibitemShut {NoStop}%
\bibitem [{\citenamefont {Chiril{\u{a}}}\ \emph {et~al.}(2011)\citenamefont
  {Chiril{\u{a}}}, \citenamefont {Buecheler}, \citenamefont {Pianezzi},
  \citenamefont {Bloesch}, \citenamefont {Gretener}, \citenamefont {Uhl},
  \citenamefont {Fella}, \citenamefont {Kranz}, \citenamefont {Perrenoud},
  \citenamefont {Seyrling} \emph {et~al.}}]{chirilua2011highly}%
  \BibitemOpen
  \bibfield  {author} {\bibinfo {author} {\bibfnamefont {A.}~\bibnamefont
  {Chiril{\u{a}}}}, \bibinfo {author} {\bibfnamefont {S.}~\bibnamefont
  {Buecheler}}, \bibinfo {author} {\bibfnamefont {F.}~\bibnamefont {Pianezzi}},
  \bibinfo {author} {\bibfnamefont {P.}~\bibnamefont {Bloesch}}, \bibinfo
  {author} {\bibfnamefont {C.}~\bibnamefont {Gretener}}, \bibinfo {author}
  {\bibfnamefont {A.~R.}\ \bibnamefont {Uhl}}, \bibinfo {author} {\bibfnamefont
  {C.}~\bibnamefont {Fella}}, \bibinfo {author} {\bibfnamefont
  {L.}~\bibnamefont {Kranz}}, \bibinfo {author} {\bibfnamefont
  {J.}~\bibnamefont {Perrenoud}}, \bibinfo {author} {\bibfnamefont
  {S.}~\bibnamefont {Seyrling}},  \emph {et~al.},\ }\href@noop {} {\bibfield
  {journal} {\bibinfo  {journal} {Nature materials}\ }\textbf {\bibinfo
  {volume} {10}},\ \bibinfo {pages} {857} (\bibinfo {year} {2011})}\BibitemShut
  {NoStop}%
\bibitem [{\citenamefont {Chen}\ \emph {et~al.}(2015)\citenamefont {Chen},
  \citenamefont {Wu}, \citenamefont {Yue}, \citenamefont {Liu}, \citenamefont
  {Zhang}, \citenamefont {Yang}, \citenamefont {Chen}, \citenamefont {Bi},
  \citenamefont {Ashraful}, \citenamefont {Gr{\"a}tzel} \emph
  {et~al.}}]{chen2015efficient}%
  \BibitemOpen
  \bibfield  {author} {\bibinfo {author} {\bibfnamefont {W.}~\bibnamefont
  {Chen}}, \bibinfo {author} {\bibfnamefont {Y.}~\bibnamefont {Wu}}, \bibinfo
  {author} {\bibfnamefont {Y.}~\bibnamefont {Yue}}, \bibinfo {author}
  {\bibfnamefont {J.}~\bibnamefont {Liu}}, \bibinfo {author} {\bibfnamefont
  {W.}~\bibnamefont {Zhang}}, \bibinfo {author} {\bibfnamefont
  {X.}~\bibnamefont {Yang}}, \bibinfo {author} {\bibfnamefont {H.}~\bibnamefont
  {Chen}}, \bibinfo {author} {\bibfnamefont {E.}~\bibnamefont {Bi}}, \bibinfo
  {author} {\bibfnamefont {I.}~\bibnamefont {Ashraful}}, \bibinfo {author}
  {\bibfnamefont {M.}~\bibnamefont {Gr{\"a}tzel}},  \emph {et~al.},\
  }\href@noop {} {\bibfield  {journal} {\bibinfo  {journal} {Science}\ }\textbf
  {\bibinfo {volume} {350}},\ \bibinfo {pages} {944} (\bibinfo {year}
  {2015})}\BibitemShut {NoStop}%
\bibitem [{\citenamefont {Liu}\ and\ \citenamefont
  {Kelly}(2014)}]{liu2014perovskite}%
  \BibitemOpen
  \bibfield  {author} {\bibinfo {author} {\bibfnamefont {D.}~\bibnamefont
  {Liu}}\ and\ \bibinfo {author} {\bibfnamefont {T.~L.}\ \bibnamefont
  {Kelly}},\ }\href@noop {} {\bibfield  {journal} {\bibinfo  {journal} {Nature
  photonics}\ }\textbf {\bibinfo {volume} {8}},\ \bibinfo {pages} {133}
  (\bibinfo {year} {2014})}\BibitemShut {NoStop}%
\bibitem [{\citenamefont {Chen}\ \emph {et~al.}(2013)\citenamefont {Chen},
  \citenamefont {Zhou}, \citenamefont {Hong}, \citenamefont {Luo},
  \citenamefont {Duan}, \citenamefont {Wang}, \citenamefont {Liu},
  \citenamefont {Li},\ and\ \citenamefont {Yang}}]{chen2013planar}%
  \BibitemOpen
  \bibfield  {author} {\bibinfo {author} {\bibfnamefont {Q.}~\bibnamefont
  {Chen}}, \bibinfo {author} {\bibfnamefont {H.}~\bibnamefont {Zhou}}, \bibinfo
  {author} {\bibfnamefont {Z.}~\bibnamefont {Hong}}, \bibinfo {author}
  {\bibfnamefont {S.}~\bibnamefont {Luo}}, \bibinfo {author} {\bibfnamefont
  {H.-S.}\ \bibnamefont {Duan}}, \bibinfo {author} {\bibfnamefont {H.-H.}\
  \bibnamefont {Wang}}, \bibinfo {author} {\bibfnamefont {Y.}~\bibnamefont
  {Liu}}, \bibinfo {author} {\bibfnamefont {G.}~\bibnamefont {Li}}, \ and\
  \bibinfo {author} {\bibfnamefont {Y.}~\bibnamefont {Yang}},\ }\href@noop {}
  {\bibfield  {journal} {\bibinfo  {journal} {Journal of the American Chemical
  Society}\ }\textbf {\bibinfo {volume} {136}},\ \bibinfo {pages} {622}
  (\bibinfo {year} {2013})}\BibitemShut {NoStop}%
\bibitem [{\citenamefont {Docampo}\ \emph {et~al.}(2013)\citenamefont
  {Docampo}, \citenamefont {Ball}, \citenamefont {Darwich}, \citenamefont
  {Eperon},\ and\ \citenamefont {Snaith}}]{docampo2013efficient}%
  \BibitemOpen
  \bibfield  {author} {\bibinfo {author} {\bibfnamefont {P.}~\bibnamefont
  {Docampo}}, \bibinfo {author} {\bibfnamefont {J.~M.}\ \bibnamefont {Ball}},
  \bibinfo {author} {\bibfnamefont {M.}~\bibnamefont {Darwich}}, \bibinfo
  {author} {\bibfnamefont {G.~E.}\ \bibnamefont {Eperon}}, \ and\ \bibinfo
  {author} {\bibfnamefont {H.~J.}\ \bibnamefont {Snaith}},\ }\href@noop {}
  {\bibfield  {journal} {\bibinfo  {journal} {Nature communications}\ }\textbf
  {\bibinfo {volume} {4}} (\bibinfo {year} {2013})}\BibitemShut {NoStop}%
\bibitem [{\citenamefont {Green}\ \emph {et~al.}(2014)\citenamefont {Green},
  \citenamefont {Ho-Baillie},\ and\ \citenamefont
  {Snaith}}]{green2014emergence}%
  \BibitemOpen
  \bibfield  {author} {\bibinfo {author} {\bibfnamefont {M.~A.}\ \bibnamefont
  {Green}}, \bibinfo {author} {\bibfnamefont {A.}~\bibnamefont {Ho-Baillie}}, \
  and\ \bibinfo {author} {\bibfnamefont {H.~J.}\ \bibnamefont {Snaith}},\
  }\href@noop {} {\bibfield  {journal} {\bibinfo  {journal} {Nature Photonics}\
  }\textbf {\bibinfo {volume} {8}},\ \bibinfo {pages} {506} (\bibinfo {year}
  {2014})}\BibitemShut {NoStop}%
\bibitem [{\citenamefont {Berry}\ \emph {et~al.}(2015)\citenamefont {Berry},
  \citenamefont {Buonassisi}, \citenamefont {Egger}, \citenamefont {Hodes},
  \citenamefont {Kronik}, \citenamefont {Loo}, \citenamefont {Lubomirsky},
  \citenamefont {Marder}, \citenamefont {Mastai}, \citenamefont {Miller} \emph
  {et~al.}}]{berry2015hybrid}%
  \BibitemOpen
  \bibfield  {author} {\bibinfo {author} {\bibfnamefont {J.}~\bibnamefont
  {Berry}}, \bibinfo {author} {\bibfnamefont {T.}~\bibnamefont {Buonassisi}},
  \bibinfo {author} {\bibfnamefont {D.~A.}\ \bibnamefont {Egger}}, \bibinfo
  {author} {\bibfnamefont {G.}~\bibnamefont {Hodes}}, \bibinfo {author}
  {\bibfnamefont {L.}~\bibnamefont {Kronik}}, \bibinfo {author} {\bibfnamefont
  {Y.-L.}\ \bibnamefont {Loo}}, \bibinfo {author} {\bibfnamefont
  {I.}~\bibnamefont {Lubomirsky}}, \bibinfo {author} {\bibfnamefont {S.~R.}\
  \bibnamefont {Marder}}, \bibinfo {author} {\bibfnamefont {Y.}~\bibnamefont
  {Mastai}}, \bibinfo {author} {\bibfnamefont {J.~S.}\ \bibnamefont {Miller}},
  \emph {et~al.},\ }\href@noop {} {\bibfield  {journal} {\bibinfo  {journal}
  {Advanced Materials}\ }\textbf {\bibinfo {volume} {27}},\ \bibinfo {pages}
  {5102} (\bibinfo {year} {2015})}\BibitemShut {NoStop}%
\bibitem [{\citenamefont {Niu}\ \emph {et~al.}(2015)\citenamefont {Niu},
  \citenamefont {Guo},\ and\ \citenamefont {Wang}}]{niu2015review}%
  \BibitemOpen
  \bibfield  {author} {\bibinfo {author} {\bibfnamefont {G.}~\bibnamefont
  {Niu}}, \bibinfo {author} {\bibfnamefont {X.}~\bibnamefont {Guo}}, \ and\
  \bibinfo {author} {\bibfnamefont {L.}~\bibnamefont {Wang}},\ }\href@noop {}
  {\bibfield  {journal} {\bibinfo  {journal} {Journal of Materials Chemistry
  A}\ }\textbf {\bibinfo {volume} {3}},\ \bibinfo {pages} {8970} (\bibinfo
  {year} {2015})}\BibitemShut {NoStop}%
\bibitem [{\citenamefont {Zhang}\ \emph {et~al.}(2015)\citenamefont {Zhang},
  \citenamefont {Chen}, \citenamefont {Xu}, \citenamefont {Xiang},
  \citenamefont {Gong}, \citenamefont {Walsh},\ and\ \citenamefont
  {Wei}}]{zhang2015intrinsic}%
  \BibitemOpen
  \bibfield  {author} {\bibinfo {author} {\bibfnamefont {Y.-Y.}\ \bibnamefont
  {Zhang}}, \bibinfo {author} {\bibfnamefont {S.}~\bibnamefont {Chen}},
  \bibinfo {author} {\bibfnamefont {P.}~\bibnamefont {Xu}}, \bibinfo {author}
  {\bibfnamefont {H.}~\bibnamefont {Xiang}}, \bibinfo {author} {\bibfnamefont
  {X.-G.}\ \bibnamefont {Gong}}, \bibinfo {author} {\bibfnamefont
  {A.}~\bibnamefont {Walsh}}, \ and\ \bibinfo {author} {\bibfnamefont {S.-H.}\
  \bibnamefont {Wei}},\ }\href@noop {} {\bibfield  {journal} {\bibinfo
  {journal} {arXiv preprint arXiv:1506.01301}\ } (\bibinfo {year}
  {2015})}\BibitemShut {NoStop}%
\bibitem [{\citenamefont {Conings}\ \emph {et~al.}(2015)\citenamefont
  {Conings}, \citenamefont {Drijkoningen}, \citenamefont {Gauquelin},
  \citenamefont {Babayigit}, \citenamefont {D'Haen}, \citenamefont
  {D'Olieslaeger}, \citenamefont {Ethirajan}, \citenamefont {Verbeeck},
  \citenamefont {Manca}, \citenamefont {Mosconi} \emph
  {et~al.}}]{conings2015intrinsic}%
  \BibitemOpen
  \bibfield  {author} {\bibinfo {author} {\bibfnamefont {B.}~\bibnamefont
  {Conings}}, \bibinfo {author} {\bibfnamefont {J.}~\bibnamefont
  {Drijkoningen}}, \bibinfo {author} {\bibfnamefont {N.}~\bibnamefont
  {Gauquelin}}, \bibinfo {author} {\bibfnamefont {A.}~\bibnamefont
  {Babayigit}}, \bibinfo {author} {\bibfnamefont {J.}~\bibnamefont {D'Haen}},
  \bibinfo {author} {\bibfnamefont {L.}~\bibnamefont {D'Olieslaeger}}, \bibinfo
  {author} {\bibfnamefont {A.}~\bibnamefont {Ethirajan}}, \bibinfo {author}
  {\bibfnamefont {J.}~\bibnamefont {Verbeeck}}, \bibinfo {author}
  {\bibfnamefont {J.}~\bibnamefont {Manca}}, \bibinfo {author} {\bibfnamefont
  {E.}~\bibnamefont {Mosconi}},  \emph {et~al.},\ }\href@noop {} {\bibfield
  {journal} {\bibinfo  {journal} {Advanced Energy Materials}\ }\textbf
  {\bibinfo {volume} {5}} (\bibinfo {year} {2015})}\BibitemShut {NoStop}%
\bibitem [{\citenamefont {Kim}\ \emph {et~al.}(2012)\citenamefont {Kim},
  \citenamefont {Lee}, \citenamefont {Im}, \citenamefont {Lee}, \citenamefont
  {Moehl}, \citenamefont {Marchioro}, \citenamefont {Moon}, \citenamefont
  {Humphry-Baker}, \citenamefont {Yum}, \citenamefont {Moser} \emph
  {et~al.}}]{kim2012lead}%
  \BibitemOpen
  \bibfield  {author} {\bibinfo {author} {\bibfnamefont {H.-S.}\ \bibnamefont
  {Kim}}, \bibinfo {author} {\bibfnamefont {C.-R.}\ \bibnamefont {Lee}},
  \bibinfo {author} {\bibfnamefont {J.-H.}\ \bibnamefont {Im}}, \bibinfo
  {author} {\bibfnamefont {K.-B.}\ \bibnamefont {Lee}}, \bibinfo {author}
  {\bibfnamefont {T.}~\bibnamefont {Moehl}}, \bibinfo {author} {\bibfnamefont
  {A.}~\bibnamefont {Marchioro}}, \bibinfo {author} {\bibfnamefont {S.-J.}\
  \bibnamefont {Moon}}, \bibinfo {author} {\bibfnamefont {R.}~\bibnamefont
  {Humphry-Baker}}, \bibinfo {author} {\bibfnamefont {J.-H.}\ \bibnamefont
  {Yum}}, \bibinfo {author} {\bibfnamefont {J.~E.}\ \bibnamefont {Moser}},
  \emph {et~al.},\ }\href@noop {} {\bibfield  {journal} {\bibinfo  {journal}
  {Scientific reports}\ }\textbf {\bibinfo {volume} {2}},\ \bibinfo {pages}
  {591} (\bibinfo {year} {2012})}\BibitemShut {NoStop}%
\bibitem [{\citenamefont {Stoumpos}\ \emph
  {et~al.}(2013{\natexlab{a}})\citenamefont {Stoumpos}, \citenamefont
  {Malliakas},\ and\ \citenamefont {Kanatzidis}}]{stoumpos2013semiconducting}%
  \BibitemOpen
  \bibfield  {author} {\bibinfo {author} {\bibfnamefont {C.~C.}\ \bibnamefont
  {Stoumpos}}, \bibinfo {author} {\bibfnamefont {C.~D.}\ \bibnamefont
  {Malliakas}}, \ and\ \bibinfo {author} {\bibfnamefont {M.~G.}\ \bibnamefont
  {Kanatzidis}},\ }\href@noop {} {\bibfield  {journal} {\bibinfo  {journal}
  {Inorganic chemistry}\ }\textbf {\bibinfo {volume} {52}},\ \bibinfo {pages}
  {9019} (\bibinfo {year} {2013}{\natexlab{a}})}\BibitemShut {NoStop}%
\bibitem [{\citenamefont {Kitazawa}\ \emph {et~al.}(2002)\citenamefont
  {Kitazawa}, \citenamefont {Watanabe},\ and\ \citenamefont
  {Nakamura}}]{kitazawa2002optical}%
  \BibitemOpen
  \bibfield  {author} {\bibinfo {author} {\bibfnamefont {N.}~\bibnamefont
  {Kitazawa}}, \bibinfo {author} {\bibfnamefont {Y.}~\bibnamefont {Watanabe}},
  \ and\ \bibinfo {author} {\bibfnamefont {Y.}~\bibnamefont {Nakamura}},\
  }\href@noop {} {\bibfield  {journal} {\bibinfo  {journal} {Journal of
  materials science}\ }\textbf {\bibinfo {volume} {37}},\ \bibinfo {pages}
  {3585} (\bibinfo {year} {2002})}\BibitemShut {NoStop}%
\bibitem [{\citenamefont {Shockley}\ and\ \citenamefont
  {Queisser}(1961)}]{shockley1961detailed}%
  \BibitemOpen
  \bibfield  {author} {\bibinfo {author} {\bibfnamefont {W.}~\bibnamefont
  {Shockley}}\ and\ \bibinfo {author} {\bibfnamefont {H.~J.}\ \bibnamefont
  {Queisser}},\ }\href@noop {} {\bibfield  {journal} {\bibinfo  {journal}
  {Journal of applied physics}\ }\textbf {\bibinfo {volume} {32}},\ \bibinfo
  {pages} {510} (\bibinfo {year} {1961})}\BibitemShut {NoStop}%
\bibitem [{\citenamefont {Snaith}\ \emph {et~al.}(2014)\citenamefont {Snaith},
  \citenamefont {Abate}, \citenamefont {Ball}, \citenamefont {Eperon},
  \citenamefont {Leijtens}, \citenamefont {Noel}, \citenamefont {Stranks},
  \citenamefont {Wang}, \citenamefont {Wojciechowski},\ and\ \citenamefont
  {Zhang}}]{snaith2014anomalous}%
  \BibitemOpen
  \bibfield  {author} {\bibinfo {author} {\bibfnamefont {H.~J.}\ \bibnamefont
  {Snaith}}, \bibinfo {author} {\bibfnamefont {A.}~\bibnamefont {Abate}},
  \bibinfo {author} {\bibfnamefont {J.~M.}\ \bibnamefont {Ball}}, \bibinfo
  {author} {\bibfnamefont {G.~E.}\ \bibnamefont {Eperon}}, \bibinfo {author}
  {\bibfnamefont {T.}~\bibnamefont {Leijtens}}, \bibinfo {author}
  {\bibfnamefont {N.~K.}\ \bibnamefont {Noel}}, \bibinfo {author}
  {\bibfnamefont {S.~D.}\ \bibnamefont {Stranks}}, \bibinfo {author}
  {\bibfnamefont {J.~T.-W.}\ \bibnamefont {Wang}}, \bibinfo {author}
  {\bibfnamefont {K.}~\bibnamefont {Wojciechowski}}, \ and\ \bibinfo {author}
  {\bibfnamefont {W.}~\bibnamefont {Zhang}},\ }\href@noop {} {\bibfield
  {journal} {\bibinfo  {journal} {The journal of physical chemistry letters}\
  }\textbf {\bibinfo {volume} {5}},\ \bibinfo {pages} {1511} (\bibinfo {year}
  {2014})}\BibitemShut {NoStop}%
\bibitem [{\citenamefont {Lee}\ \emph {et~al.}(2014)\citenamefont {Lee},
  \citenamefont {Seol}, \citenamefont {Cho},\ and\ \citenamefont
  {Park}}]{lee2014high}%
  \BibitemOpen
  \bibfield  {author} {\bibinfo {author} {\bibfnamefont {J.-W.}\ \bibnamefont
  {Lee}}, \bibinfo {author} {\bibfnamefont {D.-J.}\ \bibnamefont {Seol}},
  \bibinfo {author} {\bibfnamefont {A.-N.}\ \bibnamefont {Cho}}, \ and\
  \bibinfo {author} {\bibfnamefont {N.-G.}\ \bibnamefont {Park}},\ }\href@noop
  {} {\bibfield  {journal} {\bibinfo  {journal} {Advanced Materials}\ }\textbf
  {\bibinfo {volume} {26}},\ \bibinfo {pages} {4991} (\bibinfo {year}
  {2014})}\BibitemShut {NoStop}%
\bibitem [{\citenamefont {Eperon}\ \emph {et~al.}(2014)\citenamefont {Eperon},
  \citenamefont {Stranks}, \citenamefont {Menelaou}, \citenamefont {Johnston},
  \citenamefont {Herz},\ and\ \citenamefont
  {Snaith}}]{eperon2014formamidinium}%
  \BibitemOpen
  \bibfield  {author} {\bibinfo {author} {\bibfnamefont {G.~E.}\ \bibnamefont
  {Eperon}}, \bibinfo {author} {\bibfnamefont {S.~D.}\ \bibnamefont {Stranks}},
  \bibinfo {author} {\bibfnamefont {C.}~\bibnamefont {Menelaou}}, \bibinfo
  {author} {\bibfnamefont {M.~B.}\ \bibnamefont {Johnston}}, \bibinfo {author}
  {\bibfnamefont {L.~M.}\ \bibnamefont {Herz}}, \ and\ \bibinfo {author}
  {\bibfnamefont {H.~J.}\ \bibnamefont {Snaith}},\ }\href@noop {} {\bibfield
  {journal} {\bibinfo  {journal} {Energy \& Environmental Science}\ }\textbf
  {\bibinfo {volume} {7}},\ \bibinfo {pages} {982} (\bibinfo {year}
  {2014})}\BibitemShut {NoStop}%
\bibitem [{\citenamefont {Koh}\ \emph {et~al.}(2013)\citenamefont {Koh},
  \citenamefont {Fu}, \citenamefont {Fang}, \citenamefont {Chen}, \citenamefont
  {Sum}, \citenamefont {Mathews}, \citenamefont {Mhaisalkar}, \citenamefont
  {Boix},\ and\ \citenamefont {Baikie}}]{koh2013formamidinium}%
  \BibitemOpen
  \bibfield  {author} {\bibinfo {author} {\bibfnamefont {T.~M.}\ \bibnamefont
  {Koh}}, \bibinfo {author} {\bibfnamefont {K.}~\bibnamefont {Fu}}, \bibinfo
  {author} {\bibfnamefont {Y.}~\bibnamefont {Fang}}, \bibinfo {author}
  {\bibfnamefont {S.}~\bibnamefont {Chen}}, \bibinfo {author} {\bibfnamefont
  {T.}~\bibnamefont {Sum}}, \bibinfo {author} {\bibfnamefont {N.}~\bibnamefont
  {Mathews}}, \bibinfo {author} {\bibfnamefont {S.~G.}\ \bibnamefont
  {Mhaisalkar}}, \bibinfo {author} {\bibfnamefont {P.~P.}\ \bibnamefont
  {Boix}}, \ and\ \bibinfo {author} {\bibfnamefont {T.}~\bibnamefont
  {Baikie}},\ }\href@noop {} {\bibfield  {journal} {\bibinfo  {journal} {The
  Journal of Physical Chemistry C}\ }\textbf {\bibinfo {volume} {118}},\
  \bibinfo {pages} {16458} (\bibinfo {year} {2013})}\BibitemShut {NoStop}%
\bibitem [{\citenamefont {Loi}\ and\ \citenamefont
  {Hummelen}(2013)}]{loi2013hybrid}%
  \BibitemOpen
  \bibfield  {author} {\bibinfo {author} {\bibfnamefont {M.~A.}\ \bibnamefont
  {Loi}}\ and\ \bibinfo {author} {\bibfnamefont {J.~C.}\ \bibnamefont
  {Hummelen}},\ }\href@noop {} {\bibfield  {journal} {\bibinfo  {journal}
  {Nature materials}\ }\textbf {\bibinfo {volume} {12}},\ \bibinfo {pages}
  {1087} (\bibinfo {year} {2013})}\BibitemShut {NoStop}%
\bibitem [{\citenamefont {Yin}\ \emph {et~al.}(2015)\citenamefont {Yin},
  \citenamefont {Yang}, \citenamefont {Kang}, \citenamefont {Yan},\ and\
  \citenamefont {Wei}}]{yin2015halide}%
  \BibitemOpen
  \bibfield  {author} {\bibinfo {author} {\bibfnamefont {W.-J.}\ \bibnamefont
  {Yin}}, \bibinfo {author} {\bibfnamefont {J.-H.}\ \bibnamefont {Yang}},
  \bibinfo {author} {\bibfnamefont {J.}~\bibnamefont {Kang}}, \bibinfo {author}
  {\bibfnamefont {Y.}~\bibnamefont {Yan}}, \ and\ \bibinfo {author}
  {\bibfnamefont {S.-H.}\ \bibnamefont {Wei}},\ }\href@noop {} {\bibfield
  {journal} {\bibinfo  {journal} {Journal of Materials Chemistry A}\ }\textbf
  {\bibinfo {volume} {3}},\ \bibinfo {pages} {8926} (\bibinfo {year}
  {2015})}\BibitemShut {NoStop}%
\bibitem [{\citenamefont {Hao}\ \emph {et~al.}(2014)\citenamefont {Hao},
  \citenamefont {Stoumpos}, \citenamefont {Cao}, \citenamefont {Chang},\ and\
  \citenamefont {Kanatzidis}}]{hao2014lead}%
  \BibitemOpen
  \bibfield  {author} {\bibinfo {author} {\bibfnamefont {F.}~\bibnamefont
  {Hao}}, \bibinfo {author} {\bibfnamefont {C.~C.}\ \bibnamefont {Stoumpos}},
  \bibinfo {author} {\bibfnamefont {D.~H.}\ \bibnamefont {Cao}}, \bibinfo
  {author} {\bibfnamefont {R.~P.}\ \bibnamefont {Chang}}, \ and\ \bibinfo
  {author} {\bibfnamefont {M.~G.}\ \bibnamefont {Kanatzidis}},\ }\href@noop {}
  {\bibfield  {journal} {\bibinfo  {journal} {Nature Photonics}\ }\textbf
  {\bibinfo {volume} {8}},\ \bibinfo {pages} {489} (\bibinfo {year}
  {2014})}\BibitemShut {NoStop}%
\bibitem [{\citenamefont {Safdari}\ \emph {et~al.}(2015)\citenamefont
  {Safdari}, \citenamefont {Fischer}, \citenamefont {Xu}, \citenamefont
  {Kloo},\ and\ \citenamefont {Gardner}}]{safdari2015structure}%
  \BibitemOpen
  \bibfield  {author} {\bibinfo {author} {\bibfnamefont {M.}~\bibnamefont
  {Safdari}}, \bibinfo {author} {\bibfnamefont {A.}~\bibnamefont {Fischer}},
  \bibinfo {author} {\bibfnamefont {B.}~\bibnamefont {Xu}}, \bibinfo {author}
  {\bibfnamefont {L.}~\bibnamefont {Kloo}}, \ and\ \bibinfo {author}
  {\bibfnamefont {J.~M.}\ \bibnamefont {Gardner}},\ }\href@noop {} {\bibfield
  {journal} {\bibinfo  {journal} {Journal of Materials Chemistry A}\ }\textbf
  {\bibinfo {volume} {3}},\ \bibinfo {pages} {9201} (\bibinfo {year}
  {2015})}\BibitemShut {NoStop}%
\bibitem [{\citenamefont {Noel}\ \emph {et~al.}(2014)\citenamefont {Noel},
  \citenamefont {Stranks}, \citenamefont {Abate}, \citenamefont {Wehrenfennig},
  \citenamefont {Guarnera}, \citenamefont {Haghighirad}, \citenamefont
  {Sadhanala}, \citenamefont {Eperon}, \citenamefont {Pathak}, \citenamefont
  {Johnston} \emph {et~al.}}]{noel2014lead}%
  \BibitemOpen
  \bibfield  {author} {\bibinfo {author} {\bibfnamefont {N.~K.}\ \bibnamefont
  {Noel}}, \bibinfo {author} {\bibfnamefont {S.~D.}\ \bibnamefont {Stranks}},
  \bibinfo {author} {\bibfnamefont {A.}~\bibnamefont {Abate}}, \bibinfo
  {author} {\bibfnamefont {C.}~\bibnamefont {Wehrenfennig}}, \bibinfo {author}
  {\bibfnamefont {S.}~\bibnamefont {Guarnera}}, \bibinfo {author}
  {\bibfnamefont {A.-A.}\ \bibnamefont {Haghighirad}}, \bibinfo {author}
  {\bibfnamefont {A.}~\bibnamefont {Sadhanala}}, \bibinfo {author}
  {\bibfnamefont {G.~E.}\ \bibnamefont {Eperon}}, \bibinfo {author}
  {\bibfnamefont {S.~K.}\ \bibnamefont {Pathak}}, \bibinfo {author}
  {\bibfnamefont {M.~B.}\ \bibnamefont {Johnston}},  \emph {et~al.},\
  }\href@noop {} {\bibfield  {journal} {\bibinfo  {journal} {Energy \&
  Environmental Science}\ }\textbf {\bibinfo {volume} {7}},\ \bibinfo {pages}
  {3061} (\bibinfo {year} {2014})}\BibitemShut {NoStop}%
\bibitem [{\citenamefont {Yu}\ and\ \citenamefont
  {Zunger}(2012)}]{yu2012identification}%
  \BibitemOpen
  \bibfield  {author} {\bibinfo {author} {\bibfnamefont {L.}~\bibnamefont
  {Yu}}\ and\ \bibinfo {author} {\bibfnamefont {A.}~\bibnamefont {Zunger}},\
  }\href@noop {} {\bibfield  {journal} {\bibinfo  {journal} {Physical review
  letters}\ }\textbf {\bibinfo {volume} {108}},\ \bibinfo {pages} {068701}
  (\bibinfo {year} {2012})}\BibitemShut {NoStop}%
\bibitem [{\citenamefont {Yu}\ \emph {et~al.}(2013)\citenamefont {Yu},
  \citenamefont {Kokenyesi}, \citenamefont {Keszler},\ and\ \citenamefont
  {Zunger}}]{yu2013inverse}%
  \BibitemOpen
  \bibfield  {author} {\bibinfo {author} {\bibfnamefont {L.}~\bibnamefont
  {Yu}}, \bibinfo {author} {\bibfnamefont {R.~S.}\ \bibnamefont {Kokenyesi}},
  \bibinfo {author} {\bibfnamefont {D.~A.}\ \bibnamefont {Keszler}}, \ and\
  \bibinfo {author} {\bibfnamefont {A.}~\bibnamefont {Zunger}},\ }\href@noop {}
  {\bibfield  {journal} {\bibinfo  {journal} {Advanced Energy Materials}\
  }\textbf {\bibinfo {volume} {3}},\ \bibinfo {pages} {43} (\bibinfo {year}
  {2013})}\BibitemShut {NoStop}%
\bibitem [{\citenamefont {Filip}\ \emph {et~al.}(2014)\citenamefont {Filip},
  \citenamefont {Eperon}, \citenamefont {Snaith},\ and\ \citenamefont
  {Giustino}}]{filip2014steric}%
  \BibitemOpen
  \bibfield  {author} {\bibinfo {author} {\bibfnamefont {M.~R.}\ \bibnamefont
  {Filip}}, \bibinfo {author} {\bibfnamefont {G.~E.}\ \bibnamefont {Eperon}},
  \bibinfo {author} {\bibfnamefont {H.~J.}\ \bibnamefont {Snaith}}, \ and\
  \bibinfo {author} {\bibfnamefont {F.}~\bibnamefont {Giustino}},\ }\href@noop
  {} {\bibfield  {journal} {\bibinfo  {journal} {Nature communications}\
  }\textbf {\bibinfo {volume} {5}} (\bibinfo {year} {2014})}\BibitemShut
  {NoStop}%
\bibitem [{\citenamefont {Filip}\ and\ \citenamefont
  {Giustino}(2015)}]{filip2015computational}%
  \BibitemOpen
  \bibfield  {author} {\bibinfo {author} {\bibfnamefont {M.~R.}\ \bibnamefont
  {Filip}}\ and\ \bibinfo {author} {\bibfnamefont {F.}~\bibnamefont
  {Giustino}},\ }\href@noop {} {\bibfield  {journal} {\bibinfo  {journal} {The
  Journal of Physical Chemistry C}\ }\textbf {\bibinfo {volume} {120}},\
  \bibinfo {pages} {166} (\bibinfo {year} {2015})}\BibitemShut {NoStop}%
\bibitem [{\citenamefont {Castelli}\ \emph {et~al.}(2014)\citenamefont
  {Castelli}, \citenamefont {Garc{\'\i}a-Lastra}, \citenamefont {Thygesen},\
  and\ \citenamefont {Jacobsen}}]{castelli2014bandgap}%
  \BibitemOpen
  \bibfield  {author} {\bibinfo {author} {\bibfnamefont {I.~E.}\ \bibnamefont
  {Castelli}}, \bibinfo {author} {\bibfnamefont {J.~M.}\ \bibnamefont
  {Garc{\'\i}a-Lastra}}, \bibinfo {author} {\bibfnamefont {K.~S.}\ \bibnamefont
  {Thygesen}}, \ and\ \bibinfo {author} {\bibfnamefont {K.~W.}\ \bibnamefont
  {Jacobsen}},\ }\href@noop {} {\bibfield  {journal} {\bibinfo  {journal} {APL
  Materials}\ }\textbf {\bibinfo {volume} {2}},\ \bibinfo {pages} {081514}
  (\bibinfo {year} {2014})}\BibitemShut {NoStop}%
\bibitem [{\citenamefont {Zheng}\ \emph
  {et~al.}(2015{\natexlab{a}})\citenamefont {Zheng}, \citenamefont
  {Saldana-Greco}, \citenamefont {Liu},\ and\ \citenamefont
  {Rappe}}]{zheng2015material}%
  \BibitemOpen
  \bibfield  {author} {\bibinfo {author} {\bibfnamefont {F.}~\bibnamefont
  {Zheng}}, \bibinfo {author} {\bibfnamefont {D.}~\bibnamefont
  {Saldana-Greco}}, \bibinfo {author} {\bibfnamefont {S.}~\bibnamefont {Liu}},
  \ and\ \bibinfo {author} {\bibfnamefont {A.~M.}\ \bibnamefont {Rappe}},\
  }\href@noop {} {\bibfield  {journal} {\bibinfo  {journal} {The journal of
  physical chemistry letters}\ }\textbf {\bibinfo {volume} {6}},\ \bibinfo
  {pages} {4862} (\bibinfo {year} {2015}{\natexlab{a}})}\BibitemShut {NoStop}%
\bibitem [{\citenamefont {Yan}\ \emph {et~al.}(2015)\citenamefont {Yan},
  \citenamefont {Zhang}, \citenamefont {Yonggang}, \citenamefont {Yu},
  \citenamefont {Nagaraja}, \citenamefont {Mason},\ and\ \citenamefont
  {Zunger}}]{yan2015design}%
  \BibitemOpen
  \bibfield  {author} {\bibinfo {author} {\bibfnamefont {F.}~\bibnamefont
  {Yan}}, \bibinfo {author} {\bibfnamefont {X.}~\bibnamefont {Zhang}}, \bibinfo
  {author} {\bibfnamefont {G.~Y.}\ \bibnamefont {Yonggang}}, \bibinfo {author}
  {\bibfnamefont {L.}~\bibnamefont {Yu}}, \bibinfo {author} {\bibfnamefont
  {A.}~\bibnamefont {Nagaraja}}, \bibinfo {author} {\bibfnamefont {T.~O.}\
  \bibnamefont {Mason}}, \ and\ \bibinfo {author} {\bibfnamefont
  {A.}~\bibnamefont {Zunger}},\ }\href@noop {} {\bibfield  {journal} {\bibinfo
  {journal} {Nature communications}\ }\textbf {\bibinfo {volume} {6}} (\bibinfo
  {year} {2015})}\BibitemShut {NoStop}%
\bibitem [{\citenamefont {Gautier}\ \emph {et~al.}(2015)\citenamefont
  {Gautier}, \citenamefont {Zhang}, \citenamefont {Hu}, \citenamefont {Yu},
  \citenamefont {Lin}, \citenamefont {Sunde}, \citenamefont {Chon},
  \citenamefont {Poeppelmeier},\ and\ \citenamefont
  {Zunger}}]{gautier2015prediction}%
  \BibitemOpen
  \bibfield  {author} {\bibinfo {author} {\bibfnamefont {R.}~\bibnamefont
  {Gautier}}, \bibinfo {author} {\bibfnamefont {X.}~\bibnamefont {Zhang}},
  \bibinfo {author} {\bibfnamefont {L.}~\bibnamefont {Hu}}, \bibinfo {author}
  {\bibfnamefont {L.}~\bibnamefont {Yu}}, \bibinfo {author} {\bibfnamefont
  {Y.}~\bibnamefont {Lin}}, \bibinfo {author} {\bibfnamefont {T.~O.}\
  \bibnamefont {Sunde}}, \bibinfo {author} {\bibfnamefont {D.}~\bibnamefont
  {Chon}}, \bibinfo {author} {\bibfnamefont {K.~R.}\ \bibnamefont
  {Poeppelmeier}}, \ and\ \bibinfo {author} {\bibfnamefont {A.}~\bibnamefont
  {Zunger}},\ }\href@noop {} {\bibfield  {journal} {\bibinfo  {journal} {Nature
  chemistry}\ }\textbf {\bibinfo {volume} {7}},\ \bibinfo {pages} {308}
  (\bibinfo {year} {2015})}\BibitemShut {NoStop}%
\bibitem [{\citenamefont {Edri}\ \emph {et~al.}(2013)\citenamefont {Edri},
  \citenamefont {Kirmayer}, \citenamefont {Cahen},\ and\ \citenamefont
  {Hodes}}]{edri2013high}%
  \BibitemOpen
  \bibfield  {author} {\bibinfo {author} {\bibfnamefont {E.}~\bibnamefont
  {Edri}}, \bibinfo {author} {\bibfnamefont {S.}~\bibnamefont {Kirmayer}},
  \bibinfo {author} {\bibfnamefont {D.}~\bibnamefont {Cahen}}, \ and\ \bibinfo
  {author} {\bibfnamefont {G.}~\bibnamefont {Hodes}},\ }\href@noop {}
  {\bibfield  {journal} {\bibinfo  {journal} {The journal of physical chemistry
  letters}\ }\textbf {\bibinfo {volume} {4}},\ \bibinfo {pages} {897} (\bibinfo
  {year} {2013})}\BibitemShut {NoStop}%
\bibitem [{\citenamefont {Baikie}\ \emph {et~al.}(2013)\citenamefont {Baikie},
  \citenamefont {Fang}, \citenamefont {Kadro}, \citenamefont {Schreyer},
  \citenamefont {Wei}, \citenamefont {Mhaisalkar}, \citenamefont {Graetzel},\
  and\ \citenamefont {White}}]{baikie2013synthesis}%
  \BibitemOpen
  \bibfield  {author} {\bibinfo {author} {\bibfnamefont {T.}~\bibnamefont
  {Baikie}}, \bibinfo {author} {\bibfnamefont {Y.}~\bibnamefont {Fang}},
  \bibinfo {author} {\bibfnamefont {J.~M.}\ \bibnamefont {Kadro}}, \bibinfo
  {author} {\bibfnamefont {M.}~\bibnamefont {Schreyer}}, \bibinfo {author}
  {\bibfnamefont {F.}~\bibnamefont {Wei}}, \bibinfo {author} {\bibfnamefont
  {S.~G.}\ \bibnamefont {Mhaisalkar}}, \bibinfo {author} {\bibfnamefont
  {M.}~\bibnamefont {Graetzel}}, \ and\ \bibinfo {author} {\bibfnamefont
  {T.~J.}\ \bibnamefont {White}},\ }\href@noop {} {\bibfield  {journal}
  {\bibinfo  {journal} {Journal of Materials Chemistry A}\ }\textbf {\bibinfo
  {volume} {1}},\ \bibinfo {pages} {5628} (\bibinfo {year} {2013})}\BibitemShut
  {NoStop}%
\bibitem [{\citenamefont {Yin}\ \emph {et~al.}(2014{\natexlab{a}})\citenamefont
  {Yin}, \citenamefont {Shi},\ and\ \citenamefont {Yan}}]{yin2014unique}%
  \BibitemOpen
  \bibfield  {author} {\bibinfo {author} {\bibfnamefont {W.-J.}\ \bibnamefont
  {Yin}}, \bibinfo {author} {\bibfnamefont {T.}~\bibnamefont {Shi}}, \ and\
  \bibinfo {author} {\bibfnamefont {Y.}~\bibnamefont {Yan}},\ }\href@noop {}
  {\bibfield  {journal} {\bibinfo  {journal} {Advanced Materials}\ }\textbf
  {\bibinfo {volume} {26}},\ \bibinfo {pages} {4653} (\bibinfo {year}
  {2014}{\natexlab{a}})}\BibitemShut {NoStop}%
\bibitem [{\citenamefont {B}(1992)}]{Dirac1992}%
  \BibitemOpen
  \bibfield  {author} {\bibinfo {author} {\bibfnamefont {T.}~\bibnamefont
  {B}},\ }\href@noop {} {\bibfield  {journal} {\bibinfo  {journal} {Springer
  Berlin Heidelberg}\ ,\ \bibinfo {pages} {1864}} (\bibinfo {year}
  {1992})}\BibitemShut {NoStop}%
\bibitem [{\citenamefont {Giorgi}\ \emph {et~al.}(2013)\citenamefont {Giorgi},
  \citenamefont {Fujisawa}, \citenamefont {Segawa},\ and\ \citenamefont
  {Yamashita}}]{giorgi2013small}%
  \BibitemOpen
  \bibfield  {author} {\bibinfo {author} {\bibfnamefont {G.}~\bibnamefont
  {Giorgi}}, \bibinfo {author} {\bibfnamefont {J.-I.}\ \bibnamefont
  {Fujisawa}}, \bibinfo {author} {\bibfnamefont {H.}~\bibnamefont {Segawa}}, \
  and\ \bibinfo {author} {\bibfnamefont {K.}~\bibnamefont {Yamashita}},\
  }\href@noop {} {\bibfield  {journal} {\bibinfo  {journal} {The journal of
  physical chemistry letters}\ }\textbf {\bibinfo {volume} {4}},\ \bibinfo
  {pages} {4213} (\bibinfo {year} {2013})}\BibitemShut {NoStop}%
\bibitem [{\citenamefont {Wei}\ and\ \citenamefont
  {Zunger}(1997)}]{wei1997electronic}%
  \BibitemOpen
  \bibfield  {author} {\bibinfo {author} {\bibfnamefont {S.-H.}\ \bibnamefont
  {Wei}}\ and\ \bibinfo {author} {\bibfnamefont {A.}~\bibnamefont {Zunger}},\
  }\href@noop {} {\bibfield  {journal} {\bibinfo  {journal} {Physical Review
  B}\ }\textbf {\bibinfo {volume} {55}},\ \bibinfo {pages} {13605} (\bibinfo
  {year} {1997})}\BibitemShut {NoStop}%
\bibitem [{\citenamefont {Miyata}\ \emph {et~al.}(2015)\citenamefont {Miyata},
  \citenamefont {Mitioglu}, \citenamefont {Plochocka}, \citenamefont
  {Portugall}, \citenamefont {Wang}, \citenamefont {Stranks}, \citenamefont
  {Snaith},\ and\ \citenamefont {Nicholas}}]{miyata2015direct}%
  \BibitemOpen
  \bibfield  {author} {\bibinfo {author} {\bibfnamefont {A.}~\bibnamefont
  {Miyata}}, \bibinfo {author} {\bibfnamefont {A.}~\bibnamefont {Mitioglu}},
  \bibinfo {author} {\bibfnamefont {P.}~\bibnamefont {Plochocka}}, \bibinfo
  {author} {\bibfnamefont {O.}~\bibnamefont {Portugall}}, \bibinfo {author}
  {\bibfnamefont {J.~T.-W.}\ \bibnamefont {Wang}}, \bibinfo {author}
  {\bibfnamefont {S.~D.}\ \bibnamefont {Stranks}}, \bibinfo {author}
  {\bibfnamefont {H.~J.}\ \bibnamefont {Snaith}}, \ and\ \bibinfo {author}
  {\bibfnamefont {R.~J.}\ \bibnamefont {Nicholas}},\ }\href@noop {} {\bibfield
  {journal} {\bibinfo  {journal} {Nature Physics}\ }\textbf {\bibinfo {volume}
  {11}},\ \bibinfo {pages} {582} (\bibinfo {year} {2015})}\BibitemShut
  {NoStop}%
\bibitem [{\citenamefont {Lin}\ \emph {et~al.}(2015)\citenamefont {Lin},
  \citenamefont {Armin}, \citenamefont {Nagiri}, \citenamefont {Burn},\ and\
  \citenamefont {Meredith}}]{lin2015electro}%
  \BibitemOpen
  \bibfield  {author} {\bibinfo {author} {\bibfnamefont {Q.}~\bibnamefont
  {Lin}}, \bibinfo {author} {\bibfnamefont {A.}~\bibnamefont {Armin}}, \bibinfo
  {author} {\bibfnamefont {R.~C.~R.}\ \bibnamefont {Nagiri}}, \bibinfo {author}
  {\bibfnamefont {P.~L.}\ \bibnamefont {Burn}}, \ and\ \bibinfo {author}
  {\bibfnamefont {P.}~\bibnamefont {Meredith}},\ }\href@noop {} {\bibfield
  {journal} {\bibinfo  {journal} {Nature Photonics}\ }\textbf {\bibinfo
  {volume} {9}},\ \bibinfo {pages} {106} (\bibinfo {year} {2015})}\BibitemShut
  {NoStop}%
\bibitem [{\citenamefont {Yamada}\ \emph {et~al.}(2015)\citenamefont {Yamada},
  \citenamefont {Nakamura}, \citenamefont {Endo}, \citenamefont {Wakamiya},\
  and\ \citenamefont {Kanemitsu}}]{yamada2015photoelectronic}%
  \BibitemOpen
  \bibfield  {author} {\bibinfo {author} {\bibfnamefont {Y.}~\bibnamefont
  {Yamada}}, \bibinfo {author} {\bibfnamefont {T.}~\bibnamefont {Nakamura}},
  \bibinfo {author} {\bibfnamefont {M.}~\bibnamefont {Endo}}, \bibinfo {author}
  {\bibfnamefont {A.}~\bibnamefont {Wakamiya}}, \ and\ \bibinfo {author}
  {\bibfnamefont {Y.}~\bibnamefont {Kanemitsu}},\ }\href@noop {} {\bibfield
  {journal} {\bibinfo  {journal} {IEEE Journal of Photovoltaics}\ }\textbf
  {\bibinfo {volume} {5}},\ \bibinfo {pages} {401} (\bibinfo {year}
  {2015})}\BibitemShut {NoStop}%
\bibitem [{\citenamefont {Even}\ \emph {et~al.}(2014)\citenamefont {Even},
  \citenamefont {Pedesseau},\ and\ \citenamefont {Katan}}]{even2014analysis}%
  \BibitemOpen
  \bibfield  {author} {\bibinfo {author} {\bibfnamefont {J.}~\bibnamefont
  {Even}}, \bibinfo {author} {\bibfnamefont {L.}~\bibnamefont {Pedesseau}}, \
  and\ \bibinfo {author} {\bibfnamefont {C.}~\bibnamefont {Katan}},\
  }\href@noop {} {\bibfield  {journal} {\bibinfo  {journal} {The Journal of
  Physical Chemistry C}\ }\textbf {\bibinfo {volume} {118}},\ \bibinfo {pages}
  {11566} (\bibinfo {year} {2014})}\BibitemShut {NoStop}%
\bibitem [{\citenamefont {Men{\'e}ndez-Proupin}\ \emph
  {et~al.}(2014)\citenamefont {Men{\'e}ndez-Proupin}, \citenamefont {Palacios},
  \citenamefont {Wahn{\'o}n},\ and\ \citenamefont {Conesa}}]{menendez2014self}%
  \BibitemOpen
  \bibfield  {author} {\bibinfo {author} {\bibfnamefont {E.}~\bibnamefont
  {Men{\'e}ndez-Proupin}}, \bibinfo {author} {\bibfnamefont {P.}~\bibnamefont
  {Palacios}}, \bibinfo {author} {\bibfnamefont {P.}~\bibnamefont
  {Wahn{\'o}n}}, \ and\ \bibinfo {author} {\bibfnamefont {J.}~\bibnamefont
  {Conesa}},\ }\href@noop {} {\bibfield  {journal} {\bibinfo  {journal}
  {Physical Review B}\ }\textbf {\bibinfo {volume} {90}},\ \bibinfo {pages}
  {045207} (\bibinfo {year} {2014})}\BibitemShut {NoStop}%
\bibitem [{\citenamefont {Koutselas}\ \emph {et~al.}(1996)\citenamefont
  {Koutselas}, \citenamefont {Ducasse},\ and\ \citenamefont
  {Papavassiliou}}]{koutselas1996electronic}%
  \BibitemOpen
  \bibfield  {author} {\bibinfo {author} {\bibfnamefont {I.}~\bibnamefont
  {Koutselas}}, \bibinfo {author} {\bibfnamefont {L.}~\bibnamefont {Ducasse}},
  \ and\ \bibinfo {author} {\bibfnamefont {G.~C.}\ \bibnamefont
  {Papavassiliou}},\ }\href@noop {} {\bibfield  {journal} {\bibinfo  {journal}
  {Journal of Physics: Condensed Matter}\ }\textbf {\bibinfo {volume} {8}},\
  \bibinfo {pages} {1217} (\bibinfo {year} {1996})}\BibitemShut {NoStop}%
\bibitem [{\citenamefont {Hirasawa}\ \emph
  {et~al.}(1994{\natexlab{a}})\citenamefont {Hirasawa}, \citenamefont
  {Ishihara},\ and\ \citenamefont {Goto}}]{hirasawa1994exciton}%
  \BibitemOpen
  \bibfield  {author} {\bibinfo {author} {\bibfnamefont {M.}~\bibnamefont
  {Hirasawa}}, \bibinfo {author} {\bibfnamefont {T.}~\bibnamefont {Ishihara}},
  \ and\ \bibinfo {author} {\bibfnamefont {T.}~\bibnamefont {Goto}},\
  }\href@noop {} {\bibfield  {journal} {\bibinfo  {journal} {Journal of the
  Physical Society of Japan}\ }\textbf {\bibinfo {volume} {63}},\ \bibinfo
  {pages} {3870} (\bibinfo {year} {1994}{\natexlab{a}})}\BibitemShut {NoStop}%
\bibitem [{\citenamefont {Xing}\ \emph {et~al.}(2013)\citenamefont {Xing},
  \citenamefont {Mathews}, \citenamefont {Sun}, \citenamefont {Lim},
  \citenamefont {Lam}, \citenamefont {Gr{\"a}tzel}, \citenamefont
  {Mhaisalkar},\ and\ \citenamefont {Sum}}]{xing2013long}%
  \BibitemOpen
  \bibfield  {author} {\bibinfo {author} {\bibfnamefont {G.}~\bibnamefont
  {Xing}}, \bibinfo {author} {\bibfnamefont {N.}~\bibnamefont {Mathews}},
  \bibinfo {author} {\bibfnamefont {S.}~\bibnamefont {Sun}}, \bibinfo {author}
  {\bibfnamefont {S.~S.}\ \bibnamefont {Lim}}, \bibinfo {author} {\bibfnamefont
  {Y.~M.}\ \bibnamefont {Lam}}, \bibinfo {author} {\bibfnamefont
  {M.}~\bibnamefont {Gr{\"a}tzel}}, \bibinfo {author} {\bibfnamefont
  {S.}~\bibnamefont {Mhaisalkar}}, \ and\ \bibinfo {author} {\bibfnamefont
  {T.~C.}\ \bibnamefont {Sum}},\ }\href@noop {} {\bibfield  {journal} {\bibinfo
   {journal} {Science}\ }\textbf {\bibinfo {volume} {342}},\ \bibinfo {pages}
  {344} (\bibinfo {year} {2013})}\BibitemShut {NoStop}%
\bibitem [{\citenamefont {Stranks}\ \emph {et~al.}(2013)\citenamefont
  {Stranks}, \citenamefont {Eperon}, \citenamefont {Grancini}, \citenamefont
  {Menelaou}, \citenamefont {Alcocer}, \citenamefont {Leijtens}, \citenamefont
  {Herz}, \citenamefont {Petrozza},\ and\ \citenamefont
  {Snaith}}]{stranks2013electron}%
  \BibitemOpen
  \bibfield  {author} {\bibinfo {author} {\bibfnamefont {S.~D.}\ \bibnamefont
  {Stranks}}, \bibinfo {author} {\bibfnamefont {G.~E.}\ \bibnamefont {Eperon}},
  \bibinfo {author} {\bibfnamefont {G.}~\bibnamefont {Grancini}}, \bibinfo
  {author} {\bibfnamefont {C.}~\bibnamefont {Menelaou}}, \bibinfo {author}
  {\bibfnamefont {M.~J.}\ \bibnamefont {Alcocer}}, \bibinfo {author}
  {\bibfnamefont {T.}~\bibnamefont {Leijtens}}, \bibinfo {author}
  {\bibfnamefont {L.~M.}\ \bibnamefont {Herz}}, \bibinfo {author}
  {\bibfnamefont {A.}~\bibnamefont {Petrozza}}, \ and\ \bibinfo {author}
  {\bibfnamefont {H.~J.}\ \bibnamefont {Snaith}},\ }\href@noop {} {\bibfield
  {journal} {\bibinfo  {journal} {Science}\ }\textbf {\bibinfo {volume}
  {342}},\ \bibinfo {pages} {341} (\bibinfo {year} {2013})}\BibitemShut
  {NoStop}%
\bibitem [{\citenamefont {Tanaka}\ \emph {et~al.}(2003)\citenamefont {Tanaka},
  \citenamefont {Takahashi}, \citenamefont {Ban}, \citenamefont {Kondo},
  \citenamefont {Uchida},\ and\ \citenamefont {Miura}}]{tanaka2003comparative}%
  \BibitemOpen
  \bibfield  {author} {\bibinfo {author} {\bibfnamefont {K.}~\bibnamefont
  {Tanaka}}, \bibinfo {author} {\bibfnamefont {T.}~\bibnamefont {Takahashi}},
  \bibinfo {author} {\bibfnamefont {T.}~\bibnamefont {Ban}}, \bibinfo {author}
  {\bibfnamefont {T.}~\bibnamefont {Kondo}}, \bibinfo {author} {\bibfnamefont
  {K.}~\bibnamefont {Uchida}}, \ and\ \bibinfo {author} {\bibfnamefont
  {N.}~\bibnamefont {Miura}},\ }\href@noop {} {\bibfield  {journal} {\bibinfo
  {journal} {Solid state communications}\ }\textbf {\bibinfo {volume} {127}},\
  \bibinfo {pages} {619} (\bibinfo {year} {2003})}\BibitemShut {NoStop}%
\bibitem [{\citenamefont {Zhang}\ \emph {et~al.}(1998)\citenamefont {Zhang},
  \citenamefont {Wei}, \citenamefont {Zunger},\ and\ \citenamefont
  {Katayama-Yoshida}}]{zhang1998defect}%
  \BibitemOpen
  \bibfield  {author} {\bibinfo {author} {\bibfnamefont {S.}~\bibnamefont
  {Zhang}}, \bibinfo {author} {\bibfnamefont {S.-H.}\ \bibnamefont {Wei}},
  \bibinfo {author} {\bibfnamefont {A.}~\bibnamefont {Zunger}}, \ and\ \bibinfo
  {author} {\bibfnamefont {H.}~\bibnamefont {Katayama-Yoshida}},\ }\href@noop
  {} {\bibfield  {journal} {\bibinfo  {journal} {Physical Review B}\ }\textbf
  {\bibinfo {volume} {57}},\ \bibinfo {pages} {9642} (\bibinfo {year}
  {1998})}\BibitemShut {NoStop}%
\bibitem [{\citenamefont {Lany}\ and\ \citenamefont
  {Zunger}(2005)}]{lany2005anion}%
  \BibitemOpen
  \bibfield  {author} {\bibinfo {author} {\bibfnamefont {S.}~\bibnamefont
  {Lany}}\ and\ \bibinfo {author} {\bibfnamefont {A.}~\bibnamefont {Zunger}},\
  }\href@noop {} {\bibfield  {journal} {\bibinfo  {journal} {Physical Review
  B}\ }\textbf {\bibinfo {volume} {72}},\ \bibinfo {pages} {035215} (\bibinfo
  {year} {2005})}\BibitemShut {NoStop}%
\bibitem [{\citenamefont {Brandt}\ \emph {et~al.}(2015)\citenamefont {Brandt},
  \citenamefont {Stevanovi{\'c}}, \citenamefont {Ginley},\ and\ \citenamefont
  {Buonassisi}}]{brandt2015identifying}%
  \BibitemOpen
  \bibfield  {author} {\bibinfo {author} {\bibfnamefont {R.~E.}\ \bibnamefont
  {Brandt}}, \bibinfo {author} {\bibfnamefont {V.}~\bibnamefont
  {Stevanovi{\'c}}}, \bibinfo {author} {\bibfnamefont {D.~S.}\ \bibnamefont
  {Ginley}}, \ and\ \bibinfo {author} {\bibfnamefont {T.}~\bibnamefont
  {Buonassisi}},\ }\href@noop {} {\bibfield  {journal} {\bibinfo  {journal}
  {MRS Communications}\ }\textbf {\bibinfo {volume} {5}},\ \bibinfo {pages}
  {265} (\bibinfo {year} {2015})}\BibitemShut {NoStop}%
\bibitem [{\citenamefont {Du}(2014)}]{du2014efficient}%
  \BibitemOpen
  \bibfield  {author} {\bibinfo {author} {\bibfnamefont {M.-H.}\ \bibnamefont
  {Du}},\ }\href@noop {} {\bibfield  {journal} {\bibinfo  {journal} {Journal of
  Materials Chemistry A}\ }\textbf {\bibinfo {volume} {2}},\ \bibinfo {pages}
  {9091} (\bibinfo {year} {2014})}\BibitemShut {NoStop}%
\bibitem [{\citenamefont {Yin}\ \emph {et~al.}(2014{\natexlab{b}})\citenamefont
  {Yin}, \citenamefont {Shi},\ and\ \citenamefont {Yan}}]{yin2014unusual}%
  \BibitemOpen
  \bibfield  {author} {\bibinfo {author} {\bibfnamefont {W.-J.}\ \bibnamefont
  {Yin}}, \bibinfo {author} {\bibfnamefont {T.}~\bibnamefont {Shi}}, \ and\
  \bibinfo {author} {\bibfnamefont {Y.}~\bibnamefont {Yan}},\ }\href@noop {}
  {\bibfield  {journal} {\bibinfo  {journal} {Applied Physics Letters}\
  }\textbf {\bibinfo {volume} {104}},\ \bibinfo {pages} {063903} (\bibinfo
  {year} {2014}{\natexlab{b}})}\BibitemShut {NoStop}%
\bibitem [{\citenamefont {Persson}\ and\ \citenamefont
  {Zunger}(2003)}]{persson2003anomalous}%
  \BibitemOpen
  \bibfield  {author} {\bibinfo {author} {\bibfnamefont {C.}~\bibnamefont
  {Persson}}\ and\ \bibinfo {author} {\bibfnamefont {A.}~\bibnamefont
  {Zunger}},\ }\href@noop {} {\bibfield  {journal} {\bibinfo  {journal}
  {Physical Review Letters}\ }\textbf {\bibinfo {volume} {91}},\ \bibinfo
  {pages} {266401} (\bibinfo {year} {2003})}\BibitemShut {NoStop}%
\bibitem [{\citenamefont {Swarnkar}\ \emph {et~al.}(2016)\citenamefont
  {Swarnkar}, \citenamefont {Marshall}, \citenamefont {Sanehira}, \citenamefont
  {Chernomordik}, \citenamefont {Moore}, \citenamefont {Christians},
  \citenamefont {Chakrabarti},\ and\ \citenamefont
  {Luther}}]{swarnkar2016quantum}%
  \BibitemOpen
  \bibfield  {author} {\bibinfo {author} {\bibfnamefont {A.}~\bibnamefont
  {Swarnkar}}, \bibinfo {author} {\bibfnamefont {A.~R.}\ \bibnamefont
  {Marshall}}, \bibinfo {author} {\bibfnamefont {E.~M.}\ \bibnamefont
  {Sanehira}}, \bibinfo {author} {\bibfnamefont {B.~D.}\ \bibnamefont
  {Chernomordik}}, \bibinfo {author} {\bibfnamefont {D.~T.}\ \bibnamefont
  {Moore}}, \bibinfo {author} {\bibfnamefont {J.~A.}\ \bibnamefont
  {Christians}}, \bibinfo {author} {\bibfnamefont {T.}~\bibnamefont
  {Chakrabarti}}, \ and\ \bibinfo {author} {\bibfnamefont {J.~M.}\ \bibnamefont
  {Luther}},\ }\href@noop {} {\bibfield  {journal} {\bibinfo  {journal}
  {Science}\ }\textbf {\bibinfo {volume} {354}},\ \bibinfo {pages} {92}
  (\bibinfo {year} {2016})}\BibitemShut {NoStop}%
\bibitem [{\citenamefont {Beal}\ \emph {et~al.}(2016)\citenamefont {Beal},
  \citenamefont {Slotcavage}, \citenamefont {Leijtens}, \citenamefont
  {Bowring}, \citenamefont {Belisle}, \citenamefont {Nguyen}, \citenamefont
  {Burkhard}, \citenamefont {Hoke},\ and\ \citenamefont
  {McGehee}}]{beal2016cesium}%
  \BibitemOpen
  \bibfield  {author} {\bibinfo {author} {\bibfnamefont {R.~E.}\ \bibnamefont
  {Beal}}, \bibinfo {author} {\bibfnamefont {D.~J.}\ \bibnamefont
  {Slotcavage}}, \bibinfo {author} {\bibfnamefont {T.}~\bibnamefont
  {Leijtens}}, \bibinfo {author} {\bibfnamefont {A.~R.}\ \bibnamefont
  {Bowring}}, \bibinfo {author} {\bibfnamefont {R.~A.}\ \bibnamefont
  {Belisle}}, \bibinfo {author} {\bibfnamefont {W.~H.}\ \bibnamefont {Nguyen}},
  \bibinfo {author} {\bibfnamefont {G.~F.}\ \bibnamefont {Burkhard}}, \bibinfo
  {author} {\bibfnamefont {E.~T.}\ \bibnamefont {Hoke}}, \ and\ \bibinfo
  {author} {\bibfnamefont {M.~D.}\ \bibnamefont {McGehee}},\ }\href@noop {}
  {\bibfield  {journal} {\bibinfo  {journal} {The journal of physical chemistry
  letters}\ }\textbf {\bibinfo {volume} {7}},\ \bibinfo {pages} {746} (\bibinfo
  {year} {2016})}\BibitemShut {NoStop}%
\bibitem [{\citenamefont {M{\O}LLER}(1957)}]{moller1957phase}%
  \BibitemOpen
  \bibfield  {author} {\bibinfo {author} {\bibfnamefont {C.~K.}\ \bibnamefont
  {M{\O}LLER}},\ }\href@noop {} {\  (\bibinfo {year} {1957})}\BibitemShut
  {NoStop}%
\bibitem [{\citenamefont {Stoumpos}\ \emph
  {et~al.}(2013{\natexlab{b}})\citenamefont {Stoumpos}, \citenamefont
  {Malliakas}, \citenamefont {Peters}, \citenamefont {Liu}, \citenamefont
  {Sebastian}, \citenamefont {Im}, \citenamefont {Chasapis}, \citenamefont
  {Wibowo}, \citenamefont {Chung}, \citenamefont {Freeman} \emph
  {et~al.}}]{stoumpos2013crystal}%
  \BibitemOpen
  \bibfield  {author} {\bibinfo {author} {\bibfnamefont {C.~C.}\ \bibnamefont
  {Stoumpos}}, \bibinfo {author} {\bibfnamefont {C.~D.}\ \bibnamefont
  {Malliakas}}, \bibinfo {author} {\bibfnamefont {J.~A.}\ \bibnamefont
  {Peters}}, \bibinfo {author} {\bibfnamefont {Z.}~\bibnamefont {Liu}},
  \bibinfo {author} {\bibfnamefont {M.}~\bibnamefont {Sebastian}}, \bibinfo
  {author} {\bibfnamefont {J.}~\bibnamefont {Im}}, \bibinfo {author}
  {\bibfnamefont {T.~C.}\ \bibnamefont {Chasapis}}, \bibinfo {author}
  {\bibfnamefont {A.~C.}\ \bibnamefont {Wibowo}}, \bibinfo {author}
  {\bibfnamefont {D.~Y.}\ \bibnamefont {Chung}}, \bibinfo {author}
  {\bibfnamefont {A.~J.}\ \bibnamefont {Freeman}},  \emph {et~al.},\
  }\href@noop {} {\bibfield  {journal} {\bibinfo  {journal} {Crystal Growth \&
  Design}\ }\textbf {\bibinfo {volume} {13}},\ \bibinfo {pages} {2722}
  (\bibinfo {year} {2013}{\natexlab{b}})}\BibitemShut {NoStop}%
\bibitem [{\citenamefont {Kuok}(1992)}]{kuok1992raman}%
  \BibitemOpen
  \bibfield  {author} {\bibinfo {author} {\bibfnamefont {M.}~\bibnamefont
  {Kuok}},\ }\href@noop {} {\bibfield  {journal} {\bibinfo  {journal} {Journal
  of Raman spectroscopy}\ }\textbf {\bibinfo {volume} {23}},\ \bibinfo {pages}
  {225} (\bibinfo {year} {1992})}\BibitemShut {NoStop}%
\bibitem [{\citenamefont {Schwarz}\ \emph {et~al.}(1996)\citenamefont
  {Schwarz}, \citenamefont {Wagner}, \citenamefont {Syassen},\ and\
  \citenamefont {Hillebrecht}}]{schwarz1996effect}%
  \BibitemOpen
  \bibfield  {author} {\bibinfo {author} {\bibfnamefont {U.}~\bibnamefont
  {Schwarz}}, \bibinfo {author} {\bibfnamefont {F.}~\bibnamefont {Wagner}},
  \bibinfo {author} {\bibfnamefont {K.}~\bibnamefont {Syassen}}, \ and\
  \bibinfo {author} {\bibfnamefont {H.}~\bibnamefont {Hillebrecht}},\
  }\href@noop {} {\bibfield  {journal} {\bibinfo  {journal} {Physical Review
  B}\ }\textbf {\bibinfo {volume} {53}},\ \bibinfo {pages} {12545} (\bibinfo
  {year} {1996})}\BibitemShut {NoStop}%
\bibitem [{\citenamefont {Eperon}\ \emph {et~al.}(2015)\citenamefont {Eperon},
  \citenamefont {Patern{\`o}}, \citenamefont {Sutton}, \citenamefont
  {Zampetti}, \citenamefont {Haghighirad}, \citenamefont {Cacialli},\ and\
  \citenamefont {Snaith}}]{eperon2015inorganic}%
  \BibitemOpen
  \bibfield  {author} {\bibinfo {author} {\bibfnamefont {G.~E.}\ \bibnamefont
  {Eperon}}, \bibinfo {author} {\bibfnamefont {G.~M.}\ \bibnamefont
  {Patern{\`o}}}, \bibinfo {author} {\bibfnamefont {R.~J.}\ \bibnamefont
  {Sutton}}, \bibinfo {author} {\bibfnamefont {A.}~\bibnamefont {Zampetti}},
  \bibinfo {author} {\bibfnamefont {A.~A.}\ \bibnamefont {Haghighirad}},
  \bibinfo {author} {\bibfnamefont {F.}~\bibnamefont {Cacialli}}, \ and\
  \bibinfo {author} {\bibfnamefont {H.~J.}\ \bibnamefont {Snaith}},\
  }\href@noop {} {\bibfield  {journal} {\bibinfo  {journal} {Journal of
  Materials Chemistry A}\ }\textbf {\bibinfo {volume} {3}},\ \bibinfo {pages}
  {19688} (\bibinfo {year} {2015})}\BibitemShut {NoStop}%
\bibitem [{\citenamefont {Sabba}\ \emph {et~al.}(2015)\citenamefont {Sabba},
  \citenamefont {Mulmudi}, \citenamefont {Prabhakar}, \citenamefont
  {Krishnamoorthy}, \citenamefont {Baikie}, \citenamefont {Boix}, \citenamefont
  {Mhaisalkar},\ and\ \citenamefont {Mathews}}]{sabba2015impact}%
  \BibitemOpen
  \bibfield  {author} {\bibinfo {author} {\bibfnamefont {D.}~\bibnamefont
  {Sabba}}, \bibinfo {author} {\bibfnamefont {H.~K.}\ \bibnamefont {Mulmudi}},
  \bibinfo {author} {\bibfnamefont {R.~R.}\ \bibnamefont {Prabhakar}}, \bibinfo
  {author} {\bibfnamefont {T.}~\bibnamefont {Krishnamoorthy}}, \bibinfo
  {author} {\bibfnamefont {T.}~\bibnamefont {Baikie}}, \bibinfo {author}
  {\bibfnamefont {P.~P.}\ \bibnamefont {Boix}}, \bibinfo {author}
  {\bibfnamefont {S.}~\bibnamefont {Mhaisalkar}}, \ and\ \bibinfo {author}
  {\bibfnamefont {N.}~\bibnamefont {Mathews}},\ }\href@noop {} {\bibfield
  {journal} {\bibinfo  {journal} {The Journal of Physical Chemistry C}\
  }\textbf {\bibinfo {volume} {119}},\ \bibinfo {pages} {1763} (\bibinfo {year}
  {2015})}\BibitemShut {NoStop}%
\bibitem [{\citenamefont {Kumar}\ \emph {et~al.}(2014)\citenamefont {Kumar},
  \citenamefont {Dharani}, \citenamefont {Leong}, \citenamefont {Boix},
  \citenamefont {Prabhakar}, \citenamefont {Baikie}, \citenamefont {Shi},
  \citenamefont {Ding}, \citenamefont {Ramesh}, \citenamefont {Asta} \emph
  {et~al.}}]{kumar2014lead}%
  \BibitemOpen
  \bibfield  {author} {\bibinfo {author} {\bibfnamefont {M.~H.}\ \bibnamefont
  {Kumar}}, \bibinfo {author} {\bibfnamefont {S.}~\bibnamefont {Dharani}},
  \bibinfo {author} {\bibfnamefont {W.~L.}\ \bibnamefont {Leong}}, \bibinfo
  {author} {\bibfnamefont {P.~P.}\ \bibnamefont {Boix}}, \bibinfo {author}
  {\bibfnamefont {R.~R.}\ \bibnamefont {Prabhakar}}, \bibinfo {author}
  {\bibfnamefont {T.}~\bibnamefont {Baikie}}, \bibinfo {author} {\bibfnamefont
  {C.}~\bibnamefont {Shi}}, \bibinfo {author} {\bibfnamefont {H.}~\bibnamefont
  {Ding}}, \bibinfo {author} {\bibfnamefont {R.}~\bibnamefont {Ramesh}},
  \bibinfo {author} {\bibfnamefont {M.}~\bibnamefont {Asta}},  \emph {et~al.},\
  }\href@noop {} {\bibfield  {journal} {\bibinfo  {journal} {Advanced
  Materials}\ }\textbf {\bibinfo {volume} {26}},\ \bibinfo {pages} {7122}
  (\bibinfo {year} {2014})}\BibitemShut {NoStop}%
\bibitem [{\citenamefont {Krishnamoorthy}\ \emph {et~al.}(2015)\citenamefont
  {Krishnamoorthy}, \citenamefont {Ding}, \citenamefont {Yan}, \citenamefont
  {Leong}, \citenamefont {Baikie}, \citenamefont {Zhang}, \citenamefont
  {Sherburne}, \citenamefont {Li}, \citenamefont {Asta}, \citenamefont
  {Mathews} \emph {et~al.}}]{krishnamoorthy2015lead}%
  \BibitemOpen
  \bibfield  {author} {\bibinfo {author} {\bibfnamefont {T.}~\bibnamefont
  {Krishnamoorthy}}, \bibinfo {author} {\bibfnamefont {H.}~\bibnamefont
  {Ding}}, \bibinfo {author} {\bibfnamefont {C.}~\bibnamefont {Yan}}, \bibinfo
  {author} {\bibfnamefont {W.~L.}\ \bibnamefont {Leong}}, \bibinfo {author}
  {\bibfnamefont {T.}~\bibnamefont {Baikie}}, \bibinfo {author} {\bibfnamefont
  {Z.}~\bibnamefont {Zhang}}, \bibinfo {author} {\bibfnamefont
  {M.}~\bibnamefont {Sherburne}}, \bibinfo {author} {\bibfnamefont
  {S.}~\bibnamefont {Li}}, \bibinfo {author} {\bibfnamefont {M.}~\bibnamefont
  {Asta}}, \bibinfo {author} {\bibfnamefont {N.}~\bibnamefont {Mathews}},
  \emph {et~al.},\ }\href@noop {} {\bibfield  {journal} {\bibinfo  {journal}
  {Journal of Materials Chemistry A}\ }\textbf {\bibinfo {volume} {3}},\
  \bibinfo {pages} {23829} (\bibinfo {year} {2015})}\BibitemShut {NoStop}%
\bibitem [{\citenamefont {Stoumpos}\ \emph {et~al.}(2015)\citenamefont
  {Stoumpos}, \citenamefont {Frazer}, \citenamefont {Clark}, \citenamefont
  {Kim}, \citenamefont {Rhim}, \citenamefont {Freeman}, \citenamefont
  {Ketterson}, \citenamefont {Jang},\ and\ \citenamefont
  {Kanatzidis}}]{stoumpos2015hybrid}%
  \BibitemOpen
  \bibfield  {author} {\bibinfo {author} {\bibfnamefont {C.~C.}\ \bibnamefont
  {Stoumpos}}, \bibinfo {author} {\bibfnamefont {L.}~\bibnamefont {Frazer}},
  \bibinfo {author} {\bibfnamefont {D.~J.}\ \bibnamefont {Clark}}, \bibinfo
  {author} {\bibfnamefont {Y.~S.}\ \bibnamefont {Kim}}, \bibinfo {author}
  {\bibfnamefont {S.~H.}\ \bibnamefont {Rhim}}, \bibinfo {author}
  {\bibfnamefont {A.~J.}\ \bibnamefont {Freeman}}, \bibinfo {author}
  {\bibfnamefont {J.~B.}\ \bibnamefont {Ketterson}}, \bibinfo {author}
  {\bibfnamefont {J.~I.}\ \bibnamefont {Jang}}, \ and\ \bibinfo {author}
  {\bibfnamefont {M.~G.}\ \bibnamefont {Kanatzidis}},\ }\href@noop {}
  {\bibfield  {journal} {\bibinfo  {journal} {Journal of the American Chemical
  Society}\ }\textbf {\bibinfo {volume} {137}},\ \bibinfo {pages} {6804}
  (\bibinfo {year} {2015})}\BibitemShut {NoStop}%
\bibitem [{\citenamefont {Koh}\ \emph {et~al.}(2015)\citenamefont {Koh},
  \citenamefont {Krishnamoorthy}, \citenamefont {Yantara}, \citenamefont {Shi},
  \citenamefont {Leong}, \citenamefont {Boix}, \citenamefont {Grimsdale},
  \citenamefont {Mhaisalkar},\ and\ \citenamefont
  {Mathews}}]{koh2015formamidinium}%
  \BibitemOpen
  \bibfield  {author} {\bibinfo {author} {\bibfnamefont {T.~M.}\ \bibnamefont
  {Koh}}, \bibinfo {author} {\bibfnamefont {T.}~\bibnamefont {Krishnamoorthy}},
  \bibinfo {author} {\bibfnamefont {N.}~\bibnamefont {Yantara}}, \bibinfo
  {author} {\bibfnamefont {C.}~\bibnamefont {Shi}}, \bibinfo {author}
  {\bibfnamefont {W.~L.}\ \bibnamefont {Leong}}, \bibinfo {author}
  {\bibfnamefont {P.~P.}\ \bibnamefont {Boix}}, \bibinfo {author}
  {\bibfnamefont {A.~C.}\ \bibnamefont {Grimsdale}}, \bibinfo {author}
  {\bibfnamefont {S.~G.}\ \bibnamefont {Mhaisalkar}}, \ and\ \bibinfo {author}
  {\bibfnamefont {N.}~\bibnamefont {Mathews}},\ }\href@noop {} {\bibfield
  {journal} {\bibinfo  {journal} {Journal of Materials Chemistry A}\ }\textbf
  {\bibinfo {volume} {3}},\ \bibinfo {pages} {14996} (\bibinfo {year}
  {2015})}\BibitemShut {NoStop}%
\bibitem [{\citenamefont {Selbach}\ \emph {et~al.}(2008)\citenamefont
  {Selbach}, \citenamefont {Einarsrud},\ and\ \citenamefont
  {Grande}}]{selbach2008thermodynamic}%
  \BibitemOpen
  \bibfield  {author} {\bibinfo {author} {\bibfnamefont {S.~M.}\ \bibnamefont
  {Selbach}}, \bibinfo {author} {\bibfnamefont {M.-A.}\ \bibnamefont
  {Einarsrud}}, \ and\ \bibinfo {author} {\bibfnamefont {T.}~\bibnamefont
  {Grande}},\ }\href@noop {} {\bibfield  {journal} {\bibinfo  {journal}
  {Chemistry of Materials}\ }\textbf {\bibinfo {volume} {21}},\ \bibinfo
  {pages} {169} (\bibinfo {year} {2008})}\BibitemShut {NoStop}%
\bibitem [{\citenamefont {Xu}\ \emph {et~al.}(2005)\citenamefont {Xu},
  \citenamefont {Navrotsky}, \citenamefont {Su},\ and\ \citenamefont
  {Balmer}}]{xu2005perovskite}%
  \BibitemOpen
  \bibfield  {author} {\bibinfo {author} {\bibfnamefont {H.}~\bibnamefont
  {Xu}}, \bibinfo {author} {\bibfnamefont {A.}~\bibnamefont {Navrotsky}},
  \bibinfo {author} {\bibfnamefont {Y.}~\bibnamefont {Su}}, \ and\ \bibinfo
  {author} {\bibfnamefont {M.~L.}\ \bibnamefont {Balmer}},\ }\href@noop {}
  {\bibfield  {journal} {\bibinfo  {journal} {Chemistry of materials}\ }\textbf
  {\bibinfo {volume} {17}},\ \bibinfo {pages} {1880} (\bibinfo {year}
  {2005})}\BibitemShut {NoStop}%
\bibitem [{\citenamefont {Scragg}\ \emph {et~al.}(2011)\citenamefont {Scragg},
  \citenamefont {Ericson}, \citenamefont {Kubart}, \citenamefont {Edoff},\ and\
  \citenamefont {Platzer-Bjo?rkman}}]{scragg2011chemical}%
  \BibitemOpen
  \bibfield  {author} {\bibinfo {author} {\bibfnamefont {J.~J.}\ \bibnamefont
  {Scragg}}, \bibinfo {author} {\bibfnamefont {T.}~\bibnamefont {Ericson}},
  \bibinfo {author} {\bibfnamefont {T.}~\bibnamefont {Kubart}}, \bibinfo
  {author} {\bibfnamefont {M.}~\bibnamefont {Edoff}}, \ and\ \bibinfo {author}
  {\bibfnamefont {C.}~\bibnamefont {Platzer-Bjo?rkman}},\ }\href@noop {}
  {\bibfield  {journal} {\bibinfo  {journal} {Chemistry of Materials}\ }\textbf
  {\bibinfo {volume} {23}},\ \bibinfo {pages} {4625} (\bibinfo {year}
  {2011})}\BibitemShut {NoStop}%
\bibitem [{\citenamefont {Biswas}\ \emph
  {et~al.}(2010{\natexlab{a}})\citenamefont {Biswas}, \citenamefont {Lany},\
  and\ \citenamefont {Zunger}}]{biswas2010electronic}%
  \BibitemOpen
  \bibfield  {author} {\bibinfo {author} {\bibfnamefont {K.}~\bibnamefont
  {Biswas}}, \bibinfo {author} {\bibfnamefont {S.}~\bibnamefont {Lany}}, \ and\
  \bibinfo {author} {\bibfnamefont {A.}~\bibnamefont {Zunger}},\ }\href@noop {}
  {\bibfield  {journal} {\bibinfo  {journal} {Applied Physics Letters}\
  }\textbf {\bibinfo {volume} {96}},\ \bibinfo {pages} {201902} (\bibinfo
  {year} {2010}{\natexlab{a}})}\BibitemShut {NoStop}%
\bibitem [{\citenamefont {Binek}\ \emph {et~al.}(2015)\citenamefont {Binek},
  \citenamefont {Hanusch}, \citenamefont {Docampo},\ and\ \citenamefont
  {Bein}}]{binek2015stabilization}%
  \BibitemOpen
  \bibfield  {author} {\bibinfo {author} {\bibfnamefont {A.}~\bibnamefont
  {Binek}}, \bibinfo {author} {\bibfnamefont {F.~C.}\ \bibnamefont {Hanusch}},
  \bibinfo {author} {\bibfnamefont {P.}~\bibnamefont {Docampo}}, \ and\
  \bibinfo {author} {\bibfnamefont {T.}~\bibnamefont {Bein}},\ }\href@noop {}
  {\bibfield  {journal} {\bibinfo  {journal} {The journal of physical chemistry
  letters}\ }\textbf {\bibinfo {volume} {6}},\ \bibinfo {pages} {1249}
  (\bibinfo {year} {2015})}\BibitemShut {NoStop}%
\bibitem [{\citenamefont {Li}\ \emph {et~al.}(2008)\citenamefont {Li},
  \citenamefont {Lu}, \citenamefont {Ding}, \citenamefont {Feng}, \citenamefont
  {Gao},\ and\ \citenamefont {Guo}}]{li2008formability}%
  \BibitemOpen
  \bibfield  {author} {\bibinfo {author} {\bibfnamefont {C.}~\bibnamefont
  {Li}}, \bibinfo {author} {\bibfnamefont {X.}~\bibnamefont {Lu}}, \bibinfo
  {author} {\bibfnamefont {W.}~\bibnamefont {Ding}}, \bibinfo {author}
  {\bibfnamefont {L.}~\bibnamefont {Feng}}, \bibinfo {author} {\bibfnamefont
  {Y.}~\bibnamefont {Gao}}, \ and\ \bibinfo {author} {\bibfnamefont
  {Z.}~\bibnamefont {Guo}},\ }\href@noop {} {\bibfield  {journal} {\bibinfo
  {journal} {Acta Crystallographica Section B: Structural Science}\ }\textbf
  {\bibinfo {volume} {64}},\ \bibinfo {pages} {702} (\bibinfo {year}
  {2008})}\BibitemShut {NoStop}%
\bibitem [{\citenamefont {Yokokawa}\ \emph {et~al.}(1989)\citenamefont
  {Yokokawa}, \citenamefont {Kawada},\ and\ \citenamefont
  {Dokiya}}]{yokokawa1989thermodynamic}%
  \BibitemOpen
  \bibfield  {author} {\bibinfo {author} {\bibfnamefont {H.}~\bibnamefont
  {Yokokawa}}, \bibinfo {author} {\bibfnamefont {T.}~\bibnamefont {Kawada}}, \
  and\ \bibinfo {author} {\bibfnamefont {M.}~\bibnamefont {Dokiya}},\
  }\href@noop {} {\bibfield  {journal} {\bibinfo  {journal} {Journal of the
  American Ceramic Society}\ }\textbf {\bibinfo {volume} {72}},\ \bibinfo
  {pages} {152} (\bibinfo {year} {1989})}\BibitemShut {NoStop}%
\bibitem [{\citenamefont {Motta}\ \emph {et~al.}(2015)\citenamefont {Motta},
  \citenamefont {El-Mellouhi}, \citenamefont {Kais}, \citenamefont {Tabet},
  \citenamefont {Alharbi},\ and\ \citenamefont {Sanvito}}]{motta2015revealing}%
  \BibitemOpen
  \bibfield  {author} {\bibinfo {author} {\bibfnamefont {C.}~\bibnamefont
  {Motta}}, \bibinfo {author} {\bibfnamefont {F.}~\bibnamefont {El-Mellouhi}},
  \bibinfo {author} {\bibfnamefont {S.}~\bibnamefont {Kais}}, \bibinfo {author}
  {\bibfnamefont {N.}~\bibnamefont {Tabet}}, \bibinfo {author} {\bibfnamefont
  {F.}~\bibnamefont {Alharbi}}, \ and\ \bibinfo {author} {\bibfnamefont
  {S.}~\bibnamefont {Sanvito}},\ }\href@noop {} {\bibfield  {journal} {\bibinfo
   {journal} {Nature communications}\ }\textbf {\bibinfo {volume} {6}}
  (\bibinfo {year} {2015})}\BibitemShut {NoStop}%
\bibitem [{\citenamefont {Zheng}\ \emph
  {et~al.}(2015{\natexlab{b}})\citenamefont {Zheng}, \citenamefont {Tan},
  \citenamefont {Liu},\ and\ \citenamefont {Rappe}}]{zheng2015rashba}%
  \BibitemOpen
  \bibfield  {author} {\bibinfo {author} {\bibfnamefont {F.}~\bibnamefont
  {Zheng}}, \bibinfo {author} {\bibfnamefont {L.~Z.}\ \bibnamefont {Tan}},
  \bibinfo {author} {\bibfnamefont {S.}~\bibnamefont {Liu}}, \ and\ \bibinfo
  {author} {\bibfnamefont {A.~M.}\ \bibnamefont {Rappe}},\ }\href@noop {}
  {\bibfield  {journal} {\bibinfo  {journal} {Nano letters}\ }\textbf {\bibinfo
  {volume} {15}},\ \bibinfo {pages} {7794} (\bibinfo {year}
  {2015}{\natexlab{b}})}\BibitemShut {NoStop}%
\bibitem [{\citenamefont {Hoffmann}(1987)}]{hoffmann1987chemistry}%
  \BibitemOpen
  \bibfield  {author} {\bibinfo {author} {\bibfnamefont {R.}~\bibnamefont
  {Hoffmann}},\ }\href@noop {} {\bibfield  {journal} {\bibinfo  {journal}
  {Angewandte Chemie International Edition in English}\ }\textbf {\bibinfo
  {volume} {26}},\ \bibinfo {pages} {846} (\bibinfo {year} {1987})}\BibitemShut
  {NoStop}%
\bibitem [{\citenamefont {Noh}\ \emph {et~al.}(2013)\citenamefont {Noh},
  \citenamefont {Im}, \citenamefont {Heo}, \citenamefont {Mandal},\ and\
  \citenamefont {Seok}}]{noh2013chemical}%
  \BibitemOpen
  \bibfield  {author} {\bibinfo {author} {\bibfnamefont {J.~H.}\ \bibnamefont
  {Noh}}, \bibinfo {author} {\bibfnamefont {S.~H.}\ \bibnamefont {Im}},
  \bibinfo {author} {\bibfnamefont {J.~H.}\ \bibnamefont {Heo}}, \bibinfo
  {author} {\bibfnamefont {T.~N.}\ \bibnamefont {Mandal}}, \ and\ \bibinfo
  {author} {\bibfnamefont {S.~I.}\ \bibnamefont {Seok}},\ }\href@noop {}
  {\bibfield  {journal} {\bibinfo  {journal} {Nano letters}\ }\textbf {\bibinfo
  {volume} {13}},\ \bibinfo {pages} {1764} (\bibinfo {year}
  {2013})}\BibitemShut {NoStop}%
\bibitem [{\citenamefont {Kulkarni}\ \emph {et~al.}(2014)\citenamefont
  {Kulkarni}, \citenamefont {Baikie}, \citenamefont {Boix}, \citenamefont
  {Yantara}, \citenamefont {Mathews},\ and\ \citenamefont
  {Mhaisalkar}}]{kulkarni2014band}%
  \BibitemOpen
  \bibfield  {author} {\bibinfo {author} {\bibfnamefont {S.~A.}\ \bibnamefont
  {Kulkarni}}, \bibinfo {author} {\bibfnamefont {T.}~\bibnamefont {Baikie}},
  \bibinfo {author} {\bibfnamefont {P.~P.}\ \bibnamefont {Boix}}, \bibinfo
  {author} {\bibfnamefont {N.}~\bibnamefont {Yantara}}, \bibinfo {author}
  {\bibfnamefont {N.}~\bibnamefont {Mathews}}, \ and\ \bibinfo {author}
  {\bibfnamefont {S.}~\bibnamefont {Mhaisalkar}},\ }\href@noop {} {\bibfield
  {journal} {\bibinfo  {journal} {Journal of Materials Chemistry A}\ }\textbf
  {\bibinfo {volume} {2}},\ \bibinfo {pages} {9221} (\bibinfo {year}
  {2014})}\BibitemShut {NoStop}%
\bibitem [{\citenamefont {Suarez}\ \emph {et~al.}(2014)\citenamefont {Suarez},
  \citenamefont {Gonzalez-Pedro}, \citenamefont {Ripolles}, \citenamefont
  {Sanchez}, \citenamefont {Otero},\ and\ \citenamefont
  {Mora-Sero}}]{suarez2014recombination}%
  \BibitemOpen
  \bibfield  {author} {\bibinfo {author} {\bibfnamefont {B.}~\bibnamefont
  {Suarez}}, \bibinfo {author} {\bibfnamefont {V.}~\bibnamefont
  {Gonzalez-Pedro}}, \bibinfo {author} {\bibfnamefont {T.~S.}\ \bibnamefont
  {Ripolles}}, \bibinfo {author} {\bibfnamefont {R.~S.}\ \bibnamefont
  {Sanchez}}, \bibinfo {author} {\bibfnamefont {L.}~\bibnamefont {Otero}}, \
  and\ \bibinfo {author} {\bibfnamefont {I.}~\bibnamefont {Mora-Sero}},\
  }\href@noop {} {\bibfield  {journal} {\bibinfo  {journal} {The journal of
  physical chemistry letters}\ }\textbf {\bibinfo {volume} {5}},\ \bibinfo
  {pages} {1628} (\bibinfo {year} {2014})}\BibitemShut {NoStop}%
\bibitem [{\citenamefont {D¡¯Innocenzo}\ \emph {et~al.}(2014)\citenamefont
  {D¡¯Innocenzo}, \citenamefont {Grancini}, \citenamefont {Alcocer},
  \citenamefont {Kandada}, \citenamefont {Stranks}, \citenamefont {Lee},
  \citenamefont {Lanzani}, \citenamefont {Snaith},\ and\ \citenamefont
  {Petrozza}}]{d2014excitons}%
  \BibitemOpen
  \bibfield  {author} {\bibinfo {author} {\bibfnamefont {V.}~\bibnamefont
  {D¡¯Innocenzo}}, \bibinfo {author} {\bibfnamefont {G.}~\bibnamefont
  {Grancini}}, \bibinfo {author} {\bibfnamefont {M.~J.}\ \bibnamefont
  {Alcocer}}, \bibinfo {author} {\bibfnamefont {A.~R.~S.}\ \bibnamefont
  {Kandada}}, \bibinfo {author} {\bibfnamefont {S.~D.}\ \bibnamefont
  {Stranks}}, \bibinfo {author} {\bibfnamefont {M.~M.}\ \bibnamefont {Lee}},
  \bibinfo {author} {\bibfnamefont {G.}~\bibnamefont {Lanzani}}, \bibinfo
  {author} {\bibfnamefont {H.~J.}\ \bibnamefont {Snaith}}, \ and\ \bibinfo
  {author} {\bibfnamefont {A.}~\bibnamefont {Petrozza}},\ }\href@noop {}
  {\bibfield  {journal} {\bibinfo  {journal} {Nature communications}\ }\textbf
  {\bibinfo {volume} {5}} (\bibinfo {year} {2014})}\BibitemShut {NoStop}%
\bibitem [{\citenamefont {Hirasawa}\ \emph
  {et~al.}(1994{\natexlab{b}})\citenamefont {Hirasawa}, \citenamefont
  {Ishihara}, \citenamefont {Goto}, \citenamefont {Uchida},\ and\ \citenamefont
  {Miura}}]{hirasawa1994magnetoabsorption}%
  \BibitemOpen
  \bibfield  {author} {\bibinfo {author} {\bibfnamefont {M.}~\bibnamefont
  {Hirasawa}}, \bibinfo {author} {\bibfnamefont {T.}~\bibnamefont {Ishihara}},
  \bibinfo {author} {\bibfnamefont {T.}~\bibnamefont {Goto}}, \bibinfo {author}
  {\bibfnamefont {K.}~\bibnamefont {Uchida}}, \ and\ \bibinfo {author}
  {\bibfnamefont {N.}~\bibnamefont {Miura}},\ }\href@noop {} {\bibfield
  {journal} {\bibinfo  {journal} {Physica B: Condensed Matter}\ }\textbf
  {\bibinfo {volume} {201}},\ \bibinfo {pages} {427} (\bibinfo {year}
  {1994}{\natexlab{b}})}\BibitemShut {NoStop}%
\bibitem [{\citenamefont {Jiang}\ \emph {et~al.}(2015)\citenamefont {Jiang},
  \citenamefont {Rebollar}, \citenamefont {Gong}, \citenamefont {Piacentino},
  \citenamefont {Zheng},\ and\ \citenamefont {Xu}}]{jiang2015pseudohalide}%
  \BibitemOpen
  \bibfield  {author} {\bibinfo {author} {\bibfnamefont {Q.}~\bibnamefont
  {Jiang}}, \bibinfo {author} {\bibfnamefont {D.}~\bibnamefont {Rebollar}},
  \bibinfo {author} {\bibfnamefont {J.}~\bibnamefont {Gong}}, \bibinfo {author}
  {\bibfnamefont {E.~L.}\ \bibnamefont {Piacentino}}, \bibinfo {author}
  {\bibfnamefont {C.}~\bibnamefont {Zheng}}, \ and\ \bibinfo {author}
  {\bibfnamefont {T.}~\bibnamefont {Xu}},\ }\href@noop {} {\bibfield  {journal}
  {\bibinfo  {journal} {Angewandte Chemie}\ }\textbf {\bibinfo {volume}
  {127}},\ \bibinfo {pages} {7727} (\bibinfo {year} {2015})}\BibitemShut
  {NoStop}%
\bibitem [{\citenamefont {Nagane}\ \emph {et~al.}(2014)\citenamefont {Nagane},
  \citenamefont {Bansode}, \citenamefont {Game}, \citenamefont {Chhatre},\ and\
  \citenamefont {Ogale}}]{nagane2014ch}%
  \BibitemOpen
  \bibfield  {author} {\bibinfo {author} {\bibfnamefont {S.}~\bibnamefont
  {Nagane}}, \bibinfo {author} {\bibfnamefont {U.}~\bibnamefont {Bansode}},
  \bibinfo {author} {\bibfnamefont {O.}~\bibnamefont {Game}}, \bibinfo {author}
  {\bibfnamefont {S.}~\bibnamefont {Chhatre}}, \ and\ \bibinfo {author}
  {\bibfnamefont {S.}~\bibnamefont {Ogale}},\ }\href@noop {} {\bibfield
  {journal} {\bibinfo  {journal} {Chemical Communications}\ }\textbf {\bibinfo
  {volume} {50}},\ \bibinfo {pages} {9741} (\bibinfo {year}
  {2014})}\BibitemShut {NoStop}%
\bibitem [{\citenamefont {Hendon}\ \emph {et~al.}(2015)\citenamefont {Hendon},
  \citenamefont {Yang}, \citenamefont {Burton},\ and\ \citenamefont
  {Walsh}}]{hendon2015assessment}%
  \BibitemOpen
  \bibfield  {author} {\bibinfo {author} {\bibfnamefont {C.~H.}\ \bibnamefont
  {Hendon}}, \bibinfo {author} {\bibfnamefont {R.~X.}\ \bibnamefont {Yang}},
  \bibinfo {author} {\bibfnamefont {L.~A.}\ \bibnamefont {Burton}}, \ and\
  \bibinfo {author} {\bibfnamefont {A.}~\bibnamefont {Walsh}},\ }\href@noop {}
  {\bibfield  {journal} {\bibinfo  {journal} {Journal of Materials Chemistry
  A}\ }\textbf {\bibinfo {volume} {3}},\ \bibinfo {pages} {9067} (\bibinfo
  {year} {2015})}\BibitemShut {NoStop}%
\bibitem [{\citenamefont {Biswas}\ \emph
  {et~al.}(2010{\natexlab{b}})\citenamefont {Biswas}, \citenamefont {Lany},\
  and\ \citenamefont {Zunger}}]{Biswas2010}%
  \BibitemOpen
  \bibfield  {author} {\bibinfo {author} {\bibfnamefont {K.}~\bibnamefont
  {Biswas}}, \bibinfo {author} {\bibfnamefont {S.}~\bibnamefont {Lany}}, \ and\
  \bibinfo {author} {\bibfnamefont {A.}~\bibnamefont {Zunger}},\ }\href@noop {}
  {\bibfield  {journal} {\bibinfo  {journal} {Applied Physics Letters}\
  }\textbf {\bibinfo {volume} {96}},\ \bibinfo {pages} {94} (\bibinfo {year}
  {2010}{\natexlab{b}})}\BibitemShut {NoStop}%
\bibitem [{\citenamefont {Kresse}\ and\ \citenamefont
  {Furthm{\"u}ller}(1996)}]{kresse1996efficient}%
  \BibitemOpen
  \bibfield  {author} {\bibinfo {author} {\bibfnamefont {G.}~\bibnamefont
  {Kresse}}\ and\ \bibinfo {author} {\bibfnamefont {J.}~\bibnamefont
  {Furthm{\"u}ller}},\ }\href@noop {} {\bibfield  {journal} {\bibinfo
  {journal} {Physical review B}\ }\textbf {\bibinfo {volume} {54}},\ \bibinfo
  {pages} {11169} (\bibinfo {year} {1996})}\BibitemShut {NoStop}%
\bibitem [{\citenamefont {Bl{\"o}chl}(1994)}]{blochl1994projector}%
  \BibitemOpen
  \bibfield  {author} {\bibinfo {author} {\bibfnamefont {P.~E.}\ \bibnamefont
  {Bl{\"o}chl}},\ }\href@noop {} {\bibfield  {journal} {\bibinfo  {journal}
  {Physical Review B}\ }\textbf {\bibinfo {volume} {50}},\ \bibinfo {pages}
  {17953} (\bibinfo {year} {1994})}\BibitemShut {NoStop}%
\bibitem [{\citenamefont {Perdew}\ \emph {et~al.}(1996)\citenamefont {Perdew},
  \citenamefont {Burke},\ and\ \citenamefont
  {Ernzerhof}}]{perdew1996generalized}%
  \BibitemOpen
  \bibfield  {author} {\bibinfo {author} {\bibfnamefont {J.~P.}\ \bibnamefont
  {Perdew}}, \bibinfo {author} {\bibfnamefont {K.}~\bibnamefont {Burke}}, \
  and\ \bibinfo {author} {\bibfnamefont {M.}~\bibnamefont {Ernzerhof}},\
  }\href@noop {} {\bibfield  {journal} {\bibinfo  {journal} {Physical review
  letters}\ }\textbf {\bibinfo {volume} {77}},\ \bibinfo {pages} {3865}
  (\bibinfo {year} {1996})}\BibitemShut {NoStop}%
\bibitem [{\citenamefont {Klime{\v{s}}}\ \emph {et~al.}(2009)\citenamefont
  {Klime{\v{s}}}, \citenamefont {Bowler},\ and\ \citenamefont
  {Michaelides}}]{klimevs2009chemical}%
  \BibitemOpen
  \bibfield  {author} {\bibinfo {author} {\bibfnamefont {J.}~\bibnamefont
  {Klime{\v{s}}}}, \bibinfo {author} {\bibfnamefont {D.~R.}\ \bibnamefont
  {Bowler}}, \ and\ \bibinfo {author} {\bibfnamefont {A.}~\bibnamefont
  {Michaelides}},\ }\href@noop {} {\bibfield  {journal} {\bibinfo  {journal}
  {Journal of Physics: Condensed Matter}\ }\textbf {\bibinfo {volume} {22}},\
  \bibinfo {pages} {022201} (\bibinfo {year} {2009})}\BibitemShut {NoStop}%
\bibitem [{\citenamefont {Grimme}(2006)}]{grimme2006semiempirical}%
  \BibitemOpen
  \bibfield  {author} {\bibinfo {author} {\bibfnamefont {S.}~\bibnamefont
  {Grimme}},\ }\href@noop {} {\bibfield  {journal} {\bibinfo  {journal}
  {Journal of computational chemistry}\ }\textbf {\bibinfo {volume} {27}},\
  \bibinfo {pages} {1787} (\bibinfo {year} {2006})}\BibitemShut {NoStop}%
\bibitem [{\citenamefont {Krukau}\ \emph {et~al.}(2006)\citenamefont {Krukau},
  \citenamefont {Vydrov}, \citenamefont {Izmaylov},\ and\ \citenamefont
  {Scuseria}}]{krukau2006influence}%
  \BibitemOpen
  \bibfield  {author} {\bibinfo {author} {\bibfnamefont {A.~V.}\ \bibnamefont
  {Krukau}}, \bibinfo {author} {\bibfnamefont {O.~A.}\ \bibnamefont {Vydrov}},
  \bibinfo {author} {\bibfnamefont {A.~F.}\ \bibnamefont {Izmaylov}}, \ and\
  \bibinfo {author} {\bibfnamefont {G.~E.}\ \bibnamefont {Scuseria}},\
  }\href@noop {} {\bibfield  {journal} {\bibinfo  {journal} {The Journal of
  chemical physics}\ }\textbf {\bibinfo {volume} {125}},\ \bibinfo {pages}
  {224106} (\bibinfo {year} {2006})}\BibitemShut {NoStop}%
\bibitem [{\citenamefont {Poglitsch}\ and\ \citenamefont
  {Weber}(1987)}]{poglitsch1987dynamic}%
  \BibitemOpen
  \bibfield  {author} {\bibinfo {author} {\bibfnamefont {A.}~\bibnamefont
  {Poglitsch}}\ and\ \bibinfo {author} {\bibfnamefont {D.}~\bibnamefont
  {Weber}},\ }\href@noop {} {\bibfield  {journal} {\bibinfo  {journal} {The
  Journal of chemical physics}\ }\textbf {\bibinfo {volume} {87}},\ \bibinfo
  {pages} {6373} (\bibinfo {year} {1987})}\BibitemShut {NoStop}%
\bibitem [{\citenamefont {Ryu}\ \emph {et~al.}(2014)\citenamefont {Ryu},
  \citenamefont {Noh}, \citenamefont {Jeon}, \citenamefont {Kim}, \citenamefont
  {Yang}, \citenamefont {Seo},\ and\ \citenamefont {Seok}}]{ryu2014voltage}%
  \BibitemOpen
  \bibfield  {author} {\bibinfo {author} {\bibfnamefont {S.}~\bibnamefont
  {Ryu}}, \bibinfo {author} {\bibfnamefont {J.~H.}\ \bibnamefont {Noh}},
  \bibinfo {author} {\bibfnamefont {N.~J.}\ \bibnamefont {Jeon}}, \bibinfo
  {author} {\bibfnamefont {Y.~C.}\ \bibnamefont {Kim}}, \bibinfo {author}
  {\bibfnamefont {W.~S.}\ \bibnamefont {Yang}}, \bibinfo {author}
  {\bibfnamefont {J.}~\bibnamefont {Seo}}, \ and\ \bibinfo {author}
  {\bibfnamefont {S.~I.}\ \bibnamefont {Seok}},\ }\href@noop {} {\bibfield
  {journal} {\bibinfo  {journal} {Energy \& Environmental Science}\ }\textbf
  {\bibinfo {volume} {7}},\ \bibinfo {pages} {2614} (\bibinfo {year}
  {2014})}\BibitemShut {NoStop}%
\bibitem [{\citenamefont {Frost}\ \emph {et~al.}(2014)\citenamefont {Frost},
  \citenamefont {Butler}, \citenamefont {Brivio}, \citenamefont {Hendon},
  \citenamefont {Van~Schilfgaarde},\ and\ \citenamefont
  {Walsh}}]{frost2014atomistic}%
  \BibitemOpen
  \bibfield  {author} {\bibinfo {author} {\bibfnamefont {J.~M.}\ \bibnamefont
  {Frost}}, \bibinfo {author} {\bibfnamefont {K.~T.}\ \bibnamefont {Butler}},
  \bibinfo {author} {\bibfnamefont {F.}~\bibnamefont {Brivio}}, \bibinfo
  {author} {\bibfnamefont {C.~H.}\ \bibnamefont {Hendon}}, \bibinfo {author}
  {\bibfnamefont {M.}~\bibnamefont {Van~Schilfgaarde}}, \ and\ \bibinfo
  {author} {\bibfnamefont {A.}~\bibnamefont {Walsh}},\ }\href@noop {}
  {\bibfield  {journal} {\bibinfo  {journal} {Nano letters}\ }\textbf {\bibinfo
  {volume} {14}},\ \bibinfo {pages} {2584} (\bibinfo {year}
  {2014})}\BibitemShut {NoStop}%
\bibitem [{\citenamefont {Wasylishen}\ \emph {et~al.}(1985)\citenamefont
  {Wasylishen}, \citenamefont {Knop},\ and\ \citenamefont
  {Macdonald}}]{wasylishen1985cation}%
  \BibitemOpen
  \bibfield  {author} {\bibinfo {author} {\bibfnamefont {R.~E.}\ \bibnamefont
  {Wasylishen}}, \bibinfo {author} {\bibfnamefont {O.}~\bibnamefont {Knop}}, \
  and\ \bibinfo {author} {\bibfnamefont {J.~B.}\ \bibnamefont {Macdonald}},\
  }\href@noop {} {\bibfield  {journal} {\bibinfo  {journal} {Solid state
  communications}\ }\textbf {\bibinfo {volume} {56}},\ \bibinfo {pages} {581}
  (\bibinfo {year} {1985})}\BibitemShut {NoStop}%
\bibitem [{\citenamefont {Ishida}\ \emph {et~al.}(1995)\citenamefont {Ishida},
  \citenamefont {Maeda}, \citenamefont {Hirano}, \citenamefont {Fujimoto},
  \citenamefont {Kubozono}, \citenamefont {Kashino},\ and\ \citenamefont
  {Emura}}]{ishida1995exafs}%
  \BibitemOpen
  \bibfield  {author} {\bibinfo {author} {\bibfnamefont {H.}~\bibnamefont
  {Ishida}}, \bibinfo {author} {\bibfnamefont {H.}~\bibnamefont {Maeda}},
  \bibinfo {author} {\bibfnamefont {A.}~\bibnamefont {Hirano}}, \bibinfo
  {author} {\bibfnamefont {T.}~\bibnamefont {Fujimoto}}, \bibinfo {author}
  {\bibfnamefont {Y.}~\bibnamefont {Kubozono}}, \bibinfo {author}
  {\bibfnamefont {S.}~\bibnamefont {Kashino}}, \ and\ \bibinfo {author}
  {\bibfnamefont {S.}~\bibnamefont {Emura}},\ }\href@noop {} {\bibfield
  {journal} {\bibinfo  {journal} {Zeitschrift f{\"u}r Naturforschung A}\
  }\textbf {\bibinfo {volume} {50}},\ \bibinfo {pages} {876} (\bibinfo {year}
  {1995})}\BibitemShut {NoStop}%
\bibitem [{\citenamefont {Ishida}\ \emph {et~al.}(1982)\citenamefont {Ishida},
  \citenamefont {Ikeda},\ and\ \citenamefont {Nakamura}}]{ishida1982pre}%
  \BibitemOpen
  \bibfield  {author} {\bibinfo {author} {\bibfnamefont {H.}~\bibnamefont
  {Ishida}}, \bibinfo {author} {\bibfnamefont {R.}~\bibnamefont {Ikeda}}, \
  and\ \bibinfo {author} {\bibfnamefont {D.}~\bibnamefont {Nakamura}},\
  }\href@noop {} {\bibfield  {journal} {\bibinfo  {journal} {physica status
  solidi (a)}\ }\textbf {\bibinfo {volume} {70}},\ \bibinfo {pages} {K151}
  (\bibinfo {year} {1982})}\BibitemShut {NoStop}%
\bibitem [{\citenamefont {Umari}\ \emph {et~al.}(2014)\citenamefont {Umari},
  \citenamefont {Mosconi},\ and\ \citenamefont
  {De~Angelis}}]{umari2014relativistic}%
  \BibitemOpen
  \bibfield  {author} {\bibinfo {author} {\bibfnamefont {P.}~\bibnamefont
  {Umari}}, \bibinfo {author} {\bibfnamefont {E.}~\bibnamefont {Mosconi}}, \
  and\ \bibinfo {author} {\bibfnamefont {F.}~\bibnamefont {De~Angelis}},\
  }\href@noop {} {\bibfield  {journal} {\bibinfo  {journal} {Scientific
  reports}\ }\textbf {\bibinfo {volume} {4}} (\bibinfo {year}
  {2014})}\BibitemShut {NoStop}%
\bibitem [{\citenamefont {Shannon}(1976)}]{shannon1976revised}%
  \BibitemOpen
  \bibfield  {author} {\bibinfo {author} {\bibfnamefont {R.~t.}\ \bibnamefont
  {Shannon}},\ }\href@noop {} {\bibfield  {journal} {\bibinfo  {journal} {Acta
  Crystallographica Section A: Crystal Physics, Diffraction, Theoretical and
  General Crystallography}\ }\textbf {\bibinfo {volume} {32}},\ \bibinfo
  {pages} {751} (\bibinfo {year} {1976})}\BibitemShut {NoStop}%
\bibitem [{\citenamefont {Madsen}\ and\ \citenamefont
  {Singh}(2006)}]{madsen2006boltztrap}%
  \BibitemOpen
  \bibfield  {author} {\bibinfo {author} {\bibfnamefont {G.~K.}\ \bibnamefont
  {Madsen}}\ and\ \bibinfo {author} {\bibfnamefont {D.~J.}\ \bibnamefont
  {Singh}},\ }\href@noop {} {\bibfield  {journal} {\bibinfo  {journal}
  {Computer Physics Communications}\ }\textbf {\bibinfo {volume} {175}},\
  \bibinfo {pages} {67} (\bibinfo {year} {2006})}\BibitemShut {NoStop}%
\bibitem [{\citenamefont {Umari}\ and\ \citenamefont
  {Pasquarello}(2002)}]{umari2002ab}%
  \BibitemOpen
  \bibfield  {author} {\bibinfo {author} {\bibfnamefont {P.}~\bibnamefont
  {Umari}}\ and\ \bibinfo {author} {\bibfnamefont {A.}~\bibnamefont
  {Pasquarello}},\ }\href@noop {} {\bibfield  {journal} {\bibinfo  {journal}
  {Physical review letters}\ }\textbf {\bibinfo {volume} {89}},\ \bibinfo
  {pages} {157602} (\bibinfo {year} {2002})}\BibitemShut {NoStop}%
\bibitem [{\citenamefont {Souza}\ \emph {et~al.}(2002)\citenamefont {Souza},
  \citenamefont {{\'I}{\~n}iguez},\ and\ \citenamefont
  {Vanderbilt}}]{souza2002first}%
  \BibitemOpen
  \bibfield  {author} {\bibinfo {author} {\bibfnamefont {I.}~\bibnamefont
  {Souza}}, \bibinfo {author} {\bibfnamefont {J.}~\bibnamefont
  {{\'I}{\~n}iguez}}, \ and\ \bibinfo {author} {\bibfnamefont {D.}~\bibnamefont
  {Vanderbilt}},\ }\href@noop {} {\bibfield  {journal} {\bibinfo  {journal}
  {Physical review letters}\ }\textbf {\bibinfo {volume} {89}},\ \bibinfo
  {pages} {117602} (\bibinfo {year} {2002})}\BibitemShut {NoStop}%
\end{thebibliography}%

\end{document}


\title{\textbf{Supplemental Material} for ``Functionality-directed Screening of Pb-free Hybrid Organic-inorganic Perovskites with Desired Intrinsic Photovoltaic Functionalities''}


\author{Dongwen Yang}
\affiliation{College of Materials Science and Engineering and Key Laboratory of Automobile Materials of MOE, Jilin University, Changchun 130012, China}
\author{Jian Lv}
\affiliation{College of Materials Science and Engineering and Key Laboratory of Automobile Materials of MOE, Jilin University, Changchun 130012, China}
\author{Xingang Zhao}
\author{Qiaoling Xu}
\author{Yuhao Fu}
\affiliation{College of Materials Science and Engineering and Key Laboratory of Automobile Materials of MOE, Jilin University, Changchun 130012, China}
\author{Yiqiang Zhan}
\affiliation{State Key Laboratory of ASIC and System, Department of Microelectronics, SIST, Fudan University, Shanghai 200433, China}
\author{Alex Zunger}
\email{ alex.zunger@gmail.com}
\affiliation{University of Colorado and Renewable and Sustainable Energy Institute, Boulder, Colorado 80309, USA}
\author{Lijun Zhang}
\email{lijun_zhang@jlu.edu.cn}
\affiliation{College of Materials Science and Engineering and Key Laboratory of Automobile Materials of MOE, Jilin University, Changchun 130012, China}

\date{\today}

\maketitle

\pagebreak
\clearpage

\vspace*{\fill}
\begin{figure}[ht]
\includegraphics[width=3.5in]{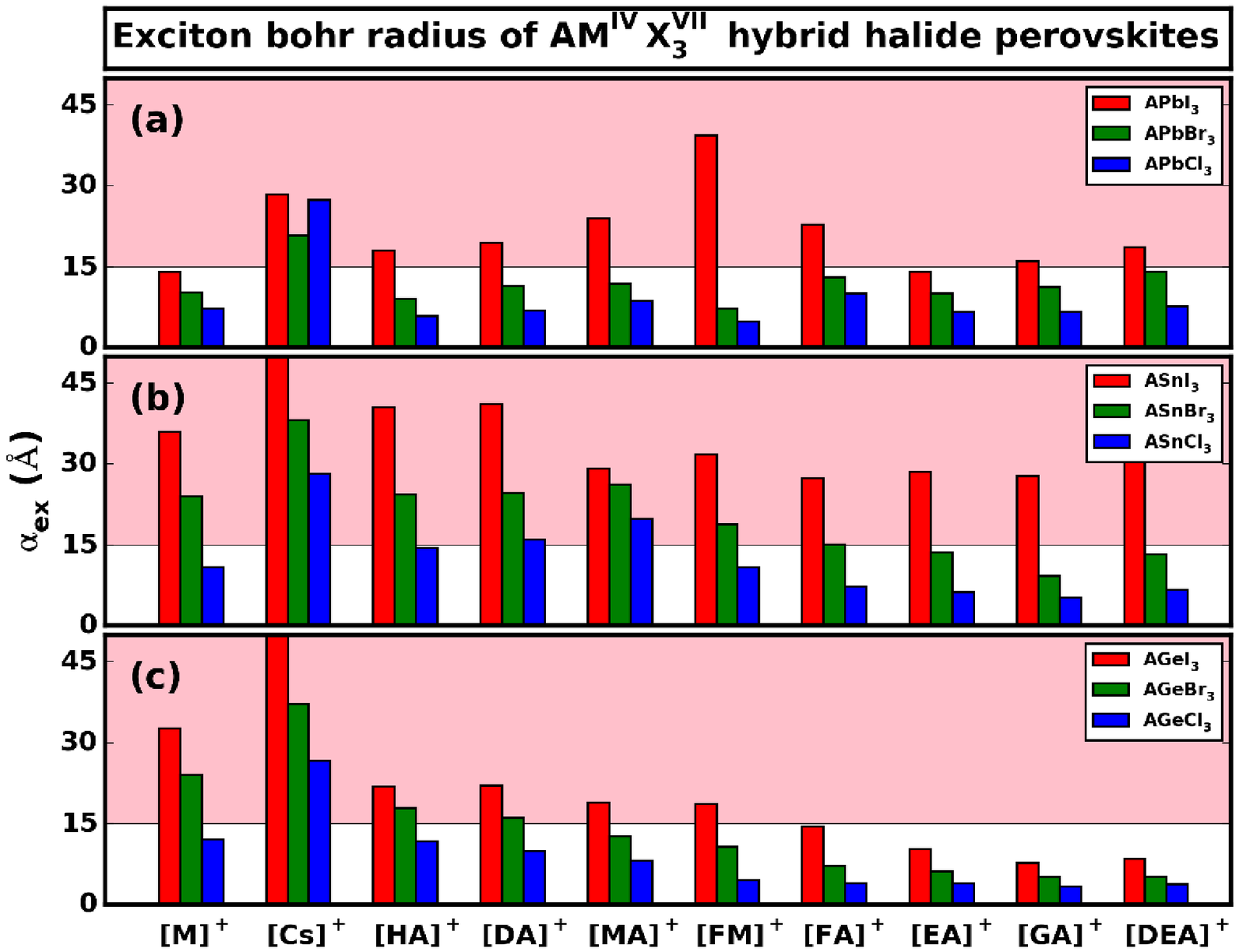}
\centering
\caption{Calculated exciton Bohr radii of (a) Pb, (b) Sn and (c) Ge based AM$^{IV}$X$^{VII}_3$
 perovskites with the hydrogen-like Wannier-Mott model. Shaded areas indicate the criterion applied ($\alpha_{ex}$ $>$ 1.5 nm) for the materials screening.}
\end{figure}
\vfill
\pagebreak
\clearpage
%
\vspace*{\fill}
\begin{figure}[ht]
\includegraphics[width=3.5in]{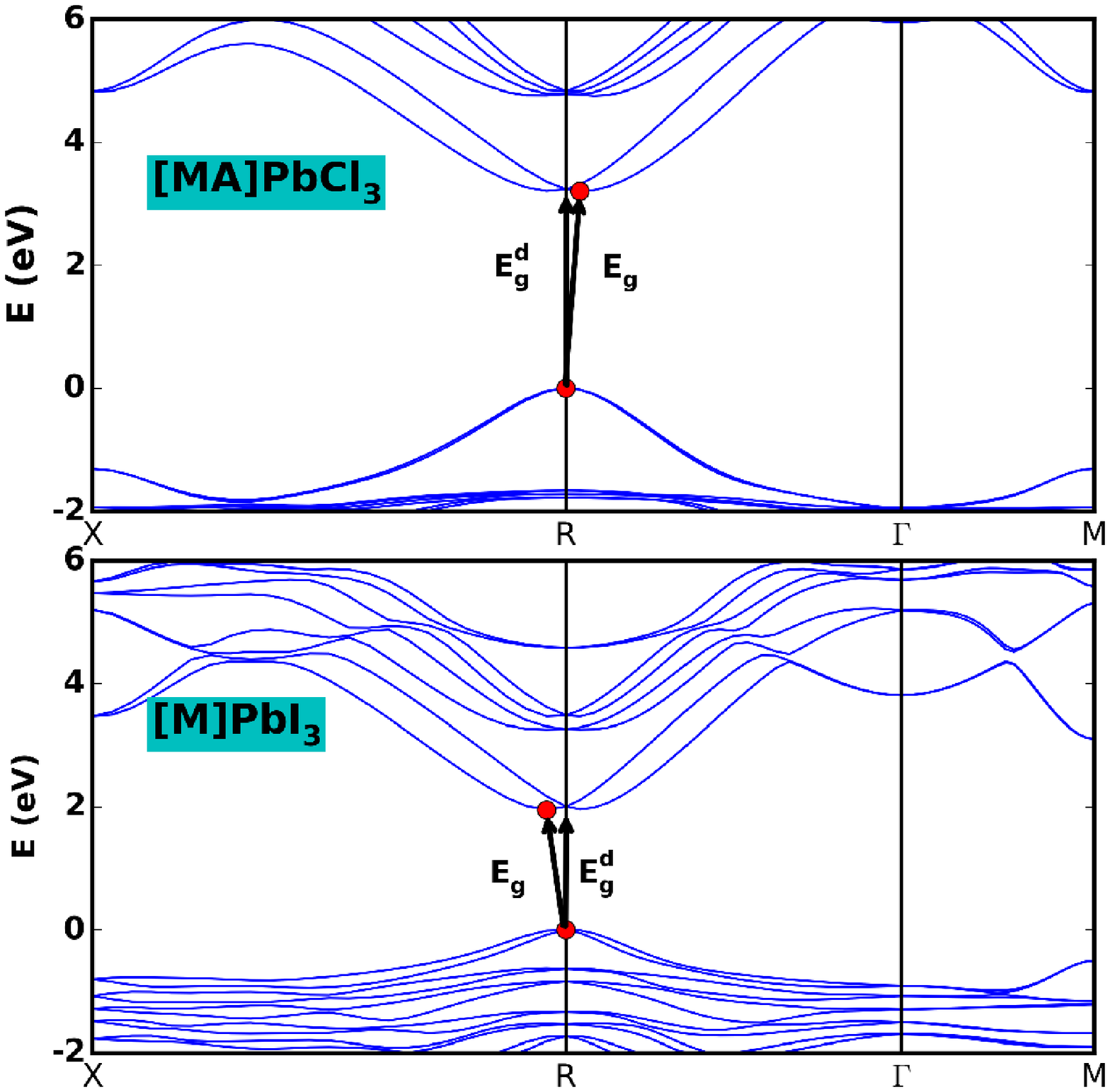}
\centering
\caption{Band structures of the selected AM$^{IV}$X$^{VII}_3$ perovskites having indirect band gaps. The valence band maximum and conduction band minimum are marked by red circles. The actual band gaps (E$_{g}$) and direct band gaps ( E$_{g}^{d}$) are indicated.}
\end{figure}
\vfill
\pagebreak
\clearpage
\vspace*{\fill}
\begin{figure}[ht]
\includegraphics[width=3.5in]{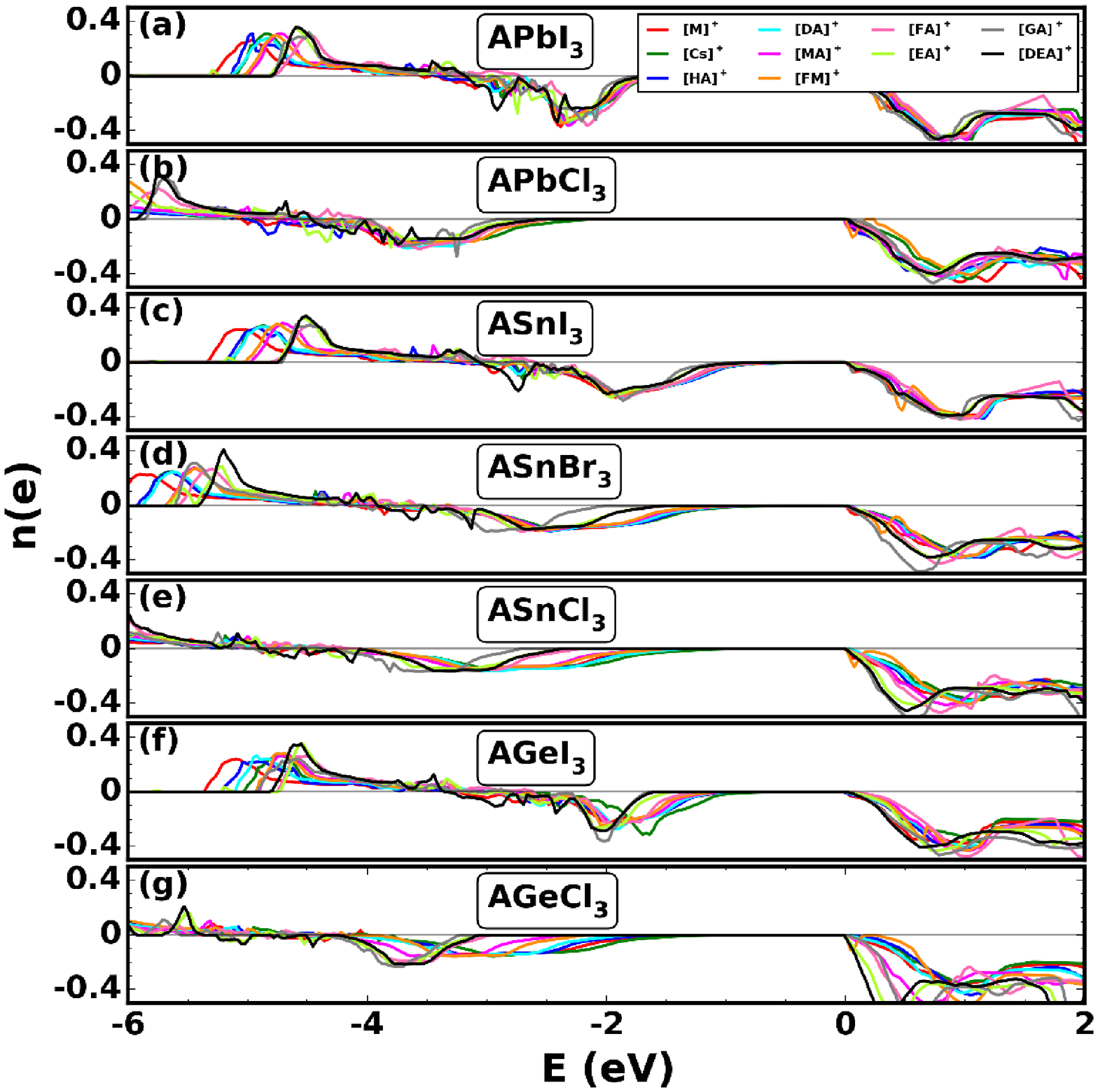}
\centering
\caption{
Crystal orbital overlap populations (COOP) of the AM$^{IV}$X$^{VII}_3$ perovskites. For comparison, the CBM of each material is set to energy zero.}
\end{figure}
\vfill
\pagebreak
\clearpage
\vspace*{\fill}
\begin{figure}[ht]
\includegraphics[width=3.5in]{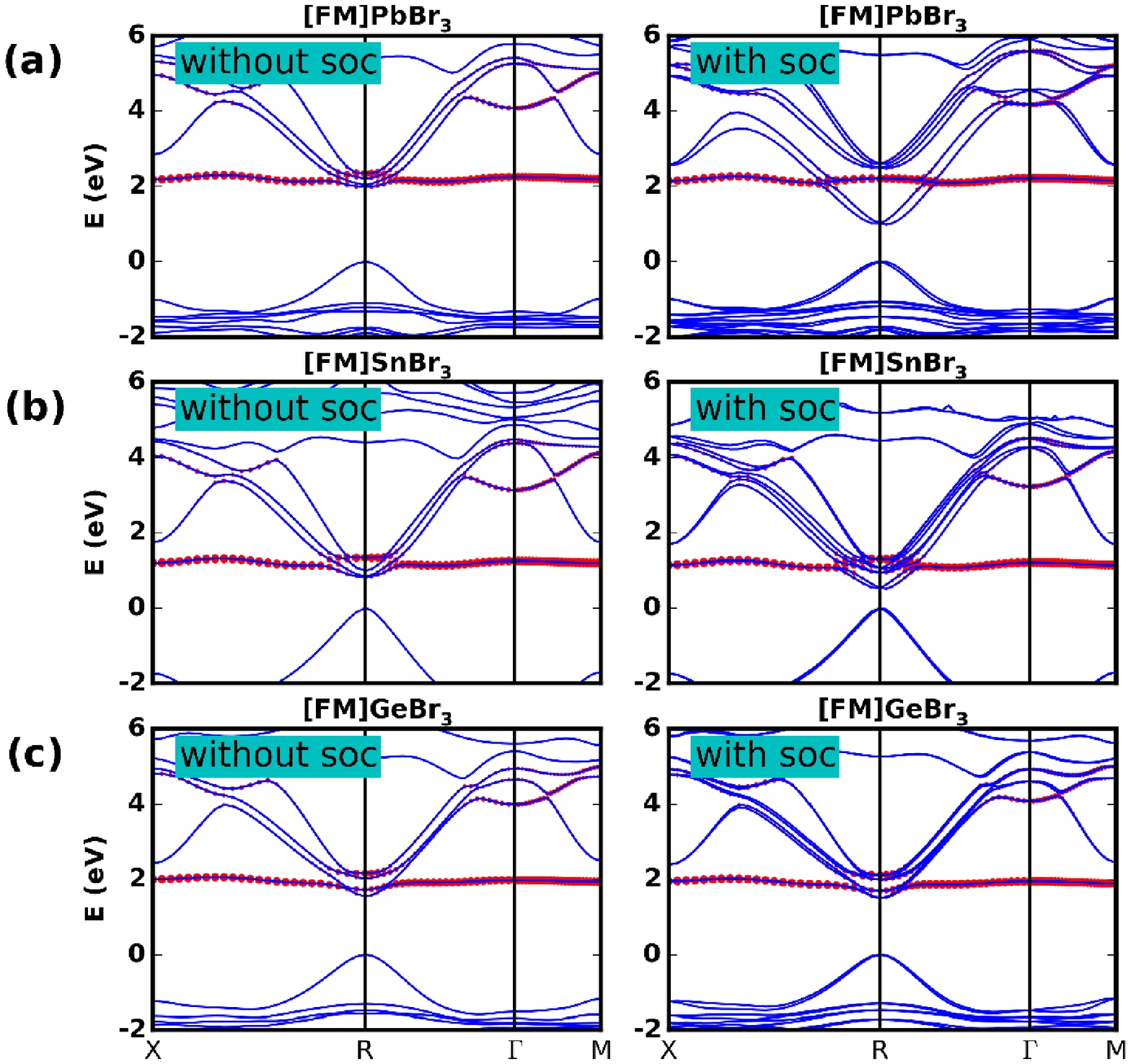}
\centering
\caption{
 Band structures of (a) [FM]PbBr$_3$, (b) [FM]SnBr$_3$, (c) [FM]GeBr$_3$ calculated without (left panels) and with (right panels) the spin-orbit coupling (SOC) effect. 
The orbital projections onto the FM molecule are indicated by red circles.}
\end{figure}
\vfill
\pagebreak
\clearpage
\vspace*{\fill}
\begin{figure}[ht]
\includegraphics[width=3.5in]{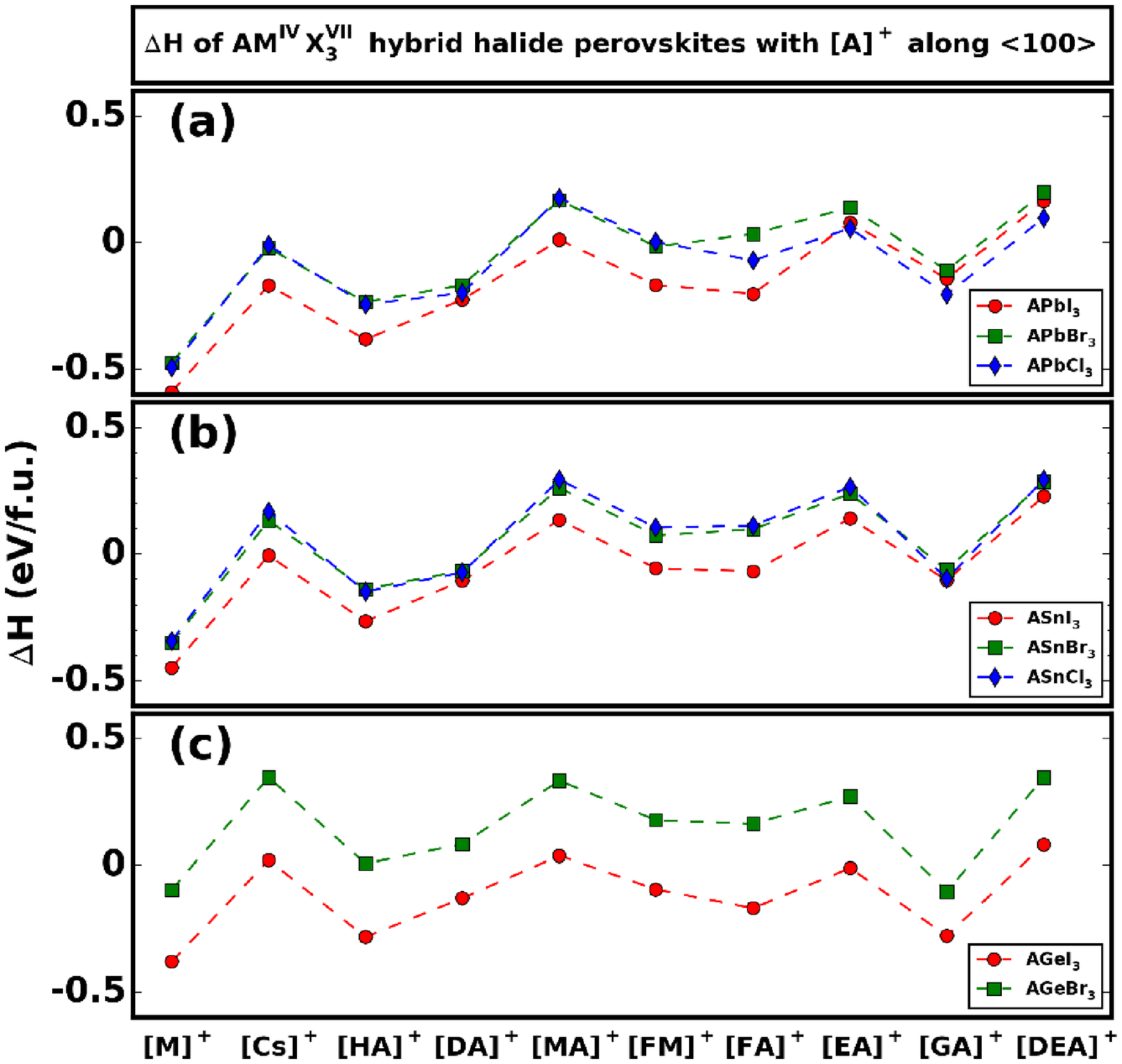}
\centering
\caption{
Calculated decomposition enthalpies $\Delta$H of the AM$^{IV}$X$^{VII}_3$ perovskites with the principal axes of small molecules aligned along the $<$ 100 $>$ direction, 
with respect to decomposed products of AX$^{VII}$+M$^{IV}$X$^{VII}_2$. }
\end{figure}
\vfill
\pagebreak
\clearpage
\vspace*{\fill}
\begin{figure}[ht]
\includegraphics[width=3.5in]{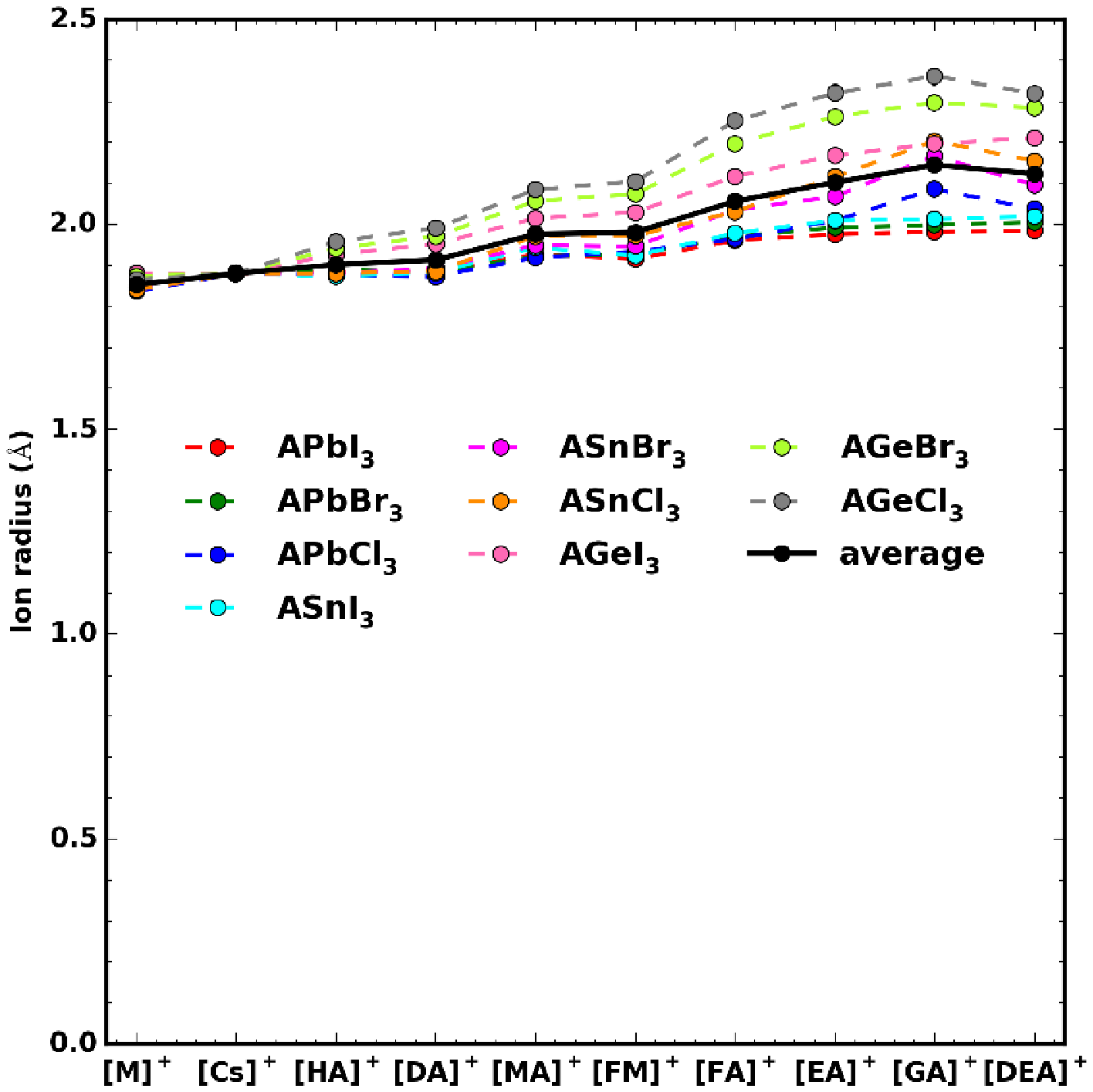}
\centering
\caption{
Evaluation of the steric sizes of organic molecular cations within the idealized solid-sphere model 
[see the Experimental Section (iv)]. }
\end{figure}
\vfill
\pagebreak
\clearpage

\begin{table}
\label{Table S1}
\caption{
Calculated decomposition enthalpies ($\Delta$H) of the candidate AM$^{IV}$X$^{VII}_3$ perovskites with respect to decomposed products of AX$^{IV}$+M$^{IV}$X$^{VII}_2$. 
The compounds passing the screening (with $\Delta$H larger than -0.1 eV/f.u.) are marked by green shading. }
\centering
\begin{tabular}{?c?c?c?c?c?c?c?c?c?c?p{102cm}?}
\Xhline{1pt}
$\Delta$H (eV) &  PbI$_3$ & PbBr$_3$ & PbCl$_3$ & SnI$_3$ & SnBr$_3$ & SnCl$_3$ & GeI$_3$ & GeBr$_3$ & GeCl$_3$  \\
\Xhline{1pt}
[M]$^{+}$  & \cellcolor{white!0} -0.59  & \cellcolor{white!0} -0.48  &  \cellcolor{white!0} -0.49  
&  \cellcolor{white!0} -0.45 & \cellcolor{white!0} -0.35  & \cellcolor{white!0} -0.34  
& \cellcolor{white!0} -0.38  & \cellcolor{white!0} -0.10  & \cellcolor{white!0} -- \\
\Xhline{1pt}
[Cs]$^{+}$  & \cellcolor{white!0} -0.17  & \cellcolor{green!100} -0.02  &  \cellcolor{green!100} -0.01  
&  \cellcolor{green!100} -0.01 & \cellcolor{green!100} 0.13  & \cellcolor{green!100} 0.17  
& \cellcolor{green!100} 0.02  & \cellcolor{green!100} 0.35  & \cellcolor{white!0} -- \\ 
\Xhline{1pt}
[HA]$^{+}$  & \cellcolor{white!0} -0.45  & \cellcolor{white!0} -0.26  &  \cellcolor{white!0} -0.29  
&  \cellcolor{white!0} -0.32 & \cellcolor{white!0} -0.23  & \cellcolor{white!0} -0.23  
& \cellcolor{white!0} -0.30  & \cellcolor{green!100} -0.03  & \cellcolor{white!0} -- \\
\Xhline{1pt}
[DA]$^{+}$  & \cellcolor{white!0} -0.25  & \cellcolor{white!0} -0.19  &  \cellcolor{white!0} -0.21  
&  \cellcolor{white!0} -0.12 & \cellcolor{green!100} -0.09  & \cellcolor{green!100} -0.09  
& \cellcolor{white!0} -0.13  & \cellcolor{green!100} 0.07  & \cellcolor{white!0} -- \\ 
\Xhline{1pt}
[MA]$^{+}$  & \cellcolor{green!100} -0.02  & \cellcolor{green!100} 0.12  &  \cellcolor{green!100} 0.12  
&  \cellcolor{green!100} 0.10 & \cellcolor{green!100} 0.21  & \cellcolor{green!100} 0.24  
& \cellcolor{green!100} 0.03  & \cellcolor{green!100} 0.33  & \cellcolor{white!0} -- \\ 
\Xhline{1pt}
[FM]$^{+}$  & \cellcolor{white!0} -0.21  & \cellcolor{green!100} -0.09  &  \cellcolor{white!0} -0.11  
&  \cellcolor{white!0} -0.11 & \cellcolor{green!100} -0.02  & \cellcolor{green!100}  -0.01  
& \cellcolor{white!0} -0.18  & \cellcolor{green!100} 0.08  & \cellcolor{white!0} -- \\ 
\Xhline{1pt}
[FA]$^{+}$  & \cellcolor{green!100} -0.10  & \cellcolor{green!100} -0.01  &  \cellcolor{green!100} -0.02  
&  \cellcolor{green!100} -0.03 & \cellcolor{green!100} 0.04  & \cellcolor{green!100} 0.08  
& \cellcolor{white!0} -0.16  & \cellcolor{green!100} 0.10  & \cellcolor{white!0} --  \\ 
\Xhline{1pt}
[EA]$^{+}$  & \cellcolor{green!100} 0.09  & \cellcolor{green!100} 0.16  &  \cellcolor{green!100} 0.01  
&  \cellcolor{green!100} 0.16 & \cellcolor{green!100} 0.24  & \cellcolor{green!100} 0.27  
& \cellcolor{green!100} 0.01  & \cellcolor{green!100} 0.25  & \cellcolor{white!0} --  \\ 
\Xhline{1pt}
[GA]$^{+}$  & \cellcolor{white!0} -0.14  & \cellcolor{white!0} -0.11  &  \cellcolor{white!0} -0.24 
&  \cellcolor{green!100} -0.09 & \cellcolor{green!100} -0.06  & \cellcolor{green!100} 0.02  
& \cellcolor{white!0} -0.29  & \cellcolor{green!100} -0.07  & \cellcolor{white!0} -- \\ 
\Xhline{1pt}
[DEA]$^{+}$  & \cellcolor{green!100} 0.16  & \cellcolor{green!100} 0.20  &  \cellcolor{green!100} 0.10  
&  \cellcolor{green!100} 0.22 & \cellcolor{green!100} 0.28  & \cellcolor{green!100} 0.30  
& \cellcolor{green!100} 0.06  & \cellcolor{green!100} 0.34  & \cellcolor{white!0} -- \\ 
\Xhline{1pt}
\end{tabular}
\end{table}

\begin{table*}
\label{Table S2}
\caption{
Calculated decomposition enthalpies ($\Delta$H) of the candidate AM$^{IV}$X$^{VII}_3$ perovskites (with the principal axes of small molecules aligned along the $<$ 100 $>$ direction).
}
\centering
\begin{tabular}{?c?c?c?c?c?c?c?c?c?c?p{102cm}?}
\Xhline{1pt}
$\Delta$H (eV) &  PbI$_3$ & PbBr$_3$ & PbCl$_3$ & SnI$_3$ & SnBr$_3$ & SnCl$_3$ & GeI$_3$ & GeBr$_3$ & GeCl$_3$  \\
\Xhline{1pt}
[M]$^{+}$  & \cellcolor{white!0} -0.59  & \cellcolor{white!0} -0.48  &  \cellcolor{white!0} -0.49
&  \cellcolor{white!0} -0.45 & \cellcolor{white!0} -0.35  & \cellcolor{white!0} -0.34
& \cellcolor{white!0} -0.38  & \cellcolor{white!0} -0.10  & \cellcolor{white!0} -- \\
\Xhline{1pt}
[Cs]$^{+}$  & \cellcolor{white!0} -0.17  & \cellcolor{white!0} -0.02  &  \cellcolor{white!0} -0.01
&  \cellcolor{white!0} -0.01 & \cellcolor{white!0} 0.13  & \cellcolor{white!0} 0.17
& \cellcolor{white!0} 0.02  & \cellcolor{white!0} 0.35  & \cellcolor{white!0} -- \\
\Xhline{1pt}
[HA]$^{+}$  & \cellcolor{white!0} -0.38  & \cellcolor{white!0} -0.23  &  \cellcolor{white!0} -0.24
&  \cellcolor{white!0} -0.27 & \cellcolor{white!0} -0.14  & \cellcolor{white!0} -0.15
& \cellcolor{white!0} -0.28  & \cellcolor{white!0} 0.01  & \cellcolor{white!0} -- \\
\Xhline{1pt}
[DA]$^{+}$  & \cellcolor{white!0} -0.23  & \cellcolor{white!0} -0.17  &  \cellcolor{white!0} -0.20
&  \cellcolor{white!0} -0.10 & \cellcolor{white!0} -0.07  & \cellcolor{white!0} -0.07
& \cellcolor{white!0} -0.13  & \cellcolor{white!0} 0.08  & \cellcolor{white!0} -- \\
\Xhline{1pt}
[MA]$^{+}$  & \cellcolor{white!0} 0.01  & \cellcolor{white!0} 0.17  &  \cellcolor{white!0} 0.18
&  \cellcolor{white!0} 0.13 & \cellcolor{white!0} 0.26  & \cellcolor{white!0} 0.29
& \cellcolor{white!0} 0.04  & \cellcolor{white!0} 0.33  & \cellcolor{white!0} -- \\
\Xhline{1pt}
[FM]$^{+}$  & \cellcolor{white!0} -0.17  & \cellcolor{white!0} -0.01  &  \cellcolor{white!0} 0.00
&  \cellcolor{white!0} -0.06 & \cellcolor{white!0} 0.07  & \cellcolor{white!0} 0.10
& \cellcolor{white!0} -0.09  & \cellcolor{white!0} 0.18  & \cellcolor{white!0} -- \\
\Xhline{1pt}
[FA]$^{+}$  & \cellcolor{white!0} -0.20  & \cellcolor{white!0} 0.03  &  \cellcolor{white!0} -0.07
&  \cellcolor{white!0} -0.07 & \cellcolor{white!0} 0.10  & \cellcolor{white!0} 0.11
& \cellcolor{white!0} -0.17  & \cellcolor{white!0} 0.16  & \cellcolor{white!0} -- \\
\Xhline{1pt}
[EA]$^{+}$  & \cellcolor{white!0} 0.08  & \cellcolor{white!0} 0.14  &  \cellcolor{white!0} 0.06
&  \cellcolor{white!0}  0.14 & \cellcolor{white!0} 0.24  & \cellcolor{white!0} 0.27
& \cellcolor{white!0} -0.01  & \cellcolor{white!0} 0.27  & \cellcolor{white!0} -- \\
\Xhline{1pt}
[GA]$^{+}$  & \cellcolor{white!0} -0.14  & \cellcolor{white!0} -0.11  &  \cellcolor{white!0} -0.20
&  \cellcolor{white!0} -0.10 & \cellcolor{white!0} -0.06  & \cellcolor{white!0} -0.10
& \cellcolor{white!0} -0.28  & \cellcolor{white!0} -0.10  & \cellcolor{white!0} -- \\
\Xhline{1pt}
[DEA]$^{+}$  & \cellcolor{white!0} 0.16  & \cellcolor{white!0} 0.20  &  \cellcolor{white!0} 0.10
&  \cellcolor{white!0} 0.23 & \cellcolor{white!0} 0.28  & \cellcolor{white!0} 0.29
& \cellcolor{white!0} 0.08  & \cellcolor{white!0} 0.35  & \cellcolor{white!0} -- \\
\Xhline{1pt}
\end{tabular}
\end{table*}

\begin{table}
\label{Table S3}
\caption{
Calculated direct band gaps (E$_{g}^{d}$) of the candidate AM$^{IV}$X$^{VII}_3$ perovskites. 
The green shading indicates the compounds passing the current DM (E$_{g}^{d}$ $<$ 2.5 eV), 
as well as the DM in Table S1. The lightblue shading indicates the compounds passing only the current DM.}
\centering
\begin{tabular}{?c?c?c?c?c?c?c?c?c?c?p{102cm}?}
\Xhline{1pt}
E$_{g}^{d}$ (eV) &  PbI$_3$ & PbBr$_3$ & PbCl$_3$ & SnI$_3$ & SnBr$_3$ & SnCl$_3$ & GeI$_3$ & GeBr$_3$ & GeCl$_3$  \\
\Xhline{1pt}
[M]$^{+}$  & \cellcolor{blue!60} 1.95  & \cellcolor{white!0} 2.71  &  \cellcolor{white!0} 3.39
&  \cellcolor{blue!60} 1.21 & \cellcolor{blue!60} 1.91  & \cellcolor{white!0} 2.68
& \cellcolor{blue!60} 1.59  & \cellcolor{white!0} 2.01  & \cellcolor{white!0} 2.54 \\
\Xhline{1pt}
[Cs]$^{+}$  & \cellcolor{blue!60} 1.30  & \cellcolor{green!100} 1.98  &  \cellcolor{white!0} 2.63
&  \cellcolor{green!100} 0.95 & \cellcolor{green!100} 1.53  & \cellcolor{green!100} 2.14
& \cellcolor{green!100} 1.15  & \cellcolor{green!100} 1.64  & \cellcolor{blue!60} 2.17 \\
\Xhline{1pt}
[HA]$^{+}$  & \cellcolor{blue!60} 1.75  & \cellcolor{white!0} 2.74  &  \cellcolor{white!0} 3.73
&  \cellcolor{blue!60} 1.15 & \cellcolor{blue!60} 1.87  & \cellcolor{white!0} 2.85
& \cellcolor{blue!60} 1.70 & \cellcolor{green!100} 2.26  & \cellcolor{white!0} 3.12 \\
\Xhline{1pt}
[DA]$^{+}$  & \cellcolor{blue!60} 1.73  & \cellcolor{white!0} 2.55  &  \cellcolor{white!0} 3.55
&  \cellcolor{blue!60} 1.17 & \cellcolor{green!100} 1.91  & \cellcolor{white!0} 2.83
& \cellcolor{blue!60} 1.78  & \cellcolor{green!100} 2.47  & \cellcolor{white!0} 3.40 \\
\Xhline{1pt}
[MA]$^{+}$  & \cellcolor{green!100} 1.55  & \cellcolor{green!100} 2.43  &  \cellcolor{white!0} 3.23
&  \cellcolor{green!100} 1.26 & \cellcolor{green!100} 2.00  & \cellcolor{white!0} 3.01
& \cellcolor{green!100} 1.98  & \cellcolor{white!0} 3.01  & \cellcolor{white!0} 4.10 \\
\Xhline{1pt}
[FM]$^{+}$  & \cellcolor{blue!60} 1.76  & \cellcolor{white!0} 2.65  &  \cellcolor{white!0} 3.49
&  \cellcolor{blue!60} 1.32 & \cellcolor{green!100} 2.16  & \cellcolor{white!0} 3.20
& \cellcolor{blue!60} 2.04  & \cellcolor{white!0} 3.08  & \cellcolor{white!0} 4.17 \\
\Xhline{1pt}
[FA]$^{+}$  & \cellcolor{green!100} 1.74  & \cellcolor{white!0} 2.60  &  \cellcolor{white!0} 3.25
&  \cellcolor{green!100} 1.21 & \cellcolor{white!0} 2.54  & \cellcolor{white!0} 3.73
& \cellcolor{blue!60} 2.36  & \cellcolor{white!0} 3.79  & \cellcolor{white!0} 4.95 \\
\Xhline{1pt}
[EA]$^{+}$  & \cellcolor{green!100} 1.86  & \cellcolor{white!0} 2.67  &  \cellcolor{white!0} 3.56
&  \cellcolor{green!100} 1.70 & \cellcolor{white!0} 3.13  & \cellcolor{white!0} 4.47
& \cellcolor{white!0} 2.61  & \cellcolor{white!0} 3.99  & \cellcolor{white!0} 5.11 \\
\Xhline{1pt}
[GA]$^{+}$  & \cellcolor{blue!60} 1.90  & \cellcolor{white!0} 2.70  &  \cellcolor{white!0} 3.99
&  \cellcolor{green!100} 1.78 & \cellcolor{white!0} 3.77  & \cellcolor{white!0} 4.91
& \cellcolor{white!0} 3.28  & \cellcolor{white!0} 4.13  & \cellcolor{white!0} 5.90 \\
\Xhline{1pt}
[DEA]$^{+}$  & \cellcolor{green!100} 1.81  & \cellcolor{white!0} 2.60  &  \cellcolor{white!0} 3.44
&  \cellcolor{green!100} 1.62 & \cellcolor{white!0} 3.19  & \cellcolor{white!0} 4.51
& \cellcolor{white!0} 2.82  & \cellcolor{white!0} 4.15  & \cellcolor{white!0} 5.07 \\

\Xhline{1pt}
\end{tabular}
\end{table}

\begin{table*}
\label{Table S4}
\caption{
Calculated electron (m$_{e}^{*}$, the upper value) and hole (m$_{h}^{*}$, 
the lower value) effective masses of the candidate AM$^{IV}$X$^{VII}_3$ perovskites. 
The green shading indicates the compounds passing the current DM (m$_{e}^{*}$ $<$ 0.5 m$_0$ and m$_{h}^{*}$ $<$ 0.5
 m$_0$), as well as the DMs in Tables S1 and S3. 
The lightblue shading indicates the compounds passing only the current DM.} 
\centering
\begin{tabular}{?c?c?c?c?c?c?c?c?c?c?p{102cm}?}
\Xhline{1pt}
m$_{e}^{*}$ & \multirow { 2}{*}{ PbI$_3$} & \multirow { 2}{*}{PbBr$_3$} & \multirow { 2}{*}{PbCl$_3$} 
& \multirow { 2}{*}{SnI$_3$} & \multirow { 2}{*}{SnBr$_3$} &\multirow { 2}{*}{ SnCl$_3$} 
& \multirow { 2}{*}{GeI$_3$} & \multirow { 2}{*}{GeBr$_3$} & \multirow { 2}{*}{GeCl$_3$}  \\
m$_{h}^{*}$ &  &  &  &  &  &  &  & & \\
\Xhline{1pt}
\multirow { 2}{*}{[M]$^{+}$}  & \cellcolor{blue!60} 0.49 & \cellcolor{white!0} 0.70 & \cellcolor{white!0} 0.92
&  \cellcolor{white!0} 0.59 & \cellcolor{white!0} 0.70  & \cellcolor{white!0} 0.73
& \cellcolor{blue!60} 0.35  & \cellcolor{white!0} 0.52  & \cellcolor{white!0} 0.57 \\
                              & \cellcolor{blue!60} 0.37 & \cellcolor{white!0} 0.36 & \cellcolor{white!0} 0.38
&  \cellcolor{white!0} 0.18 & \cellcolor{white!0} 0.20  & \cellcolor{white!0} 0.16
& \cellcolor{blue!60} 0.24  & \cellcolor{white!0} 0.10  & \cellcolor{white!0} 0.13 \\
\Xhline{1pt}
\multirow { 2}{*}{[Cs]$^{+}$}  & \cellcolor{blue!60} 0.18 & \cellcolor{green!100} 0.23 & \cellcolor{blue!60} 0.38
&  \cellcolor{green!100} 0.24 & \cellcolor{green!100} 0.32  & \cellcolor{green!100} 0.38
& \cellcolor{green!100} 0.19  & \cellcolor{green!100} 0.26  & \cellcolor{blue!60} 0.33 \\
                              & \cellcolor{blue!60} 0.22 & \cellcolor{green!100} 0.21 & \cellcolor{blue!60} 0.21
&  \cellcolor{green!100} 0.17 & \cellcolor{green!100} 0.15  & \cellcolor{green!100} 0.10
& \cellcolor{green!100} 0.21  & \cellcolor{green!100} 0.18  & \cellcolor{blue!60} 0.19 \\
\Xhline{1pt}
\multirow { 2}{*}{[HA]$^{+}$} & \cellcolor{blue!60} 0.46 & \cellcolor{white!0} 0.65 & \cellcolor{white!0} 0.85
&  \cellcolor{blue!60} 0.43 & \cellcolor{white!0} 0.55  & \cellcolor{white!0} 0.63
& \cellcolor{blue!60} 0.33  & \cellcolor{green!100} 0.38  & \cellcolor{blue!60} 0.41 \\
                              & \cellcolor{blue!60} 0.33 & \cellcolor{white!0} 0.46 & \cellcolor{white!0} 0.59
&  \cellcolor{blue!60} 0.19 & \cellcolor{white!0} 0.22  & \cellcolor{white!0} 0.25
& \cellcolor{blue!60} 0.29  & \cellcolor{green!100} 0.17  & \cellcolor{blue!60} 0.24 \\
\Xhline{1pt}
\multirow { 2}{*}{[DA]$^{+}$} & \cellcolor{blue!60} 0.47 & \cellcolor{white!0} 0.67 & \cellcolor{white!0} 0.97
&  \cellcolor{blue!60} 0.42 & \cellcolor{white!0} 0.53  & \cellcolor{white!0} 0.62
& \cellcolor{blue!60} 0.32  & \cellcolor{green!100} 0.37  & \cellcolor{blue!60} 0.47 \\
                              & \cellcolor{blue!60} 0.31 & \cellcolor{white!0} 0.36 & \cellcolor{white!0} 0.47
&  \cellcolor{blue!60} 0.19 & \cellcolor{white!0} 0.21  & \cellcolor{white!0} 0.23
& \cellcolor{blue!60} 0.27  & \cellcolor{green!100} 0.21  & \cellcolor{blue!60} 0.31 \\
\Xhline{1pt}
\multirow { 2}{*}{[MA]$^{+}$} & \cellcolor{green!100} 0.42 & \cellcolor{white!0} 0.70 & \cellcolor{white!0} 0.88
&  \cellcolor{green!100} 0.27 & \cellcolor{green!100} 0.50  & \cellcolor{white!0} 0.58
& \cellcolor{green!100} 0.38  & \cellcolor{white!0} 0.51  & \cellcolor{white!0} 0.70 \\
                              & \cellcolor{green!100} 0.18 & \cellcolor{white!0} 0.28 & \cellcolor{white!0} 0.31
&  \cellcolor{green!100} 0.18 & \cellcolor{green!100} 0.19  & \cellcolor{white!0} 0.17
& \cellcolor{green!100} 0.28  & \cellcolor{white!0} 0.31  & \cellcolor{white!0} 0.37 \\
\Xhline{1pt}
\multirow { 2}{*}{[FM]$^{+}$} & \cellcolor{white!0} 0.54 & \cellcolor{white!0} 1.12 & \cellcolor{white!0} 8.81
&  \cellcolor{white!0} 0.56 & \cellcolor{white!0} 0.66  & \cellcolor{white!0} 4.50
& \cellcolor{blue!60} 0.40  & \cellcolor{white!0} 0.57  & \cellcolor{white!0} 7.83 \\
                              & \cellcolor{white!0} 0.25 & \cellcolor{white!0} 0.36 & \cellcolor{white!0} 0.42
&  \cellcolor{white!0} 0.19 & \cellcolor{white!0} 0.21  & \cellcolor{white!0} 0.22
& \cellcolor{blue!60} 0.30  & \cellcolor{white!0} 0.34  & \cellcolor{white!0} 0.43 \\
\Xhline{1pt}
\multirow { 2}{*}{[FA]$^{+}$}  & \cellcolor{green!100} 0.44 & \cellcolor{white!0} 0.68 & \cellcolor{white!0} 0.54
&  \cellcolor{green!100} 0.33 & \cellcolor{white!0} 0.73  & \cellcolor{white!0} 0.75
& \cellcolor{blue!60} 0.38  & \cellcolor{white!0} 0.57  & \cellcolor{white!0} 0.78 \\
                              & \cellcolor{green!100} 0.23 & \cellcolor{white!0} 0.35 & \cellcolor{white!0} 0.29
&  \cellcolor{green!100} 0.25 & \cellcolor{white!0} 0.31  & \cellcolor{white!0} 0.44
& \cellcolor{blue!60} 0.37  & \cellcolor{white!0} 0.51  & \cellcolor{white!0} 0.73 \\
\Xhline{1pt}
\multirow { 2}{*}{[EA]$^{+}$}  & \cellcolor{green!100} 0.46 & \cellcolor{white!0} 0.60 & \cellcolor{white!0} 0.80
&  \cellcolor{green!100} 0.38 & \cellcolor{white!0} 0.57  & \cellcolor{white!0} 0.82
& \cellcolor{blue!60} 0.40  & \cellcolor{white!0} 0.58  & \cellcolor{white!0} 0.88 \\
                              & \cellcolor{green!100} 0.28 & \cellcolor{white!0} 0.34 & \cellcolor{white!0} 0.41
&  \cellcolor{green!100} 0.14 & \cellcolor{white!0} 0.30  & \cellcolor{white!0} 0.50
& \cellcolor{blue!60} 0.42  & \cellcolor{white!0} 0.61  & \cellcolor{white!0} 0.81 \\
\Xhline{1pt}
\multirow { 2}{*}{[GA]$^{+}$} & \cellcolor{blue!60} 0.39 & \cellcolor{white!0} 0.61 & \cellcolor{white!0} 0.68
&  \cellcolor{green!100} 0.32 & \cellcolor{white!0} 0.85  & \cellcolor{white!0} 0.91
& \cellcolor{white!0} 0.51  & \cellcolor{white!0} 0.55  & \cellcolor{white!0} 0.77 \\
                              & \cellcolor{blue!60} 0.37 & \cellcolor{white!0} 0.35 & \cellcolor{white!0} 0.51
&  \cellcolor{green!100} 0.21 & \cellcolor{white!0} 0.39  & \cellcolor{white!0} 0.54
& \cellcolor{white!0} 0.64  & \cellcolor{white!0} 0.80  & \cellcolor{white!0} 0.95 \\
\Xhline{1pt}
\multirow { 2}{*}{[DEA]$^{+}$}  & \cellcolor{green!100} 0.26 & \cellcolor{blue!60} 0.40 & \cellcolor{white!0} 0.66
&  \cellcolor{green!100} 0.30 & \cellcolor{blue!60} 0.48  & \cellcolor{white!0} 0.61
& \cellcolor{white!0} 0.41  & \cellcolor{white!0} 0.56  & \cellcolor{white!0} 0.57 \\
                              & \cellcolor{green!100} 0.33 & \cellcolor{blue!60} 0.32 & \cellcolor{white!0} 0.36
&  \cellcolor{green!100} 0.20 & \cellcolor{blue!60} 0.29  & \cellcolor{white!0} 0.46
& \cellcolor{white!0} 0.55  & \cellcolor{white!0} 0.73  & \cellcolor{white!0} 0.82 \\
\Xhline{1pt}
\end{tabular}
\end{table*}

\begin{table*}
\label{Table S5}
\caption{
Calculated exciton binding energies (E$_B$ ) of the candidate AM$^{IV}$X$^{VII}_3$ perovskites. 
The green shading indicates the compounds passing the current DM (E$_B$ $<$ 100 meV), 
as well as the DMs in Tables S1, S3 and S4. 
The lightblue shading indicates the compounds passing only the current DM.}
\centering
\begin{tabular}{?c?c?c?c?c?c?c?c?c?c?p{102cm}?}
\Xhline{1pt}
E$_B$ (meV) &  PbI$_3$ & PbBr$_3$ & PbCl$_3$ & SnI$_3$ & SnBr$_3$ & SnCl$_3$ & GeI$_3$ & GeBr$_3$ & GeCl$_3$  \\
\Xhline{1pt}
[M]$^{+}$  & \cellcolor{blue!60} 94.25  & \cellcolor{white!0} 165.80  &  \cellcolor{white!0} 274.85
           & \cellcolor{blue!60} 23.92  & \cellcolor{blue!60} 49.79   & \cellcolor{white!0}  143.66
           & \cellcolor{blue!60} 29.05  & \cellcolor{blue!60} 49.39   & \cellcolor{white!0}  117.68 \\
\Xhline{1pt}
[Cs]$^{+}$  & \cellcolor{blue!60} 43.89  & \cellcolor{green!100} 75.87  &  \cellcolor{blue!60} 68.84
            & \cellcolor{green!100} 13.73  & \cellcolor{green!100} 26.75  & \cellcolor{green!100}  48.69
            & \cellcolor{green!100} 14.29  & \cellcolor{green!100} 26.40  & \cellcolor{blue!60}  47.42 \\
\Xhline{1pt}
[HA]$^{+}$  & \cellcolor{blue!60} 73.51  & \cellcolor{white!0} 187.52 &  \cellcolor{white!0} 347.17
            & \cellcolor{blue!60}20.93  & \cellcolor{blue!60} 48.35  & \cellcolor{white!0}  109.24
            & \cellcolor{blue!60} 46.83  & \cellcolor{green!100} 75.02  & \cellcolor{white!0}   146.80 \\
\Xhline{1pt}
[DA]$^{+}$  & \cellcolor{blue!60} 67.51  & \cellcolor{white!0} 144.02 &  \cellcolor{white!0} 287.43
            & \cellcolor{blue!60}20.75  & \cellcolor{blue!60} 48.01  & \cellcolor{blue!60}  96.39
            & \cellcolor{blue!60} 48.14  & \cellcolor{green!100} 87.14  & \cellcolor{white!0}  183.32 \\
\Xhline{1pt}
[MA]$^{+}$  & \cellcolor{green!100} 52.34  & \cellcolor{white!0} 135.85  &  \cellcolor{white!0} 223.67 
            & \cellcolor{green!100}30.05  & \cellcolor{green!100} 45.93   & \cellcolor{blue!60}   83.12
            & \cellcolor{green!100} 61.63  & \cellcolor{white!0} 127.55  & \cellcolor{white!0}  247.88 \\
\Xhline{1pt}
[FM]$^{+}$  & \cellcolor{blue!60} 33.00  & \cellcolor{white!0} 230.12  &  \cellcolor{white!0} 412.57
            & \cellcolor{blue!60}28.93  & \cellcolor{blue!60} 67.96   & \cellcolor{white!0}  158.42
            & \cellcolor{blue!60} 64.59  & \cellcolor{white!0} 152.64  & \cellcolor{white!0}  460.45 \\
\Xhline{1pt}
[FA]$^{+}$  & \cellcolor{green!100} 56.75  & \cellcolor{white!0} 125.80  &  \cellcolor{white!0} 192.72
            & \cellcolor{green!100} 38.09 & \cellcolor{white!0}  101.64 & \cellcolor{white!0}   282.48
            & \cellcolor{blue!60} 94.22  & \cellcolor{white!0} 274.09  & \cellcolor{white!0}  607.20 \\
\Xhline{1pt}
[EA]$^{+}$  & \cellcolor{white!0} 109.75 & \cellcolor{white!0} 191.90  &  \cellcolor{white!0} 357.38
            &  \cellcolor{green!100}36.26  & \cellcolor{white!0} 115.80  & \cellcolor{white!0}  337.07
            & \cellcolor{white!0} 144.62 & \cellcolor{white!0} 338.33  & \cellcolor{white!0}  632.50 \\
\Xhline{1pt}
[GA]$^{+}$  & \cellcolor{blue!60} 83.06  & \cellcolor{white!0} 153.76  &  \cellcolor{white!0} 313.45
            &  \cellcolor{green!100}38.22  & \cellcolor{white!0} 200.53  & \cellcolor{white!0}  432.95
            & \cellcolor{white!0} 202.35  & \cellcolor{white!0}419.56   & \cellcolor{white!0} 772.06 \\
\Xhline{1pt}

[DEA]$^{+}$  & \cellcolor{green!100} 69.89  & \cellcolor{white!0} 116.20  &  \cellcolor{white!0} 315.05
            &  \cellcolor{green!100}30.94  & \cellcolor{white!0} 123.16  & \cellcolor{white!0}  322.60
            & \cellcolor{white!0} 188.67  & \cellcolor{white!0}408.48   & \cellcolor{white!0} 653.61 \\
\Xhline{1pt}
\end{tabular}
\end{table*}

\begin{table*}
\label{Table S6}
\caption{
Calculated exciton Bohr radii ( $\alpha_{ex}$) of the candidate AM$^{IV}$X$^{VII}_3$ perovskites. 
The green shading indicates the compounds passing the current DM ($\alpha_{ex}$ $>$ 1.5 nm), 
as well as the DMs in Tables S1, S3 and S4. The lightblue shading indicates the compounds passing only the current DM.}
\centering
\begin{tabular}{?c?c?c?c?c?c?c?c?c?c?p{102cm}?}
\Xhline{1pt}
$\alpha_{ex}$ (nm) &  PbI$_3$ & PbBr$_3$ & PbCl$_3$ & SnI$_3$ & SnBr$_3$ & SnCl$_3$ & GeI$_3$ & GeBr$_3$ & GeCl$_3$  \\
\Xhline{1pt}
[M]$^{+}$  & \cellcolor{white!0} 1.40  & \cellcolor{white!0}   1.01  &  \cellcolor{white!0} 0.73
           & \cellcolor{blue!60} 3.60  & \cellcolor{blue!60} 2.40  & \cellcolor{white!0}  1.09
           & \cellcolor{blue!60} 3.26  & \cellcolor{blue!60}   2.41  & \cellcolor{white!0}  1.20 \\
\Xhline{1pt}
[Cs]$^{+}$ & \cellcolor{blue!60}  2.83  & \cellcolor{green!100}  2.07  &  \cellcolor{blue!60} 2.74
           & \cellcolor{green!100}  5.56  & \cellcolor{green!100}3.82  & \cellcolor{green!100}  2.82
           & \cellcolor{green!100}  5.19 & \cellcolor{green!100}   3.72 & \cellcolor{blue!60}   2.66 \\
\Xhline{1pt}
[HA]$^{+}$  & \cellcolor{blue!60} 1.79   & \cellcolor{white!0}  0.91  &  \cellcolor{white!0} 0.58
           &  \cellcolor{blue!60} 4.06   & \cellcolor{blue!60}  2.43 & \cellcolor{white!0}   1.44
           & \cellcolor{blue!60}  2.18 & \cellcolor{green!100}    1.79 & \cellcolor{white!0}   1.18 \\
\Xhline{1pt}
[DA]$^{+}$  & \cellcolor{blue!60} 1.93  & \cellcolor{white!0}   1.14 &  \cellcolor{white!0}  0.68
           &  \cellcolor{blue!60} 4.12 & \cellcolor{blue!60}    2.47 & \cellcolor{blue!60}   1.60
           & \cellcolor{blue!60}  2.21 & \cellcolor{green!100}  1.61 & \cellcolor{white!0}   0.98 \\
\Xhline{1pt}
[MA]$^{+}$  & \cellcolor{green!100} 2.40  & \cellcolor{white!0}   1.19 &  \cellcolor{white!0}  0.86
           &  \cellcolor{green!100} 2.91 & \cellcolor{green!100}  2.62 & \cellcolor{blue!60}   1.99
           & \cellcolor{green!100}  1.89 & \cellcolor{white!0}    1.26 & \cellcolor{white!0}   0.82 \\
\Xhline{1pt}
[FM]$^{+}$  & \cellcolor{blue!60} 3.94  & \cellcolor{white!0}   0.72 &  \cellcolor{white!0}  0.47
           &  \cellcolor{blue!60} 3.17 & \cellcolor{blue!60}  1.88 & \cellcolor{white!0}   1.08
           & \cellcolor{blue!60}  1.86 & \cellcolor{white!0}    1.08 & \cellcolor{white!0}   0.44 \\
\Xhline{1pt}
[FA]$^{+}$  & \cellcolor{green!100} 2.29  & \cellcolor{white!0}   1.30 &  \cellcolor{white!0}  1.00
           &  \cellcolor{green!100} 2.74 & \cellcolor{blue!60}  1.50 & \cellcolor{white!0}   0.72
           & \cellcolor{white!0}  1.45 & \cellcolor{white!0}    0.70 & \cellcolor{white!0}   0.39 \\
\Xhline{1pt}
[EA]$^{+}$  & \cellcolor{white!0}  1.40  & \cellcolor{white!0}   1.00 &  \cellcolor{white!0}   0.66
           &  \cellcolor{green!100}  2.85  & \cellcolor{white!0} 1.36 & \cellcolor{white!0}   0.62
           & \cellcolor{white!0}   1.04  & \cellcolor{white!0}   0.60 & \cellcolor{white!0}   0.39 \\
\Xhline{1pt}
[GA]$^{+}$  & \cellcolor{blue!60}  1.61 & \cellcolor{white!0}    1.12 &  \cellcolor{white!0}  0.67
           &  \cellcolor{green!100}  2.78 & \cellcolor{white!0}  0.92 & \cellcolor{white!0}   0.52
           & \cellcolor{white!0}   0.77 & \cellcolor{white!0}    0.50 & \cellcolor{white!0}   0.32 \\
\Xhline{1pt}
[DEA]$^{+}$  & \cellcolor{green!100} 1.87  & \cellcolor{white!0}  1.41  &  \cellcolor{white!0}  0.75
           &  \cellcolor{green!100}  3.21  & \cellcolor{white!0}1.33   & \cellcolor{white!0}  0.67
           & \cellcolor{white!0}   0.84  & \cellcolor{white!0}  0.51 & \cellcolor{white!0}    0.37 \\
\Xhline{1pt}
\end{tabular}
\end{table*}

\begin{table*}
\label{Table S7}
\caption{
Calculated various DMs for the AM$^{IV}$X$^{VII}_3$ perovskites containing pseudo-halogen anions, 
AM$^{IV}$[BF$_4$]$_3$ and AM$^{IV}$[SCN]$_3$ with A=Cs$^+$, MA$^+$, FA$^+$ and M$^{IV}$=Pb$^{2+}$, Sn$^{2+}$.
}
\centering
\begin{tabular}{?c?c?c?c?c?c?c?c?c?c?p{102cm}?}
\Xhline{1pt}
Materials &  $\Delta$H & E$_{g}^{d}$ & m$_{e}^{*}$ & m$_{h}^{*}$ & E$_B$ (eV)& $\alpha_{ex}$ (nm)   \\
\Xhline{1pt}
 & \multicolumn{5}{r}{Compounds based on [BF$_4$]$^{-}$}  &    \\
\Xhline{1pt}
CsPb[BF$_4$]$_3$  & \cellcolor{white!0}  0.31  & \cellcolor{white!0} 8.71  &  \cellcolor{white!0} 1.60
                  &  \cellcolor{white!0} 1.43  & \cellcolor{white!0} 2.70   & \cellcolor{white!0} 0.14 \\
\Xhline{1pt}

CsSn[BF$_4$]$_3$  & \cellcolor{white!0}  0.32  & \cellcolor{white!0} 8.70  &  \cellcolor{white!0} 1.63
                  &  \cellcolor{white!0} 1.45  & \cellcolor{white!0} 2.58   & \cellcolor{white!0} 0.14 \\
\Xhline{1pt}

MAPb[BF$_4$]$_3$  & \cellcolor{white!0}  0.71  & \cellcolor{white!0} 9.43  &  \cellcolor{white!0} 6.97
                  &  \cellcolor{white!0} 7.70  & \cellcolor{white!0} 12.07   & \cellcolor{white!0}0.03  \\
\Xhline{1pt}

MASn[BF4]3        & \cellcolor{white!0}  0.53  & \cellcolor{white!0} 9.15  &  \cellcolor{white!0} 4.86
                  &  \cellcolor{white!0} 2.93 & \cellcolor{white!0}  5.89  & \cellcolor{white!0}  0.06\\ 
\Xhline{1pt}
 & \multicolumn{5}{r}{Compounds based on [SCN]$^{-}$}  &    \\
\Xhline{1pt}

CsPb[SCN]$_3$     & \cellcolor{white!0}  0.39  & \cellcolor{white!0} 4.42  &  \cellcolor{white!0} 1.82
                  &  \cellcolor{white!0} 1.12  & \cellcolor{white!0} 0.79   & \cellcolor{white!0} 0.26 \\
\Xhline{1pt}
CsSn[SCN]$_3$     & \cellcolor{white!0}  0.34  & \cellcolor{white!0} 4.97  &  \cellcolor{white!0} 1.86
                  &  \cellcolor{white!0} 1.12  & \cellcolor{white!0} 0.71   & \cellcolor{white!0} 0.28  \\
\Xhline{1pt}
MAPb[SCN]$_3$     & \cellcolor{white!0}  -0.33  & \cellcolor{white!0}4.46   &  \cellcolor{white!0} 2.38
                  &  \cellcolor{white!0} 1.20   & \cellcolor{white!0}0.93    & \cellcolor{white!0} 0.23 \\
\Xhline{1pt}
MASn[SCN]$_3$     & \cellcolor{white!0}  -0.17  & \cellcolor{white!0}4.75   &  \cellcolor{white!0} 2.59
                  &  \cellcolor{white!0}  1.01  & \cellcolor{white!0}0.72    & \cellcolor{white!0} 0.27  \\
\Xhline{1pt}

FAPb[SCN]$_3$     & \cellcolor{white!0}  -0.47  & \cellcolor{white!0}4.61   &  \cellcolor{white!0} 1.74
                  &  \cellcolor{white!0}  1.24  & \cellcolor{white!0}0.81    & \cellcolor{white!0} 0.26 \\
\Xhline{1pt}

FASn[SCN]$_3$     & \cellcolor{white!0}  -0.38  & \cellcolor{white!0}4.68   &  \cellcolor{white!0} 2.31
                  &  \cellcolor{white!0} 1.02 & \cellcolor{white!0}  0.69  & \cellcolor{white!0}    0.28\\
\Xhline{1pt}

\end{tabular}
\end{table*}
